\def\paperversion{tr}
\def\anonymoussubmission{1} \newif\ifsinglecolumn\singlecolumnfalse
\newif\ifwidemargins\widemarginsfalse
\newif\ifwarning\warningfalse
\newif\ifshowcomments\showcommentsfalse
\newif\ifblinded\blindedfalse
\newif\ifshowlinenums\showlinenumsfalse
\newif\ifreport\reportfalse
\newif\ifcopyrightspace\copyrightspacefalse
\newif\ifacknowledgments\acknowledgmentsfalse
\newif\ifshowpagenumbers\showpagenumberstrue
\newif\iffinalformat\finalformatfalse
\newif\ifweb\webfalse
\newif\ifexternalize\externalizetrue

\ifx\anonymoussubmission\undefined
\else
  \if\anonymoussubmission 1
    \blindedtrue
  \else
    \blindedfalse
  \fi
\fi
\IfFileExists{.blinded}{\blindedtrue}

\ifx\paperversion\xxxxundefined
\PackageError{paperversions}{*** No valid document version was specified.
Macro paperversions must be defined as one of (markup, draft,
local, submission, final, web, tr, trdraft)}
\fi

\def\xxversion{\csname xx\paperversion\endcsname}
\newif\ifsawversion\sawversionfalse

\ifcase\xxversion\relax
    \widemarginstrue
    \singlecolumntrue
    \warningtrue
    \showlinenumstrue
    \sawversiontrue
    \blindedfalse
\or \warningtrue
    \showlinenumsfalse
    \showcommentstrue
    \sawversiontrue
\or \warningtrue
    \showcommentsfalse
    \blindedfalse
    \sawversiontrue
    \acknowledgmentstrue
    \showlinenumstrue
\or \sawversiontrue
    \showlinenumstrue
\or \blindedfalse
    \sawversiontrue
    \copyrightspacetrue
    \acknowledgmentstrue
    \showpagenumbersfalse
    \finalformattrue
\or \singlecolumntrue
    \blindedfalse
    \sawversiontrue
    \reporttrue
    \acknowledgmentstrue
    \webtrue
\or \blindedfalse
    \singlecolumntrue
    \showcommentstrue
    \sawversiontrue
    \reporttrue
    \showlinenumstrue
    \warningtrue
\or \blindedfalse
    \showcommentstrue
    \copyrightspacetrue
    \acknowledgmentstrue
    \sawversiontrue
    \finalformattrue
    \showlinenumstrue
    \warningtrue
\or \blindedfalse
    \sawversiontrue
    \copyrightspacetrue
    \acknowledgmentstrue
    \finalformattrue
    \webtrue
\or \singlecolumntrue
    \blindedtrue
    \acknowledgmentsfalse
    \sawversiontrue
    \reporttrue
    \webtrue
\or \blindedfalse
    \sawversiontrue
    \copyrightspacetrue
    \acknowledgmentstrue
    \finalformattrue
    \externalizefalse
    \webtrue
\fi

\ifsawversion
\else
\fi

\let\xxversion=\undefined
 
\newif\ifsecondround\secondroundtrue 

\secondroundfalse

\ifsecondround
  \showlinenumstrue
\fi

\newcommand\snd[1]{\ignorespaces
  \ifsecondround
    {\color{blue}#1}
  \else
    #1\fi
 \ignorespaces
}
 
\ifreport
    \documentclass[reprint,acmsmall,screen]{acmart}
\else
    \iffinalformat
        \ifweb
            \documentclass[acmsmall,screen]{acmart}
        \else
            \ifsecondround
                \documentclass[acmsmall,screen,review]{acmart}
            \else
                \documentclass[acmsmall,screen]{acmart}
            \fi
        \fi
    \ccsdesc[500]{Software and its engineering~Control structures}
    \ccsdesc[500]{Software and its engineering~Semantics}

    \else
        \ifblinded
            \documentclass[acmsmall,anonoymous,screen,review]{acmart}
        \else
            \documentclass[acmsmall,screen]{acmart}
        \fi
    \fi
\fi
\usepackage{acmart-overrides}
\usepackage{overrides}
 
\usepackage{graphicx,eso-pic}
 \usepackage{amsmath,amssymb,stmaryrd,tex-macros/utf8math}
\usepackage[inline]{enumitem}
\usepackage[small]{tex-macros/ttquot}
\usepackage[T1]{fontenc}
\usepackage{placeins,multirow,multicol}
\usepackage{lipsum}
\usepackage{microtype}
\usepackage{textpos}
\usepackage{stackengine}

\usepackage{wrapfig}

\usepackage{subcaption}
\usepackage{etoolbox}

\definecolor{faint-gray}{gray}{0.985}
\definecolor{light-gray}{gray}{0.75}
\definecolor{dark-gray}{gray}{0.25}
\definecolor{san-marino}{RGB}{68,108,179}
\definecolor{blue-gray}{HTML}{778ca3}
\definecolor{pomegranate}{RGB}{192,57,43}
\definecolor{pure-apple}{HTML}{6ab04c}
\definecolor{exodus-fruit}{HTML}{686de0}
\definecolor{middle-blue}{HTML}{7ED6DF}
\definecolor{soaring-eagle}{HTML}{95AFC0}
\definecolor{light-indigo}{HTML}{7158E2}
\definecolor{bright-lilac}{HTML}{CD84F1}
\definecolor{cream}{HTML}{FFFFCC}
\definecolor{confetti}{HTML}{E9D460}
\definecolor{malibu}{HTML}{6BB9F0}
\definecolor{chambray}{HTML}{3a539b}
\definecolor{scampi}{HTML}{736598}
\definecolor{studio}{HTML}{8e44ad}
\definecolor{salem}{HTML}{1e824c}

\newenvironment{centered}{\centering
}{\par
}

\usepackage{hyperref}
\hypersetup{
  colorlinks,
  urlcolor=black,
  citecolor=black,
  filecolor=black
  linkcolor=black,
}

\usepackage{cleveref}
  \crefformat{section}{\mbox{Section #2#1#3}}
  \crefformat{appendix}{\mbox{Appendix #2#1#3}}
  \crefformat{subsection}{\mbox{Section #2#1#3}}
  \crefrangeformat{section}{\mbox{Sections #3#1#4--#5#2#6}}
  \crefrangeformat{appendix}{\mbox{Appendices #3#1#4--#5#2#6}}
  \crefrangeformat{subsection}{\mbox{Sections #3#1#4--#5#2#6}}
  \crefrangeformat{subsubsection}{\mbox{Sections #3#1#4--#5#2#6}}
  \crefmultiformat{section}{\mbox{Sections #2#1#3}}{ and \mbox{#2#1#3}}
  {, \mbox{#2#1#3}}{, and \mbox{#2#1#3}}
  \crefmultiformat{appendix}{\mbox{Appendices #2#1#3}}{ and \mbox{#2#1#3}}
  {, \mbox{#2#1#3}}{, and \mbox{#2#1#3}}
  \crefmultiformat{subsection}{\mbox{Sections #2#1#3}}{ and \mbox{#2#1#3}}
  {, \mbox{#2#1#3}}{, and \mbox{#2#1#3}}
  \crefmultiformat{subsubsection}{\mbox{Sections #2#1#3}}{ and \mbox{#2#1#3}}
  {, \mbox{#2#1#3}}{, and \mbox{#2#1#3}}

  \crefname{table}{Table}{Table}

  \crefformat{figure}{Figure~#2#1#3}
  \crefrangeformat{figure}{\mbox{Figures #3#1#4--#5#2#6}}
  \crefmultiformat{figure}{\mbox{Figures #2#1#3}}{ and \mbox{#2#1#3}}
  {, \mbox{#2#1#3}}{, and \mbox{#2#1#3}}

  \crefformat{equation}{(#1)}
  \crefrangeformat{equation}{\mbox{(#1)--(#2)}}

\newsavebox\shiftbox
\newsavebox\resetbox
\newsavebox\fresetbox
\sbox\shiftbox{\hbox{\lower 0.5pt \hbox{\scalebox{0.050}[0.050]{\includegraphics{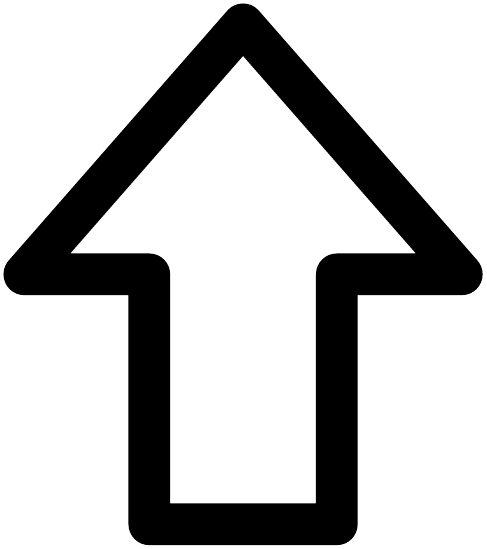}}}}}
\sbox\resetbox{\hbox{\raise 7.4pt \hbox{\scalebox{0.050}[-0.050]{\includegraphics{upward-arrow}}}}}
\sbox\fresetbox{\hbox{\lower 0.5pt \hbox{\scalebox{0.050}[0.050]{\includegraphics{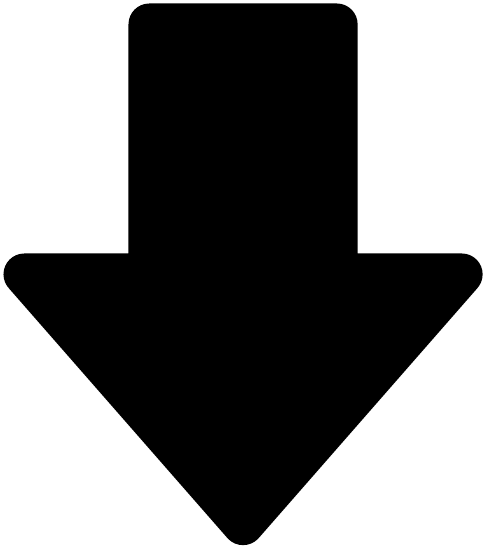}}}}}
\newcommand\shift{\usebox\shiftbox}
\newcommand\reset{\usebox\resetbox}
\newcommand\freset{\usebox\fresetbox}

\usepackage{formalism/formalism}

\definecolor{keyword-color}{HTML}{6633CC}
\definecolor{keyword-color-1}{HTML}{148F77}
\definecolor{keyword-color-2}{HTML}{1465B6}
\definecolor{keyword-color-3}{HTML}{CA6F1E}
\colorlet{codecomment-color}{blue-gray}
\colorlet{string-color}{dark-gray}
\usepackage{listings,lstautogobble}
  \let\origthelstnumber\thelstnumber
  \makeatletter
  \newcommand*\Suppressnumber{\lst@AddToHook{OnNewLine}{\let\thelstnumber\relax \advance\c@lstnumber-\@ne\relax }}
  \newcommand*\Reactivatenumber{\lst@AddToHook{OnNewLine}{\let\thelstnumber\origthelstnumber \advance\c@lstnumber\@ne\relax}}

\newcommand\codecomment[1]{\color{codecomment-color}\textrm{\itshape//~#1}
}

\definecolor{codeemph-color}{HTML}{ECF5F8}
\newcommand\codeemph[2][]{\ifthenelse{\isempty{#1}}{\colorlet{color}{black}}{\colorlet{color}{#1}}\setlength\fboxsep{1pt}\colorbox{codeemph-color}{\textcolor{color}{\mbox{#2}}}}

\lstdefinelanguage{JavaScript}{
    keywords={typeof, new, true, false, catch, function, return, null, catch, switch, var, const, let, async, await, if, in, while, do, else, case, break, from},
    ndkeywords={class, export, boolean, throw, implements, import, this},
    sensitive=false,
    comment=[l]{//},
    morecomment=[s]{/*}{*/},
    morestring=[b]',
    morestring=[b]"
}

\definecolor{csharp-keyword-color}{HTML}{000000}
\colorlet{csharp-comment-color}{blue-gray}
\lstdefinestyle{csharp-style}{
    language=[Sharp]C,
    morekeywords={yield,async,await,var},
    basicstyle={
      \fontsize{8}{8}
      \ttfamily\color{csharp-keyword-color}
    },
    keywordstyle=\color{csharp-keyword-color}\bfseries,
    identifierstyle=\color{black},
    commentstyle=\color{csharp-comment-color}\rmfamily\itshape,
    stringstyle=\color{gray},
}

\newcommand{\dadada}{\textnormal{\textcolor{black}{...}}}

\newcommand\BracketsXY{\textls[-45]{[X,\!Y]}}
\newcommand\EVAR[1][45]{\textls[-#1]{[\textcolor{keyword-color-1}{α}]}}
\newcommand\EVar[1][45]{\textls[-#1]{[α]}}
\newcommand\HVAR[1]{\textcolor{blue-gray}{\textsf{H}$_\texttt{\textls[-30]{\textsf{#1}}}$}}
\newcommand\HVar[1]{\textsf{H}$_\texttt{\textls[-30]{\textsf{#1}}}$}
\newcommand\SELFVAR{\textcolor{blue-gray}{\textsf{self}}}
\newcommand\VOID[1][35]{\textls[-#1]{\textcolor{keyword-color-1}{void}}}
\newcommand\RAISES[1][45]{\textls[-#1]{\textcolor{keyword-color}{\;\,raises\,\;}}}
\newcommand\STRING[1][45]{\textls[-#1]{String}}

\newcommand\PONG[1][35]{\textls[-#1]{\textcolor{keyword-color-2}{Pong}}}
\newcommand\ASYNC[1][35]{\textls[-#1]{\textcolor{keyword-color-2}{Async}}}
\newcommand\DEF[1][35]{\textls[-#1]{\textcolor{keyword-color}{def}}}

\makeatletter
  \renewcommand\@secfont{\bfseries\sffamily\section@raggedright\MakeTextUppercase}
  \renewcommand\@subsecfont{\bfseries\sffamily}
  \renewcommand\@subsubsecfont{\bfseries\sffamily}
  \renewcommand\@parfont{\bfseries\sffamily}
  \renewcommand\paragraph{\@startsection{paragraph}{4}{0pt}{-.5\baselineskip \@plus -2\p@ \@minus -.2\p@}{-3.5\p@}{\ACM@NRadjust{\@parfont\@adddotafter}}
  }
\makeatother

\def\thetitle{Handling Bidirectional Control Flow}
\ifreport
\title{\thetitle: Technical Report}
\else
\title{\thetitle}
\fi

\ifblinded
\else
    \author{Yizhou Zhang}
    \affiliation{
        \department{Cheriton School of Computer Science}
        \institution{University of Waterloo}
        \streetaddress{200 University Avenue West}
        \city{Waterloo}
        \state{ON}
        \postcode{N2L~3G1}
        \country{Canada}
    }
    \email{yizhou@uwaterloo.ca}

    \author{Guido Salvaneschi}
    \affiliation{
        \institution{University of St. Gallen}
        \streetaddress{Rosenbergstrasse 51}
        \city{9000 St. Gallen}
        \country{Switzerland}
    }
    \email{guido.salvaneschi@unisg.ch}

    \author{Andrew C. Myers}
    \affiliation{
        \department{Department of Computer Science}
        \institution{Cornell University}
        \streetaddress{Gates Hall}
        \city{Ithaca}
        \state{NY}
        \postcode{14853}
        \country{USA}
    }
    \email{andru@cs.cornell.edu}
\fi

\ifcopyrightspace
    \setcopyright{acmlicensed}
    \acmJournal{PACMPL}
    \acmYear{2020}
    \acmVolume{4}
    \acmNumber{OOPSLA}
    \acmArticle{139}
    \acmMonth{11}
    \acmPrice{}
    \acmDOI{10.1145/3428207}
\else
    \setcopyright{none}
    \renewcommand\footnotetextcopyrightpermission[1]{}
\fi

\ifwarning
\AtBeginDocument{
 \AddToShipoutPicture*{\put(418,750){\it\color{red}{\fbox{\large Draft---please do not distribute}}}}
}
\fi

\newsavebox\crunchedbox
\def\vcrunch#1{\savebox\crunchedbox{#1}\smash{\usebox\crunchedbox}{\vrule width 0pt height 0.9em}}

\begin{document}
\begin{abstract}
Pressed by the difficulty of writing asynchronous, event-driven
code, mainstream languages have recently been
building in support for a variety of advanced control-flow features.
Meanwhile, experimental language designs have suggested
effect handlers as a unifying solution to programmer-defined
control effects, subsuming exceptions, generators, and
async--await.
However, despite these trends, complex control flow---in
particular, control flow that exhibits a bidirectional
pattern---remains challenging to manage.

We introduce _bidirectional algebraic effects_, a new programming 
abstraction that supports bidirectional control transfer in a
more natural way.
Handlers of bidirectional effects can raise further effects to
transfer control back to the site where the initiating effect was
raised, and can use themselves to handle their own effects.
We present applications of this expressive power, which falls out
naturally as we push toward the unification of effectful
programming with object-oriented programming.
We pin down the mechanism and the unification formally using a core language that
makes generalizations to effect operations and effect handlers.

The usual propagation semantics of control effects such as exceptions
conflicts with modular reasoning in the presence of effect polymorphism---it breaks parametricity. Bidirectionality exacerbates the problem.
Hence, we set out to show the core language, which builds on the
existing tunneling semantics for algebraic effects, is not only
type-safe (no effects go unhandled), but also
abstraction-safe (no effects are _accidentally_ handled).
We devise a step-indexed logical-relations model, and construct
its parametricity and soundness proofs.
These core results are fully mechanized in Coq.
While a full-featured compiler is left to future work,
experiments show that as a first-class language feature,
bidirectional handlers can be implemented efficiently.

 \end{abstract}

\maketitle

\section{Introduction}
\label{sec:intro}

Modern software places new demands on programming languages. In particular, the need
to interact with high-latency external entities---users, file systems, databases, and
geodistributed systems---has led software to become
increasingly event-driven. Callback functions are a conventional
pattern for event-driven programming, but unconstrained callbacks
become complex and hard to reason about as applications grow.
Hence, it is currently in vogue for programming languages
to build in support for advanced control-flow transfer features like
generators and async--await. These features support more structured
programming of asynchronous, event-driven code.

Meanwhile, algebraic effects~\cite{pp2003,pp2013,eff-lang} have emerged
as a powerful alternative that allows programmers to
define their own control effects. They
subsume a wide range of features including exceptions,
generators, and async--await.
Compared to the monadic approach to effects, algebraic effects compose
naturally without requiring awkward monad transformers, and enjoy
a nice separation between the syntax (i.e., a set of effect
operations) and the semantics (i.e., handling of those operations).

However, even with these advanced language features at hand,
programmers today still find certain complex control-flow
patterns painful to manage.
As we argue, features found in mainstream languages
are not expressive enough to capture bidirectional control
transfer without losing the desirable guarantee that all effects are
handled, and existing language designs of algebraic effects
cannot readily express this bidirectionality without falling
back to patterns that algebraic effects are intended to help avoid.

To resolve these challenges, we generalize the idea of algebraic
effects.
With algebraic effects, effectful code initiates control transfer
by raising effects that propagate up the dynamic call stack to
their handlers.
With _bidirectional algebraic effects_, effect handlers can raise
subsequent effects that propagate in the opposite direction, to
the site where the initiating effect was raised.
This bidirectionality makes it easy to transmit information and
control, to and fro, between program fragments.
Accordingly, the type system requires the invocation site of an
effect operation to handle not only the initiating effect, but
also the reverse-direction effects.
{\bf All effects are guaranteed to be handled.}

The usual propagation semantics of control effects is known to
interfere with abstraction boundaries in
the presence of effect polymorphism, because higher-order functions can
intercept effects they are not supposed to handle~\cite{exceptions-pldi16,bpps2018,zm18,clmm2020}.
A possible concern might be that bidirectional propagation would
further muddle the problem, leading to effect-polymorphic abstractions
being violated in previously unidentified ways.
To address this concern, we provide bidirectional
algebraic effects with a semantics that respects abstraction
boundaries, and we rigorously substantiate this strong abstraction claim.
{\bf All effects are guaranteed not to be accidentally handled.}

Bidirectionality and the safety guarantees fall out naturally when a language
designer views algebraic effects through an object-oriented lens.
In fact, the enabling and most visible language change is a
generalization of effect operations to make them appear like
methods:
the notion of an effect
operation is extended to allow it to declare further effects its
handling code may raise---just as methods in Java can
declare exceptions their implementations may throw.
Accordingly,
handlers of bidirectional effects, which we call {\it bidirectional
handlers}, are generalized to make them appear like objects.
In particular, a {\it self}\/ handler,
analogous to the {\it self}\/ reference found in object-oriented
languages, is brought into the context of a handler definition.
Self-reference makes a bidirectional effect handler
a fixpoint definition---it can ask that its own effects be
handled by itself.

\snd{
The complexity of bidirectional control flow is innate to
many modern software applications; bidirectional algebraic effects do
not simply make this complexity disappear.
Instead, the static guarantees afforded by the type system enable
programmers to reason compositionally about bidirectional
control flow and therefore to manage complex control flow more easily.
}

\paragraph{Contributions}

The dynamic behavior of bidirectionality is attainable in many
languages in various ways---this paper does not aim to
rediscover bidirectional control flow.
Rather, the contributions consist in
\begin{enumerate*}[label=(\roman*)]
\item a language-design recipe that allows integrating
bidirectional control effects in a sound, unified, and efficient
way, addressing a variety of programming challenges, and in
\item formal developments that capture the essence of the
mechanism and that establish strong guarantees about it,
putting the mechanism on a sound theoretical footing.
\end{enumerate*}
We proceed as follows:

\begin{itemize}[topsep=1ex,leftmargin=4.5ex]
\item \cref{sec:challenges}
  examines some control-flow features in mainstream languages and
  reviews algebraic effects, identifying opportunities to improve
  support for bidirectional control transfer.

\item \cref{sec:baf-examples} demonstrates the new programming
  abstraction informally (in the setting of a typical
  object-oriented language) using its various applications,
  interspersed with discussions on design issues.

\item \cref{sec:abstraction-safety} shows that, importantly,
  an abstraction-safe mechanism for bidirectional effects allows
  programmers to reason compositionally about correctness.

\item \cref{sec:core} defines a core language, \lamlang,
  capturing the informally introduced features and unification
  from previous sections. It gives an operational semantics and a
  static semantics.

\item \cref{sec:results} continues the formal developments with a
  logical-relations model for \lamlang, culminating in proofs of
  coveted properties including type safety and parametricity.
  These formal results are fully mechanized using the Coq proof
  assistant.
  \ifblinded The mechanization is available as part of the
  supplemental material.\fi

\item \cref{sec:compile} discusses compilation issues.
  Experimental results on hand-translated examples argue for supporting
  bidirectional handlers as a first-class language feature.

\item \cref{sec:related} discusses related work in more detail, and
  \cref{sec:conclusions} concludes.

\end{itemize}

\section{Background: Async--Await, Generators, and Algebraic Effects}
\label{sec:challenges}

\snd{
Complex, asynchronous, bidirectional control flow is already a reality
for programmers today.
This section identifies real-world programming challenges
involving bidirectional control flow and shows how existing mechanisms
fall short in addressing them.
}

\paragraph{Async--Await with Promises}

An array of languages---for example,
C\#~\cite{brmmt2012},
JavaScript~\cite{ECMAScript2018},
Rust~\cite{Rust-async-proposal}, and
Swift~\cite{Swift-async-proposal}---have recently added, or are planning to add, support for async--await
and the accompanying _promises_ abstraction~\cite{promises}, also
known as _futures_ or _tasks_.

\begingroup
As an example, consider the C\# program in \cref{fig:async-csharp}.
Method \texttt{\small Http\-Get\-Json}
(lines~\ref{line:csharp-httpgetasync-begin}--\ref{line:csharp-httpgetasync-end})
sends an HTTP GET request to retrieve a web page by
asynchronously running \texttt{\small Http\-Get}
(line~\ref{line:csharp-httpget-begin}), and
\setlength{\columnsep}{3.5ex}\BeforeBeginEnvironment{wrapfigure}{\setlength{\intextsep}{2.2ex}}\begin{wrapfigure}[12]{r}{50.8ex}
\begin{minipage}{50.5ex}
\colorlet{codecomment-color}{csharp-comment-color}
\begin{lstlisting}[style=csharp-style,numbers=left]
static byte[] HttpGet(String url);(*\label{line:csharp-httpget-begin}*)
static async Task<Json> HttpGetJson(String url) {(*\label{line:csharp-httpgetasync-begin}*)
  Task<Json> t = Task.Run(() => HttpGet(url));
  byte[] bytes = await t;(*\label{line:csharp-await-1}*)
  return JsonParse(bytes);(*\label{line:csharp-implicit-wrap-return}*)
}(*\label{line:csharp-httpgetasync-end}*)
static async Task Main() {
  Task<Json> t = HttpGetJson("xyz.org");(*\label{line:csharp-call-httpgetasync}*)
  (*\dadada*) (*\codecomment{do things that do not depend on the query result}*)(*\label{line:csharp-await-irrelevant}*)
  Json json = await t;(*\,\codecomment{block execution until query terminates}*)(*\label{line:csharp-await-2}*)
  (*\dadada*)
}
\end{lstlisting}
\end{minipage}
\setlength{\abovecaptionskip}{-.2ex}
\caption{Using async--await in C\#}
\label{fig:async-csharp}
\end{wrapfigure}converts the raw bytes into JSON format.
Because "HttpGetJson" is declared "async",
calling it (line~\ref{line:csharp-call-httpgetasync}) does not block
computations that do not depend on the result of the request
(line~\ref{line:csharp-await-irrelevant}).
The programmer _awaits_ the task when they need the result to be
ready.
Sending HTTP GET requests may raise exceptions (e.g., due to connection
issues); the reasonable point for such an exception to emerge is
where the tasks are awaited (lines~\ref{line:csharp-await-1}
and~\ref{line:csharp-await-2}).
While "await" sends a signal to a task scheduler on the .NET runtime
stack, the exceptions appear to propagate in the opposite
direction, from the .NET runtime to the "await" sites.

However, existing languages that support async--await do
not enforce at compile time that exceptions raised by
asynchronous computations are handled.
The lack of this static assurance makes asynchronous programming error-prone.
For example, the C\# compiler accepts the program above
without requiring that an exception handler be
provided---if the asynchronous query does result in an exception, the
program crashes.
The situation is worse in JavaScript: an exception raised
asynchronously is silently swallowed if not otherwise caught.
Such unhandled exceptions have been identified as a common
vulnerability in JavaScript programs~\cite{azmt2018}.

\endgroup

It is no surprise that C\# and JavaScript do not check asynchronously
raised exceptions, since neither statically checks exceptions in the first place,
unlike Java~\cite{java-11ed}---which is unfortunate because in practice, most exceptions arising
from C\# code are undocumented~\cite{cm07b}. However, since
asynchronously raised exceptions do not propagate through the regular
exception mechanism, it is not clear how to do this checking even in
languages like Java that do have checked exceptions.

\paragraph{Generators}

Coroutine-style iterators, usually called generators,
are a convenient construct available in many
languages~\cite{lsas,icon,csharp2,python-lang-ref,Sather-iterators,ruby,ECMAScript2018}.
They help avoid the verbose, error-prone pattern of maintaining
complex state machines inside _iterator objects_ as seen in
Alphard~\cite{alphard-gen2} and more recently in Java.

However, a weakness of generators is that they do not allow
clients to concurrently modify the underlying collections or streams
being iterated over.
A client iterating over a priority queue might want to change the
priority of a received element; similarly,
a client iterating over a stream of database records might want to
remove one of those records from the database.
Generators in the mentioned languages lack the
expressiveness to solve these programming challenges. In such
languages, the programmer either resorts to implementing iterators as
even more complex state machines, or simply shies away from defining
powerful, reusable iterator abstractions.

\paragraph{Algebraic Effects}

Algebraic effects~\cite{pp2003,pp2013} are a powerful unifying
language feature that can express exceptions, generators,
async--await, and other related control-flow mechanisms including
coroutines and delimited control~\cite{eff-lang,koka-lang,klo2013,dehmsw2017,fklp2017,baseem2018}.
The hallmark of algebraic effects is adding support for
_signatures_ for control effects and for _handlers_ as
implementations of these signatures.

\begin{figure}
\centering
\newbox\boxa
\newbox\boxb
\newbox\boxc
\begin{lrbox}{\boxa}\begin{minipage}{24.5ex}
\begin{lstlisting}[linewidth=\textwidth]
effect Yield[X] {
  def yield(X) : void
}
\end{lstlisting}
\end{minipage}
 \end{lrbox}
\begin{lrbox}{\boxb}\begin{minipage}{34.5ex}
\begin{lstlisting}[numbers=left,linewidth=\textwidth,lineskip=0.29ex]
class Node[X] {
  var head : X
  var tail : Node[X]
  (*\dadada*)
  def iter() : (*\label{line:iterate-start}*)void raises Yield[X] {(*\label{line:iterate-raises}*)
    yield(head)(*\label{line:List-yield}*)
    if (tail != null)
      tail.iter()(*\label{line:List-l-iter}*)
  }(*\label{line:iterate-end}*)
}
\end{lstlisting}
\end{minipage}
 \end{lrbox}
\begin{lrbox}{\boxc}\begin{minipage}{24.5ex}
\begin{lstlisting}[linewidth=\textwidth]
try { node.iter() }
with yield(x) {
  print(x)
  resume()
}
\end{lstlisting}
\end{minipage}
 \end{lrbox}
\def\figa{\usebox\boxa}
\def\figb{\usebox\boxb}
\def\figc{\usebox\boxc}
\def\capa{Effect signature "Yield"}
\def\capb{Iterator raises "Yield"}
\def\capc{Client code handles "Yield"}
\savestack{\capfiga}{\subcaptionbox{\capa\label{fig:yield-sig}}{\figa}}\savestack{\capfigb}{\subcaptionbox{\capb\label{fig:iter}}{\figb}}\savestack{\capfigc}{\subcaptionbox{\capc\label{fig:iter-client}}{\figc}}\hfill
\stackon[0.9ex]{\capfigc}{\capfiga}
\hfill
\capfigb
\hfill
\setlength{\abovecaptionskip}{.5ex}
\caption{
Yielding iterators via algebraic effects
}
\label{fig:iter-as-effects}
\end{figure}

\begin{figure}
\renewcommand\thesubfigure{\arabic{subfigure}}
\def\figa{
  \begin{minipage}{6.8ex}
  \vbox to 22.0ex {
    \includegraphics[scale=0.29]{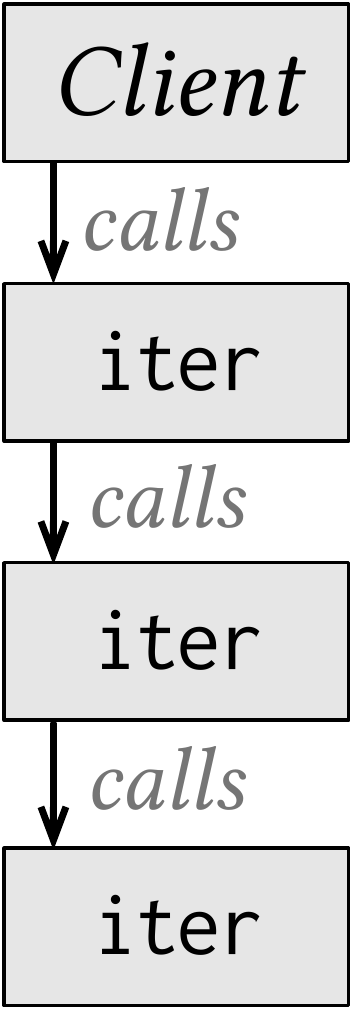}
  }
  \end{minipage}
}
\def\figb{
  \begin{minipage}{18.0ex}
  \vbox to 22.0ex {
    \includegraphics[scale=0.29]{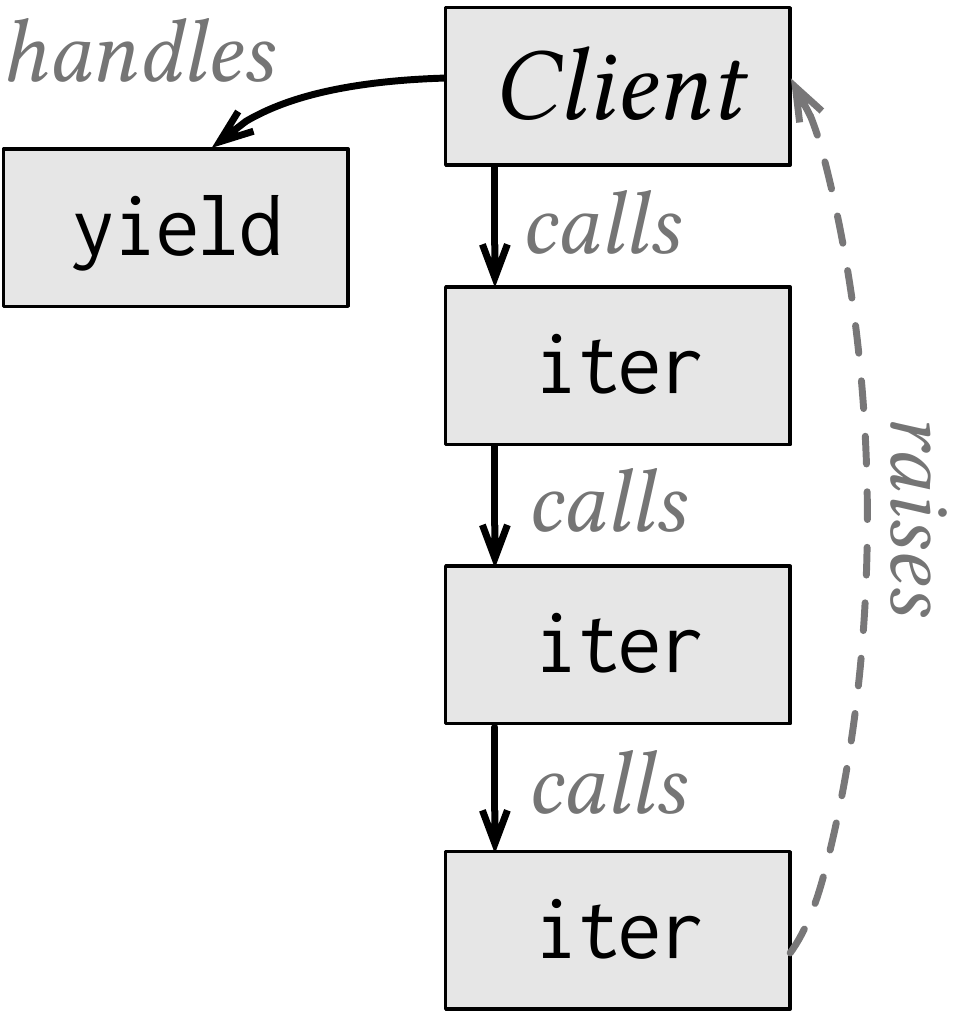}
  }
  \end{minipage}
}
\def\figc{
  \begin{minipage}{18.0ex}
  \vbox to 22.0ex {
    \includegraphics[scale=0.29]{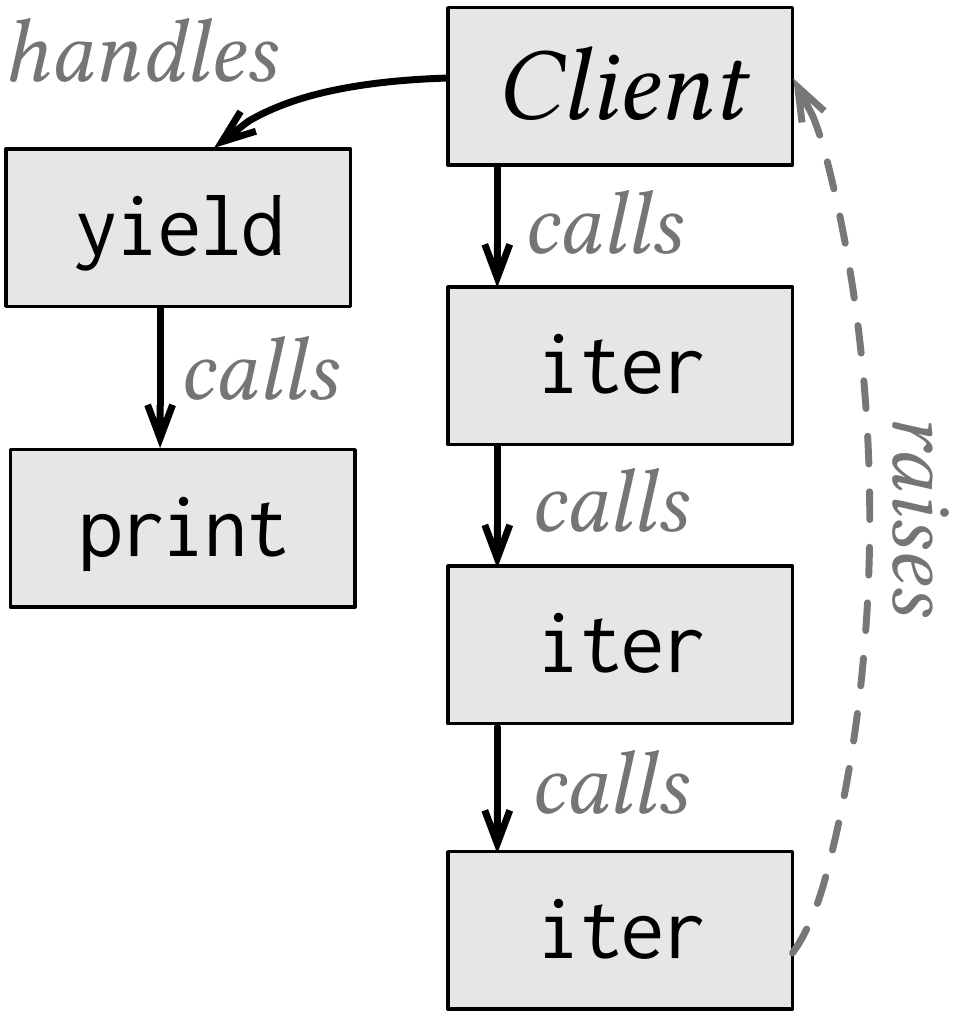}
  }
  \end{minipage}
}
\def\figd{
  \begin{minipage}{6.8ex}
  \vbox to 22.0ex {
    \includegraphics[scale=0.29]{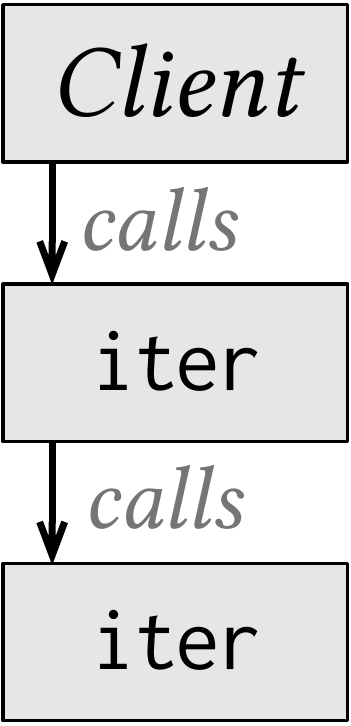}
  }
  \end{minipage}
}
\subcaptionbox{\label{fig:yield-stack-1}}{\figd}
\hfill
\subcaptionbox{\label{fig:yield-stack-2}}{\figa}
\hfill
\subcaptionbox{\label{fig:yield-stack-3}}{\figb}
\hfill
\subcaptionbox{\label{fig:yield-stack-4}}{\figc}
\hfill
\subcaptionbox{\label{fig:yield-stack-5}}{\figb}
\hfill
\subcaptionbox{\label{fig:yield-stack-6}}{\figa}
\caption{
Stack snapshots of the program in \cref{fig:iter-as-effects}.
}
\label{fig:iterate-stack}
\end{figure}

An effect signature defines one or more _operations_.
For example, the signature in \cref{fig:yield-sig}, named "Yield"
and parameterized by a type variable~"X",
contains exactly one operation, "yield".
The operation takes as argument a value of type~"X".

Lines~\ref{line:iterate-start}--\ref{line:iterate-end} of
\cref{fig:iter} uses "Yield" to
define a coroutine-style iterator for nodes in a linked list: it
recursively iterates over the tail after yielding the head.
Invoking an effect operation raises the corresponding effect:
because "iter" invokes "yield" (line~\ref{line:List-yield}),
calling "iter" can raise the "Yield" effect.
Static checking of effects requires this effect be part of the
method's type, in its "raises" clause (line~\ref{line:iterate-raises}).

The client program in \cref{fig:iter-client} traverses a chain of nodes
by calling "iter" and handling its "Yield" effect.
Effectful computations are enclosed by "try\,\textnormal{...}\,with",
followed by a handler that implements the effect operations.
Each time that the effect "yield" is raised, the recursive
iterator computation in \cref{fig:iter} is suspended and control transferred to the
handler in \cref{fig:iter-client}, which prints the yielded
element. Control then resumes in the iterator.
The resulting execution is similar to using a generator in C\#, Python, or Ruby.
The sequence of stack frames that result is shown in \cref{fig:iterate-stack}.

Handlers can resume computations suspended by the raising of
effects, by calling the resumption denoted by the special "resume"
function.
This "resume" function is essentially a delimited
continuation~\cite{evalctxts}. It takes as input the result of
the effect operation.
The call to "resume" in \cref{fig:iter-client} takes no
argument because the result type of "yield" is "void".

\snd{
\cref{fig:iterate-stack} visualizes the control flow in one iteration using stack
diagrams, with each diagram capturing the stack at a single point
in time:
\begin{enumerate}[topsep=.5ex]
\item
  The iterator has finished processing the first two elements of
  the list (hence the two "iter" frames).
\item
  A third "iter" frame is created; the iterator begins to
  process the third element.
\item
  The iterator raises a "Yield" effect.
  The effect is then caught by the client's handler.
\item
  The client prints the yielded element.
\item
  Printing finishes.
\item
  Handling of "Yield" finishes.
  Control is returned to the iterator.
\end{enumerate}
}

The handler in \cref{fig:iter-client} abbreviates
the full signature of the effect operation.
The expanded form is shown below.
We will write handlers mostly in the abbreviated syntax.

\begin{centered}
\begin{minipage}{.305\textwidth}
\begin{lstlisting}[frame=none,backgroundcolor=\color{white}]
try { node.iter() }
with yield(x : X) : void { (*\dadada*) }
\end{lstlisting}
\end{minipage}
\end{centered}

In many practical uses of algebraic effects, as in the example above, invoking "resume" is the last
action performed by an effect handler. We call this a _tail
resumption_ and call such handlers _tail-resumptive_. Not all effect
handlers are tail-resumptive; for example, exception handlers are
typically _abortive_: they do not resume the computation that raised the
effect.  Handlers that are either tail-resumptive or abortive can
be compiled to efficient code because there is no need to save stack
frames once "resume" is invoked or the handler aborts~\cite{leijen2017}.

Although algebraic effects subsume generators, they do not
address the limitations of generators outlined earlier:
handler code cannot raise an effect transferring control back to
the iterator code to perform a concurrent modification to the data structure.
Similarly, algebraic effects can express async--await, but they are
awkward when exceptions can be raised asynchronously: a handler
running asynchronous computations cannot propagate exceptions raised
by those computations back to the "await" site.
Beyond  generators and async--await, there are other
interesting control-flow applications
that algebraic effects cannot yet readily support.
What is needed is a unified mechanism that can express all these
programming challenges easily.

\section{Bidirectional Algebraic Effects, Informally}
\label{sec:baf-examples}

We generalize algebraic effects to offer the missing flexibility:
handlers of effect operations can themselves raise effects
that are handled by callers of the effect operations.
Before defining a formal semantics in \cref{sec:core},
we first introduce the mechanism informally via examples
written in a syntax similar to that of Java, Scala or Kotlin,
although the ideas could apply to many other languages, especially those
with an object-oriented flavor.

\subsection{Generators with Concurrent Modification}
\label{sec:coroutine-iterators-bidirectional}

\begin{figure}
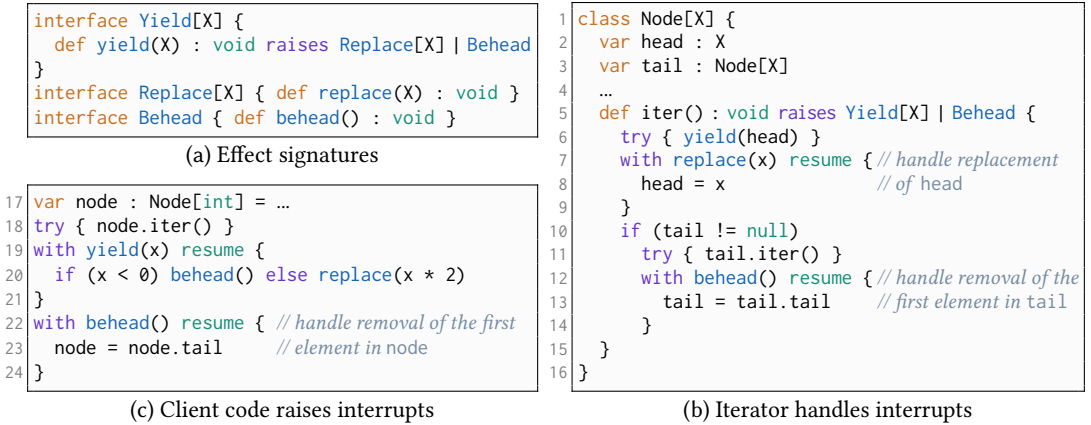

\centering
\newbox\boxa
\newbox\boxb
\newbox\boxc
\begin{lrbox}{\boxa}\begin{minipage}{44.0ex}
\begin{lstlisting}[linewidth=\textwidth,lineskip=.3ex]
interface Yield[X] {
  def yield(X) : void raises Replace[X](*\;|\;*)Behead
}
interface Replace[X] { def replace(X) : void }
interface Behead { def behead() : void }
\end{lstlisting}
\end{minipage}
 \end{lrbox}
\begin{lrbox}{\boxb}\begin{minipage}{44.2ex}
\begin{lstlisting}[numbers=left,linewidth=\textwidth,lineskip=.23ex]
class Node[X] {
  var head : X
  var tail : Node[X]
  (*\dadada*)
  def iter()(*\;*):(*\;*)void(*\RAISES*)Yield[X](*\;|\;*)Behead(*\label{line:interrupt-iter-raises}*) {
    try { yield(head) }(*\label{line:interrupt-iter-yield}*)
    with replace(x) resume {(*\,\codecomment{handle replacement}*)(*\label{line:interrupt-replace-handler-start}*)
      head = x              (*\,\codecomment{of \texttt{head}}*)
    }(*\label{line:interrupt-replace-handler-end}*)
    if (tail != null)
      try { tail.iter() }(*\label{line:interrupt-iter-recurse}*)
      with behead() resume {(*\,\codecomment{handle removal of the}*)(*\label{line:interrupt-iter-behead-handler-start}*)
        tail = tail.tail    (*\,\codecomment{first element in \texttt{tail}}*)
      }(*\label{line:interrupt-iter-behead-handler-end}*)
  }
}
\end{lstlisting}
\end{minipage}
 \end{lrbox}
\begin{lrbox}{\boxc}\begin{minipage}{44.0ex}
\begin{lstlisting}[numbers=left,firstnumber=17,linewidth=\textwidth,lineskip=.3ex]
var node : Node[int] = (*\dadada*)
try { node.iter() }
with yield(x)(*\label{line:interrupt-yield-handler-start}*) resume(*\label{line:interrupt-yield-resume}*) {
  if (x < 0) behead() else replace(x * 2)
}(*\label{line:interrupt-yield-handler-end}*)
with behead() resume { (*\codecomment{handle removal of the first}*)(*\label{line:interrupt-client-behead-handler-start}*)
  node = node.tail     (*\codecomment{element in \texttt{node}}*)
}(*\label{line:interrupt-client-behead-handler-end}*)
\end{lstlisting}
\end{minipage}
 \end{lrbox}
\def\figa{\usebox\boxa}
\def\figb{\usebox\boxb}
\def\figc{\usebox\boxc}
\def\capa{Effect signatures}
\def\capb{Iterator handles interrupts}
\def\capc{Client code raises interrupts}
\savestack{\capfiga}{\subcaptionbox{\capa\label{fig:interrupt-sig}}{\figa}}\savestack{\capfigb}{\subcaptionbox{\capb\label{fig:interrupt-iter}}{\figb}}\savestack{\capfigc}{\subcaptionbox{\capc\label{fig:interrupt-client}}{\figc}}\stackon[1.5ex]{\capfigc}{\capfiga}
\hfill
\capfigb
\setlength{\abovecaptionskip}{.5ex}
\caption{Yielding iterators with reverse-direction interrupts
for replacing and removing yielded
elements\label{fig:interruptible}}
\end{figure}

We want to extend the iterator abstraction of \cref{fig:iter-as-effects}
so that iterator clients can issue interrupts to request that the
yielded element be replaced or removed. Note that implementing iterators that support
such concurrent modifications is awkward in
standard OO languages~\cite{jmatch2}.
We start by changing the signature of "Yield", as shown in
\cref{fig:interrupt-sig}.
Apart from being defined as an "interface" (the reason for which
will soon become clear), this signature differs from the one in
\cref{fig:yield-sig} by declaring that "yield" may itself raise two
additional effects, "Replace" and "Behead", corresponding to the
two kinds of concurrent modifications that client code can request.
(A "raises" clause may include multiple effects, separated by
vertical bars.)
Allowing effect operations to declare their own "raises" clauses is
a key generalization we make to accommodate
bidirectionality.

With the modified "Yield" effect, the client code in
\cref{fig:interrupt-client} is able to remove negative integers
from a list and to double the non-negative ones even while iterating over
the list:
the handler (lines~\ref{line:interrupt-yield-handler-start}--\ref{line:interrupt-yield-handler-end})
passes to the resumption a _computation_, which invokes either
operation "behead" or "replace" based on the integer yielded.
Notice that "resume" takes as input a computation, rather than a
value, as signified by the use of curly braces instead of parentheses.

The resumption accepts a computation whose type and effects must
match the result type and effects of the effect operation.
For example,
in \cref{fig:interrupt-client}, the resumption to a "yield"
call (line~\ref{line:interrupt-yield-resume}) accepts a
computation that may raise "Replace[int]" and "Behead".

\begin{figure}
\renewcommand\thesubfigure{\arabic{subfigure}}
\def\figa{
  \begin{minipage}{6.8ex}
  \vbox to 24.0ex {
    \includegraphics[scale=0.29]{figures/iter-1.pdf}
  }
  \end{minipage}
}
\def\figb{
  \begin{minipage}{18.0ex}
  \vbox to 24.0ex {
    \includegraphics[scale=0.29]{figures/iter-2.pdf}
  }
  \end{minipage}
}
\def\figc{
  \begin{minipage}{18.0ex}
  \vbox to 24.0ex {
    \includegraphics[scale=0.29]{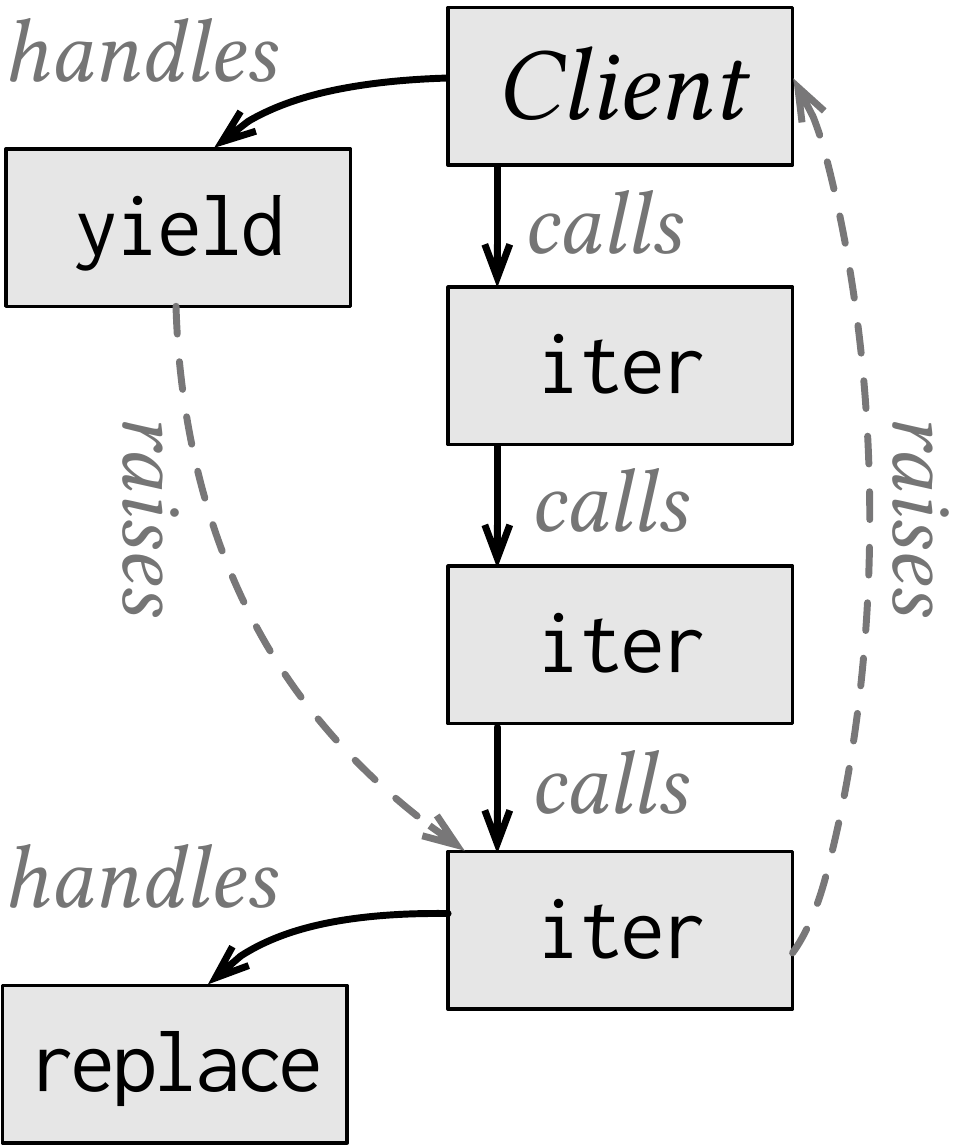}
  }
  \end{minipage}
}
\def\figd{
  \begin{minipage}{6.8ex}
  \vbox to 24.0ex {
    \includegraphics[scale=0.29]{figures/iter-0.pdf}
  }
  \end{minipage}
}
\subcaptionbox{}{\figd}
\hfill
\subcaptionbox{}{\figa}
\hfill
\subcaptionbox{\label{fig:replace-stack-3}}{\figb}
\hfill
\subcaptionbox{\label{fig:replace-stack-4}}{\figc}
\hfill
\subcaptionbox{\label{fig:replace-stack-5}}{\figb}
\hfill
\subcaptionbox{\label{fig:replace-stack-6}}{\figa}
\caption{
Stack snapshots of the program in \cref{fig:interruptible}.
}
\label{fig:interrupt-iter-stack}
\end{figure}

Meanwhile, the type system guarantees that the resumption---the
suspended computation in the iterator---contains handlers for
both effects, so that invoking operations "replace" and "behead"
can cause control to safely transfer back to the handlers in
the iterator.
The new iterator code in \cref{fig:interrupt-iter} differs from
\cref{fig:iter} in adding these two handlers.

To handle "Replace", the handler
(lines~\ref{line:interrupt-replace-handler-start}--\ref{line:interrupt-replace-handler-end},
\cref{fig:interrupt-iter}) updates the head value of the list, and
then resumes what is left off by the raising of "replace" in the
"yield" handler (line~\ref{line:interrupt-yield-resume},
\cref{fig:interrupt-client}).
What is left to be done there is to resume what is left off by the
raising of "yield" in the iterator code
(line~\ref{line:interrupt-iter-yield}, \cref{fig:interrupt-iter}).

\snd{
The stack diagrams in \cref{fig:interrupt-iter-stack} visualize the control
flow:
\begin{enumerate}[topsep=.5ex]
\item
  The iterator has finished processing the first two elements of
  the list.
\item
  A third "iter" frame is created; the iterator begins to
  process the third element.
\item
  The iterator raises a "Yield" effect.
  The effect is then caught by the client's handler.
\item
  The client issues a reverse-direction interrupt to ask that the
  third element of the list be replaced.
  This "Replace" effect is caught by the iterator.
\item
  Handling of "Replace" finishes.
  The iterator returns control to the client's "Yield" handler.
\item
  Handling of "Yield" finishes.
  The client returns control to the iterator.
\end{enumerate}
}
\noindent
Because effect "Replace" appears to propagate _down_ the stack, in the reverse
direction of "Yield", we call these bidirectional effects.

However, it is largely unnecessary to look at stack diagrams to
understand control flow.
A more meaningful interpretation is to view both "Yield"
and "Replace" as propagating outward through evaluation contexts to
callers.
A "Yield" effect raised by "iter" propagates to the caller of
"iter";
in the context of the "iter" method, this caller is represented
by "return", the return address of "iter".
Similarly, a "Replace" effect raised by the "yield" handler
propagates to the caller of "yield";
in the context of the "yield" handler, this caller is represented
by "resume", the computation to be resumed after "yield" is
handled.

In each of the handlers in \cref{fig:interruptible}, "resume"
envelops the entire handler computation.
In this common case, we allow eliding the curly braces
surrounding "resume\,\{\,\dadada\,\}".
For example, the handler on
lines~\ref{line:interrupt-yield-handler-start}--\ref{line:interrupt-yield-handler-end}
is desugarred to the following syntax:

\begin{centered}
\begin{minipage}{.305\textwidth}
\begin{lstlisting}[frame=none,backgroundcolor=\color{white}]
try { node.iter() }
with yield(x) { resume { (*\dadada*) } }
\end{lstlisting}
\end{minipage}
\end{centered}

The "Behead" interrupt must be handled differently than
"Replace", because removing a node from a linked list is a
nonlocal update---it is most appropriately done at a level that
``owns'' the current list, that is, either the preceding node in
the list (\cref{fig:interrupt-iter}) or the client code (\cref{fig:interrupt-client}).
\snd{
Notice that in this example, the client code, in addition to
the iterator code, must be prepared to handle "Behead".
The client code refers to a linked list by holding a reference to
the first node of the list.
So when the first node is ``beheaded'', it is only natural to
expect the client code to handle this event specially.
If list nodes could be accessed only indirectly (as in Java,
through a "LinkedList" object), handling of "Behead" could be
hidden from client code.
}

Instead of handling "Behead" immediately after it surfaces
from the call site of "yield" (as we did to "Replace"),
"Behead" is propagated to the call site that triggers the iteration
of the current list.
Hence, "Behead" occurs in the "raises" clause of "iter"
(line~\ref{line:interrupt-iter-raises}, \cref{fig:interrupt-iter}),
as static checking of effects entails.
Accordingly, the type system requires the two call sites of
"iter" to
deal with "Behead". Both handle "behead" by replacing the
reference to a list with its tail
(line~\ref{line:interrupt-iter-behead-handler-start}--\ref{line:interrupt-iter-behead-handler-end},
\cref{fig:interrupt-iter}
and
line~\ref{line:interrupt-client-behead-handler-start}--\ref{line:interrupt-client-behead-handler-end},
\cref{fig:interrupt-client}).

Although the control flow is complex, reasoning about it remains
tractable, especially because the static checking of
bidirectional effects offers guidance on how to program the
control flow.

\paragraph{An economy of language constructs}

As the syntax and the semantics suggest, bidirectionality makes
effect signatures and ordinary object interfaces become nearly
indistinguishable:
both effect operations and object methods can raise effects, and
effects always propagate to the caller.
This correspondence motivates their unification as a single
language construct; throughout the paper, we define effect
signatures as interfaces.
Moreover, as we introduce later, every bidirectional handler has
access to a self handler that it can use to handle effects,
analogous to how every object has access to a self (receiver) object on
which it can make method calls.

This unification is not merely a syntactic pun. We pin down
this unification in the core language (\cref{sec:core}),
which further allows methods and handlers to be defined
using the same construct:
an ordinary method definition can be viewed as an effect handler
where the entire method body is passed to a tail resumption
(cf., "return").

Nevertheless, in the surface language we distinguish them
to allow for a familiar programming experience where
"return" statements retain their idiomatic meaning in methods---that is, "return" signifies the act of returning to the
caller rather than the resumption per se, and method definitions
with a "void" return type need not have explicit "return"
statements.

\paragraph{Raising effects within handlers}
\label{sec:effects-from-within-handlers}

In existing languages, exceptions (and algebraic effects) raised
within handlers are propagated to the local context of the
handler, rather than to the handler resumption.
Bidirectional algebraic effects are compatible with this
semantics: so long as the computation raising the effect is not
passed to a handler resumption that, per the "raises" clause, can
handle the effect, the normal handling behavior is obtained.

\paragraph{Workarounds}

\snd{
One might think that an alternative to bidirectional effects
would be to make "yield" return a value of some algebraic
data type (ADT) indicating the interrupt event and for the client to
pattern-match on the returned ADT value.
Note that the "try"--"with" syntax is an entirely
cosmetic choice made to match the Java-like surface language;
in fact, algebraic-effects designs for functional languages often
use a syntax similar to pattern-matching ADTs:
a handler case-analyzes the result of an effectful computation.
Consequently, using ADTs as return types of effect operations
would not noticeably clarify the code,
but it _would_ reduce expressive power: control could not be
transferred back to the client code after "Replace" or
"Behead" were handled.
It would also be syntactically heavier-weight: one would have to
convert ADT values to algebraic effects. For comparison,
\cref{fig:ADT} shows how the "iter" code would look if "yield"
returned an ADT value.
}

A more general way is to make a "yield" handler resume with a callback value
that is the thunked handler computation.
But it also means callers of "yield" must voluntarily comply with the
contract by remembering to force the thunk---control transfer via callbacks is a pattern
algebraic effects are intended to help avoid.
Moreover, this approach can be rather inefficient;
\cref{sec:compile} explores the performance implications when
it is used to compile bidirectional effects.

\subsection{Async--Await with Exceptions}
\label{sec:async-exceptions}

We want to use algebraic effects to express both async--await and
asynchronously raised exceptions, while statically ensuring that all
exceptions are handled.

\begin{figure}
\centering
\newbox\boxa
\newbox\boxb
\begin{lrbox}{\boxa}\begin{minipage}{43.8ex}
\begin{lstlisting}[linewidth=\textwidth,lineskip=.33ex]
interface Exn[X] { def exn(X) : void }
interface Async {
  def async[X,Y](Fun2[X,Y]) : Promise[X,Y]
  def await[X,Y](Promise[X,Y]) : X(*\RAISES*)Exn[Y]
}
type Fun2[X,Y] = ()(*\;→\;*)X(*\RAISES*)Exn[Y](*\;*)|(*\;*)Async
\end{lstlisting}
\end{minipage}
 \end{lrbox}
\begin{lrbox}{\boxb}\begin{minipage}{43.0ex}
\begin{lstlisting}[linewidth=\textwidth,lineskip=.33ex]
class Promise[X,Y] {
  var state(*\;*):(*\;*)Sum[List[Awaiter[X,Y]],Fun1[X,Y]]
  Promise() { this.state = inl([]) }
}
type Awaiter[X,Y] = Fun1[X,Y](*\;→\;*)void
type Fun1[X,Y] = ()(*\;→\;*)X(*\RAISES*)Exn[Y]
\end{lstlisting}
\end{minipage}
 \end{lrbox}
\def\figa{\usebox\boxa}
\def\figb{\usebox\boxb}
\def\capa{Effect signatures "Exn" and "Async"}
\def\capb{Definition of the "Promise" structure}
\savestack{\capfiga}{\subcaptionbox{\capa\label{fig:async-sig}}{\figa}}\savestack{\capfigb}{\subcaptionbox{\capb\label{fig:async-aux}}{\figb}}\capfiga \hfill \capfigb
\caption{
Type-level definitions for expressing exceptional async--await
}
\label{fig:async-type-level}
\end{figure}

\begin{figure}
\centering
\newbox\boxc
\newbox\boxd
\newbox\boxe
\newbox\boxf
\newbox\boxg
\begin{lrbox}{\boxc}\begin{minipage}{.459\textwidth}
\begin{lstlisting}[numbers=left,lineskip=.30ex]
def httpGet(String)(*\;*):(*\;*)byte[](*\RAISES*)Exn[Http](*\;*)(*\Suppressnumber*)
(*\Reactivatenumber*)
def httpGetJson(url(*\,*):(*\,*)String)(*\;*):(*\;*)Json(*\RAISES*)
    Async(*\,*)|(*\,*)Exn[Http] { (*\codecomment{asynchronous method}*)
  val p = async(fun()(*\;*)→(*\;*)httpGet(s))(*\label{line:async-promise-1}*)
  val bytes = await(p)(*\label{line:async-await-promise-1}*)
  return jsonParse(bytes)
}(*\Suppressnumber*)
(*\Reactivatenumber*)
def main()(*\,*):(*\,*)void(*\RAISES*)Async {
  val p = async(fun()(*\,*)→(*\,*)(*\textls[-15]{httpGetJson}*)("(*\textls[-45]{\textcolor{string-color}{xyz.org}}*)"))(*\label{line:async-promise-2}*)
  (*\dadada*) (*\codecomment{do things that do not depend on the query result}*)
  try { val json = await(p); (*\dadada*) }
  with exn(http) {(*\;\dadada\;*)}(*\label{line:async-handle-exn}*)
}
\end{lstlisting}
\end{minipage}
        \end{lrbox}
\begin{lrbox}{\boxd}\begin{minipage}{.459\textwidth}
\begin{lstlisting}[numbers=left,lineskip=.30ex,firstnumber=14]
val loop = new EventLoop()
loop.run(main)
\end{lstlisting}
\end{minipage}
      \end{lrbox}
\begin{lrbox}{\boxe}\begin{minipage}{.496\textwidth}
\begin{lstlisting}[numbers=left,linewidth=\textwidth,lineskip=.20ex,firstnumber=16]
class EventLoop {
  val jobs(*\;*):(*\;*)Queue[()(*\;→\;\VOID*)]

  EventLoop() {(*\;*)this.jobs = new Queue()(*\;*)}(*\label{line:async-new-empty-queue}*)

  def run(f(*\;*):(*\;*)()(*\;→\;\VOID*)(*\RAISES*)Async)(*\;*):(*\;\VOID*) {
    handleAsync(f)(*\label{line:async-handle-main}*)
    while (true) {
      try {
        jobs.dequeue().apply() (*\codecomment{run next queued job}*)(*\label{line:async-dequeue}*)
      } with exn(nse) {
        continue (*\codecomment{queue is empty; keep polling it}*)
      }
    }
  }

  def handleAsync(f(*\,*):(*\,*)()(*\,→\,\VOID\RAISES*)Async)(*\;*):
      void {(*\;\dadada\;*)}             (*\,\codecomment{\cref{fig:async-handler}}*)

  def exec[X,Y](f2(*\;*):(*\;*)Fun2[X,Y], p(*\;*):(*\;*)Promise[X,Y])(*\;*):
      void(*\RAISES*)Async {(*\;\dadada\;*)} (*\codecomment{\cref{fig:async-exec}}*)
}
\end{lstlisting}
\end{minipage}
         \end{lrbox}
\begin{lrbox}{\boxf}\begin{minipage}{.496\textwidth}
\begin{lstlisting}[numbers=left,linewidth=\textwidth,lineskip=.25ex,firstnumber=38]
def handleAsync(f(*\;*):(*\;*)()(*\;→\;*)void(*\RAISES*)Async)(*\;*):(*\;\VOID*) {
  try { f() }
  with async(*\BracketsXY*)(f2(*\;*):(*\;*)Fun2(*\BracketsXY*))(*\;*):(*\;*)Promise(*\BracketsXY*) {(*\label{line:Async-async-begin}*)
    val p = new Promise(*\BracketsXY*)()
    __new_thread {
      handleAsync(fun()(*\;→\;*)exec(f2,(*\;*)p))(*\label{line:Async-handle-exec}*)
    }
    resume { p }
  }(*\label{line:Async-async-end}*)
  with await(*\BracketsXY*)(p(*\,*):(*\,*)Promise(*\BracketsXY*))(*\,*):(*\,*)X(*\RAISES*)Exn[Y](*\,*){(*\label{line:Async-await-begin}*)
    match (p.state) {
    | inl(awaiters) ⇒(*\label{line:await-pending-begin}*)
        awaiters.add(fun(f1)(*\;→\;*)resume { f1() })(*\label{line:await-pending-end}*)
    | inr(f1) ⇒(*\label{line:await-complete-begin}*)
        resume { f1() }(*\label{line:await-complete-end}*)
    }
  }(*\label{line:Async-await-end}*)
}
\end{lstlisting}
\end{minipage}
 \end{lrbox}
\begin{lrbox}{\boxg}\begin{minipage}{.459\textwidth}
\begin{lstlisting}[numbers=left,linewidth=\textwidth,lineskip=.25ex,firstnumber=56]
def exec[X,Y](f2(*\,*):(*\,*)Fun2[X,Y],(*\;*)p(*\,*):(*\,*)Promise[X,Y])(*\;*):
    void(*\RAISES*)Async {
  val f1(*\;*):(*\;*)Fun1[X,Y]
  try {(*\label{line:async-f-begin}*)
    val x = f2()
    f1 = fun()(*\;→\;*)x
  } with exn(y) {
    f1 = fun()(*\;→\;*)exn(y)
  }(*\label{line:async-f-end}*)
  jobs.enqueue(fun()(*\;→\;*){(*\label{line:async-callback-schedule-begin}*)
    match (p.state) {
    | inl(awaiters) ⇒
        p.state = inr(f1)(*\label{line:async-transition-complete}*)
        for (awaiter in awaiters)(*\label{line:async-awaiter-schedule-begin}*)
          jobs.enqueue(fun()(*\;→\;*)awaiter(f1))(*\label{line:async-awaiter-f1}*)(*\label{line:async-awaiter-schedule-end}*)
    | inr(_) ⇒ assert(false) (*\codecomment{impossible}*)
    }
  })(*\label{line:async-callback-schedule-end}*)
}
\end{lstlisting}
\end{minipage}
          \end{lrbox}
\def\figc{\usebox\boxc}
\def\figd{\usebox\boxd}
\def\fige{\usebox\boxe}
\def\figf{\usebox\boxf}
\def\figg{\usebox\boxg}
\def\capc{User program with effect "Async"}
\def\capd{Running "main" in an event loop}
\def\cape{Event loop}
\def\capf{"Async" handler}
\def\capg{\textls[1]{Helper: executes~"f2" and memoizes the result in~"p"}}
\savestack{\capfigc}{\subcaptionbox{\capc\label{fig:async-client}}{\figc}}\savestack{\capfigd}{\subcaptionbox{\capd\label{fig:async-runqueue}}{\figd}}\savestack{\capfige}{\subcaptionbox{\cape\label{fig:async-queue}}{\fige}}\savestack{\capfigf}{\subcaptionbox{\capf\label{fig:async-handler}}{\figf}}\savestack{\capfigg}{\subcaptionbox{\capg\label{fig:async-exec}}{\figg}}\stackon[1.92ex]{\stackon[1.92ex]{\capfigg}{\capfigd}}{\capfigc}
\hfill
\stackon[1.00ex]{\capfigf}{\capfige}
\caption{
Using and handling exceptional async--await.
Asynchronously raised exceptions are back-propagated to "await"
sites in the user program.
\snd{
Compared with the C\# program in \cref{fig:async-csharp},
the added static checking requires the user program in
\cref{fig:async-client} to handle
asynchronously raised exceptions, but otherwise
adds no essential syntactic overhead compared to
\cref{fig:async-csharp}.
\cref{fig:async-runqueue,fig:async-queue,fig:async-handler,fig:async-exec}
implements the runtime that handles asynchrony.
}
}
\label{fig:async-runtime}
\end{figure}

Exceptions are expressed through the "Exn" effect, defined in
\cref{fig:async-sig}.
Its operation "exn" takes as input a union of tags, which are
instances of singleton classes indicating particular exceptional
conditions.
For example, we use "Exn[Http]" for exceptions that occurred when processing HTTP
requests,
"Exn[NSE]" for no-such-element exceptions raised when an
empty queue is polled,
and "Exn[Http|NSE]" for the union of the two exceptional
conditions.

The "Async" effect has two operations, "async" and "await".
Both operations are parameterized by two type variables, one denoting
the result type of the asynchronously running computation, and the
other the kind of exception it may raise.
Operation "async" takes as input a computation and returns a
promise: the computation is scheduled to run by an "Async"
handler, and when it finishes, its result, which is either a
value or an exception, is memoized by the promise.
Awaiting the promise either gives back the value or raises the exception.

The "Async" signature is recursive in that operation "async"
accepts a computation whose effects can include not only "Exn[Y]"
but also "Async".
This type-level recursion is useful because it allows for
promises that await other promises, a usage pattern found in many
JavaScript and C\# programs
(including the program in \cref{fig:async-csharp}).

\paragraph{Using exceptional async--await}

The C\# program in \cref{fig:async-csharp} can be ported to
use this "Async" effect, as \cref{fig:async-client} shows.
It has the same run-time behavior, but stronger static
checking.
Because method "httpGet" may raise "Exn[Http]",
the promises on lines~\ref{line:async-promise-1}
and~\ref{line:async-promise-2} have types "Promise[byte[],Http]"
and "Promise[Json,Http]" respectively, and thus
awaiting them may raise "Exn[Http]".
The type system then requires a handler for this asynchronous
exception to be provided---all exceptions, asynchronously raised
or not, are guaranteed to be handled.

The type-level recursion in "Async" allows
invoking operation "async" with a computation that has effect
"Async" (line~\ref{line:async-promise-2}), capturing the fact
that the resulting promise awaits another one
(line~\ref{line:async-await-promise-1}).

\snd{
We remark that in \cref{fig:async-runtime}, only
\cref{fig:async-client} is user-level code, showing that we
add no essential syntactic burden compared to \cref{fig:async-csharp}.
The rest of \cref{fig:async-runtime} implements the runtime that
handles asynchrony, with probably reasonable and excusable
complexity.
}

\paragraph{The promises abstraction}

Like JavaScript promises, promises are in one of
three possible states, expressed using a "Sum" type in
\cref{fig:async-aux}.
A promise is either
\begin{enumerate*}[label={(\arabic*)}]
\item pending completion with a list of awaiters that will run
after the promise is complete,
\item is complete with a value, or
\item is complete with an exception.
\end{enumerate*}
Promises are initialized to pending completion with no awaiters.
The two completion states are expressed via a function of type
"()\;→\;X\;raises\;Exn[Y]", or "Fun1[X,Y]" for short.
The awaiters are resumptions to calls to "await"; they are
higher-order functions taking as input a function of
type "Fun1[X,Y]".
Whereas the "Async" handler in the language runtime can create
and inspect promises directly,
user programs are supposed to introduce and eliminate promises
only indirectly via operations "async" and "await".

\paragraph{Handling exceptional async--await}

Typically a scheduler for asynchronous computations exists in the
language's runtime, as is the case with JavaScript and C\#
(although an algebraic-effects encoding makes it possible for
software components to handle their own "Async" effects).
We present in
\cref{fig:async-queue,fig:async-handler,fig:async-exec} a possible implementation
of the runtime in a style similar to those of JavaScript (e.g.,
Node.js~\cite{Node}),
which maintains a queue of jobs run by an _event loop_.
Asynchronous computations of the "main" program is run in this
event loop, as \cref{fig:async-runqueue} suggests.

Initially, the queue is empty (line~\ref{line:async-new-empty-queue}),
and the "main" program is run inside an "Async" handler
(line~\ref{line:async-handle-main}) that handles all requests to
start asynchronous computations and to await their results.
New jobs are enqueued on completion of asynchronous computations
(lines~\ref{line:async-callback-schedule-begin}--\ref{line:async-callback-schedule-end}).
The queued jobs are then run in the event loop (line~\ref{line:async-dequeue}).
For simplicity, we use FIFO scheduling.

\cref{fig:async-handler} defines the "Async" handler.
To handle "async", the handler
creates a new promise,
creates a thread using a "__new_thread" intrinsic,
and returns the promise
(lines~\ref{line:Async-async-begin}--\ref{line:Async-async-end}).
The new thread executes the computation~"f2" asynchronously by
calling a helper function "exec", defined in
\cref{fig:async-exec}.
It stores the result of~"f2" into a function~"f1" that represents a
_control-stuck_ computation---invoking~"f1" either immediately
returns a value or immediately raises an exception
(lines~\ref{line:async-f-begin}--\ref{line:async-f-end}).
Lines~\ref{line:async-callback-schedule-begin}--\ref{line:async-callback-schedule-end}
then schedule the events that should happen after the
asynchronous computation's result is ready: they include
transitioning the promise into one of the two completion states
(line~\ref{line:async-transition-complete}), and
scheduling all code awaiting the promise to run
(lines~\ref{line:async-awaiter-schedule-begin}--\ref{line:async-awaiter-schedule-end}).
In the case that~"f1" is exceptional, invoking an awaiter with~"f1"
(line~\ref{line:async-awaiter-f1}) effectively causes control to
transfer to the exception handler in the user program
(line~\ref{line:async-handle-exn}).
While "exec" handles "Exn[Y]" for "f2", it does not handle its
"Async" effect. So the call to "exec" is enclosed in the very
"Async" handler being defined (line~\ref{line:Async-handle-exec}).

How to handle "await" depends on the promise's state.
To handle "await" for a promise that is still pending completion
(lines~\ref{line:await-pending-begin}--\ref{line:await-pending-end}),
the resumption to the "await" call is added to the awaiter list of the
promise.
Otherwise (lines~\ref{line:await-complete-begin}--\ref{line:await-complete-end}),
the promise must be complete, and the resumption is
invoked with the result memoized by the promise.

\paragraph{Prior encodings}

The ability to encode promise-based
async--await~\cite{koka-async,koka-lang,dehmsw2017}
speaks to the expressive power of algebraic effects,\footnote{
    In JavaScript and C\#, \texttt{async} functions implicitly wrap
    their return values in promises
    (e.g., line~\ref{line:csharp-implicit-wrap-return} in \cref{fig:async-csharp}).
    An algebraic-effects encoding does not automatically support
    this behavior, but does not preclude it either.
    This eager wrapping possibly encourages the anti-pattern of
    unnecessary promises \cite{ohdd2014,azmt2018}.
}
but encodings in existing language designs compromise on how
they accommodate exceptional computations.
Koka~\cite{koka-async,koka-lang} supports structured asynchrony via
algebraic effects, but uses an "either" monad for possible
exceptional outcomes of the "await" operation---but encoding
exceptional outcomes into monadic values is a pattern that
algebraic effects in Koka are intended to help avoid!
Unlike Koka, Multicore OCaml~\cite{dehmsw2017} does not check algebraic effects
statically.
To notify user programs about asynchronously raised
exceptions, the language adds a special "discontinue"
construct.
It is our goal to treat async--await and asynchronous exceptions
in a more unified way: both are statically checked algebraic
effects.

\subsection{Communication Protocols}
\label{sec:interprocess}

It has been shown before that algebraic effects can express
interprocess communication, but not without using a more exotic
form of effect handlers that deviates from the original
categorical interpretation of effect handlers by \citet{pp2013}.
In particular, prior work relies on _shallow handlers_ to keep
the encoding syntactically light~\cite{klo2013,lmm2017}.

\begin{figure}
\centering
\newbox\boxa
\newbox\boxb
\newbox\boxc
\begin{lrbox}{\boxa}\begin{minipage}{.305\textwidth}
\begin{lstlisting}[linewidth=\textwidth,lineskip=.30ex]
interface Ping {
  def ping() : (*\VOID\RAISES*)Pong
}
interface Pong {
  def pong() : (*\VOID\RAISES*)Ping
}
\end{lstlisting}
\end{minipage}
       \end{lrbox}
\begin{lrbox}{\boxb}\begin{minipage}{.402\textwidth}
\begin{lstlisting}[linewidth=\textwidth,numbers=left]
def pinger(i,(*\,*)N(*\;*):(*\;*)int)(*\;*):(*\;*)(*\VOID\RAISES*)Ping {
  try {
    if (i < N)
      ping()(*\label{line:comm-pinger-ping}*)
  } with pong() resume {
    pinger(i(*\,*)+(*\,*)1,(*\;*)N)(*\label{line:comm-pinger-recurse}*)
  }
}(*\Suppressnumber*)
(*\Reactivatenumber*)
def ponger() : (*\VOID\RAISES*)Pong {
  try { pong() }(*\label{line:comm-ponger-pong}*)
  with ping() resume {(*\label{line:comm-ponger-ping-handle-begin}*)
    ponger()
  }(*\label{line:comm-ponger-ping-handle-end}*)
}
\end{lstlisting}
\end{minipage}
 \end{lrbox}
\begin{lrbox}{\boxc}\begin{minipage}{.305\textwidth}
\begin{lstlisting}[linewidth=\textwidth,lineskip=.30ex,numbers=left,firstnumber=15]
try {
  pinger(0,(*\;*)50)
} with ping() resume {(*\label{line:comm-client-ping-handle-begin}*)
  ponger()(*\label{line:comm-client-call-ponger}*)
}(*\label{line:comm-client-ping-handle-end}*)
\end{lstlisting}
\end{minipage}
    \end{lrbox}
\def\figa{\usebox\boxa}
\def\figb{\usebox\boxb}
\def\figc{\usebox\boxc}
\def\figd{
  \begin{minipage}{.178\textwidth}
  \vbox to 33.4ex {
    \includegraphics[scale=0.29]{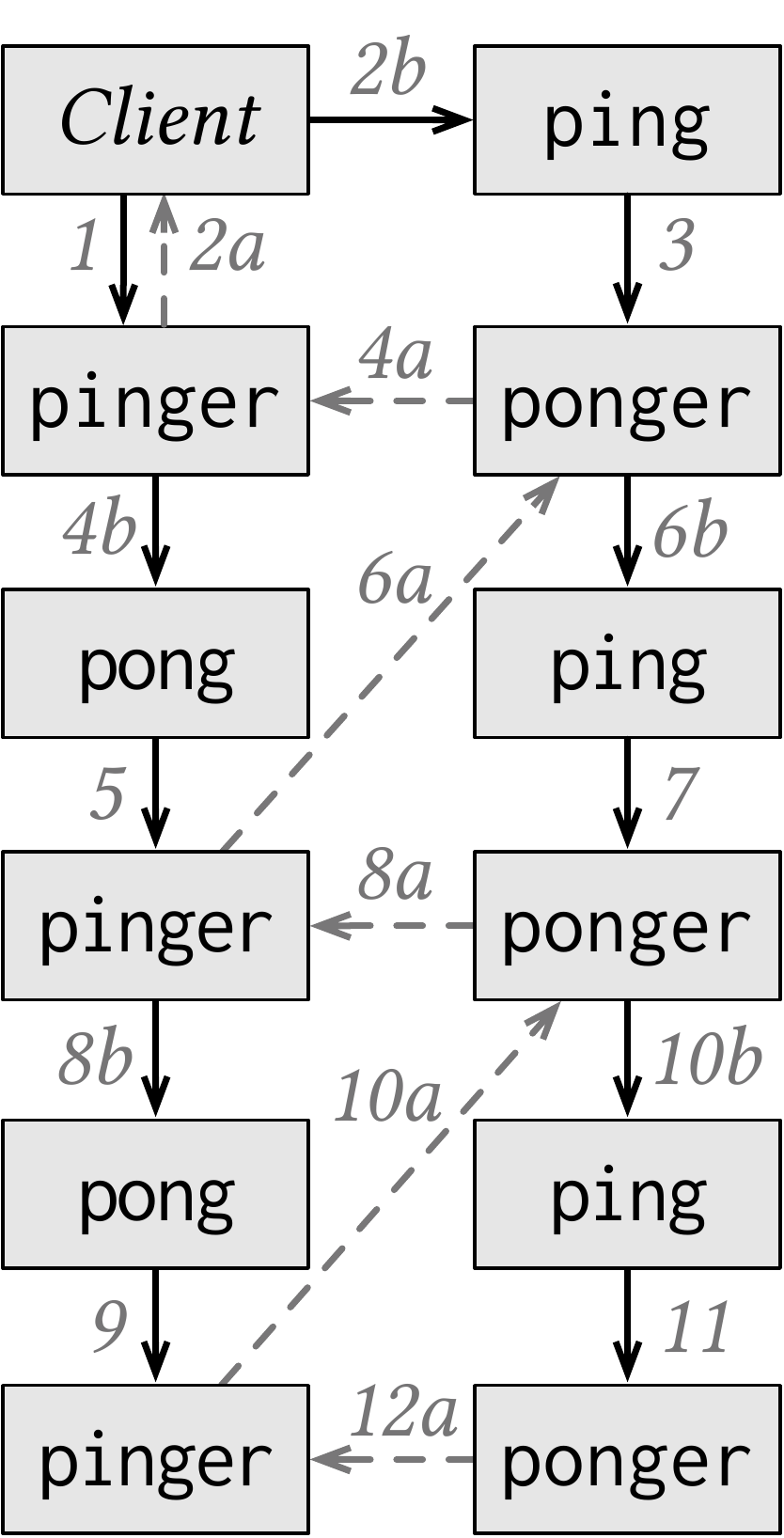}
  }
  \end{minipage}
}
\def\capa{Effect signatures}
\def\capb{Processes with effects "Ping" and "Pong"}
\def\capc{Client code}
\def\capd{Stack snapshot}
\savestack{\capfiga}{\subcaptionbox{\capa\label{fig:comm-sig}}{\figa}}\savestack{\capfigb}{\subcaptionbox{\capb\label{fig:comm-processes}}{\figb}}\savestack{\capfigc}{\subcaptionbox{\capc\label{fig:comm-client}}{\figc}}\savestack{\capfigd}{\subcaptionbox{\capd\label{fig:comm-stack}}{\figd}}\stackon[3.4ex]{\capfigc}{\capfiga}
\hfill\hspace{1ex}
\capfigb
\hfill
\capfigd
\caption{
Processes "pinger" and "ponger" send messages to each other in 
lockstep.
}
\label{fig:comm-code}
\end{figure}

Bidirectional algebraic effects offer an alternative: since we
allow all effect signatures to declare further effects, a series
of raised effects can, in general, bounce back and forth an
arbitrary number of times, turning effect signatures into
statically checked communication protocols.
We demonstrate this capability using the effect signatures "Ping"
and "Pong" (\cref{fig:comm-sig}) to obtain a pair of functions
that send messages to (i.e., raise effects at) each other in
lockstep.

Effects "Ping" and "Pong" are mutually recursive.
While invoking "ping" (resp.\ "pong") appears to one process as
sending a message, it appears to the other process as receiving the
message, as that other process must handle "ping" (resp.\ "pong").
Because "ping" (resp.\ "pong") declares it may raise "Pong"
(resp. "Ping"), a process typed with effect "Ping" (resp.\ "Pong")
should be prepared to receive a "Pong" (resp.\ "Ping") message
after sending a "Ping" (resp.\ "Pong").
The same process is allowed to do more "ping"s (resp.\ "pong"s)
on receiving the "Pong" (resp.\ "Ping").
The operations in this example do not carry a payload; a more
involved example can be found in \cref{sec:abstraction-safety}.

\cref{fig:comm-processes} shows two methods, "pinger" and
"ponger", typed with effects "Ping" and "Pong", respectively.
They are glued together by the client code in
\cref{fig:comm-client}.
At each step, "pinger" does a "ping"
(line~\ref{line:comm-pinger-ping}), and "ponger" reacts to it
by doing a "pong" (line~\ref{line:comm-ponger-pong}),
upon receiving which, "pinger" recursively calls itself
(line~\ref{line:comm-pinger-recurse}).
This interaction happens 50 times, after which the
communication ceases.
\cref{fig:comm-stack} visualizes the control flow.
As in \cref{fig:iterate-stack,fig:interrupt-iter-stack},
dashed arrows signal raising an effect to a stack frame
where its handler is found, while
solid arrows signal handler invocations and ordinary method calls.

Effects "Ping" and "Pong" can be viewed as specifying a
communication protocol, where the effect signatures choreograph
the sending and receiving of messages.
Processes statically typed with these
effects conform to the protocol dynamically.
In this sense,
bidirectional algebraic effects
offer an expressive behavioral-typing discipline resembling
session types~\cite{session-types}.

\subsubsection{Deep versus shallow}
\label{sec:deep-v-shallow}

Previous work raises the distinction between _deep handlers_
and _shallow handlers_~\cite{klo2013}.
They differ in the construction of handler resumptions:
the resumption to a deep handler contains the very handler at its
outermost layer, so subsequent effects raised in the resumption
can be handled by the same handler.
Handlers of algebraic effects, as originally introduced by
\citet{pp2013}, are deep: an effect handler is a fold (in
category-theoretic terms, a catamorphism~\cite{mfp1991}) over the
algebra of effect operations.
It has been argued that deep handlers behave more regularly and
admit easier reasoning~\cite{klo2013,lmm2017}.
However, interprocess communication has been identified as the
quintessential example where shallow handlers lead to an
easier encoding~\cite{klo2013,hl2018}.
By contrast, our encoding does not rely on shallow handlers, and
has the added benefit of capturing the sequencing of effects in
the signatures.
Nonetheless, we do not claim to settle the debate over deep vs.\
shallow; it is not clear that the two encodings faithfully reflect
each other, and other applications of shallow handlers might
emerge in the future.

\subsubsection{Deep, bidirectional handlers are recursive definitions}
\label{sec:deep-handlers-are-recursive}

The deep-handling semantics hints at recursion.
However, before bidirectional handlers, handlers have been unable
to exploit this recursion to specify that they should handle their own effects,
because handler resumptions can only accept
a value whose type matches the operation's result type.
By contrast, bidirectional handlers allow passing to "resume" a
computation that, in addition to the effects in the "raises"
clause, has the effect currently being handled.

We can use this feature to simplify the code of 
\cref{fig:comm-processes,fig:comm-client}
for a client that is privy to how "ponger" is implemented.
Notice there are two identical handlers for "Ping"---one on lines~\ref{line:comm-ponger-ping-handle-begin}--\ref{line:comm-ponger-ping-handle-end} in "ponger",
and the other on lines~\ref{line:comm-client-ping-handle-begin}--\ref{line:comm-client-ping-handle-end} in the client---meaning all "Ping" effects are handled identically.
So we should be able to obtain the same bidirectional
communication by keeping the "Ping" handler in the client code
but doing away with the one in "ponger".
The new "ponger" looks as follows:

\begin{centered}
\begin{minipage}{.485\textwidth}
\begin{lstlisting}[frame=none,backgroundcolor=\color{white}]
def ponger() : void raises Pong(*\;|\;*)Ping { pong() }
\end{lstlisting}
\end{minipage}
\end{centered}

\noindent
It has an additional "Ping" effect because we got rid of the
handler.
Despite this change, the client code in \cref{fig:comm-client}
continues to type-check.
The additional "Ping" effect raised by calling "ponger"
(line~\ref{line:comm-client-call-ponger}) is handled by the very
handler being defined.

This recursion in handling makes deep, bidirectional handlers
effectively fixpoint definitions.
While the binding structure of this fixpoint remains rather
vague at this point,
\cref{sec:fixpoint-handlers} makes it precise.

\section{Retaining Parametricity}
\label{sec:abstraction-safety}

The typical semantics for handling algebraic effects is to search for the
dynamically closest enclosing handler with a matching effect
signature; in fact, this semantics works for the examples discussed so far.
However, this semantics is known to be in conflict with modular
reasoning: higher-order, effect-polymorphic abstractions can
accidentally handle effects they are not designed to handle, breaking
abstraction boundaries~\cite{exceptions-pldi16,zm18,bpps2018}.
For example, the higher-order function "map" is declared to be
polymorphic over an effect variable~α that ranges over the
effects its argument function~"f" may raise.
The intended run-time behavior is for these effects to be
propagated to the caller of "map".

\begin{center}
\begin{minipage}{.64\textwidth}
\begin{lstlisting}[frame=none,backgroundcolor=\color{white}]
def map[X,Y,α](l(*\;:\;*)List[X], f(*\;:\;*)X(*\;→\;*)Y(*\RAISES*)α)(*\;:\;*)List[Y](*\RAISES*)α
\end{lstlisting}
\end{minipage}
\end{center}

\noindent
However, a problematic semantics could lead to "map" accidentally
handling "f"'s effects if the implementation of "map" happens to
contain an effect handler with a matching signature.

Previous work gives an alternative tunneling semantics to
algebraic effects,
addressing the accidental handling problem without giving up
the appeal of algebraic effects~\cite{zm18}.
In this section, we show that bidirectional algebraic effects
create new possibilities of accidental handling, but that
abstraction safety can be retained by adapting the tunneling
semantics.

\subsection{The Problem of Accidental Handling}
\label{sec:accidental-handling}

We call the typical algebraic-effects semantics the
_signature-based semantics_ because it identifies a propagating effect by its signature.
We show that if a signature-based semantics were used, resumptions to
bidirectional handlers could handle effects by accident.

\begin{figure}
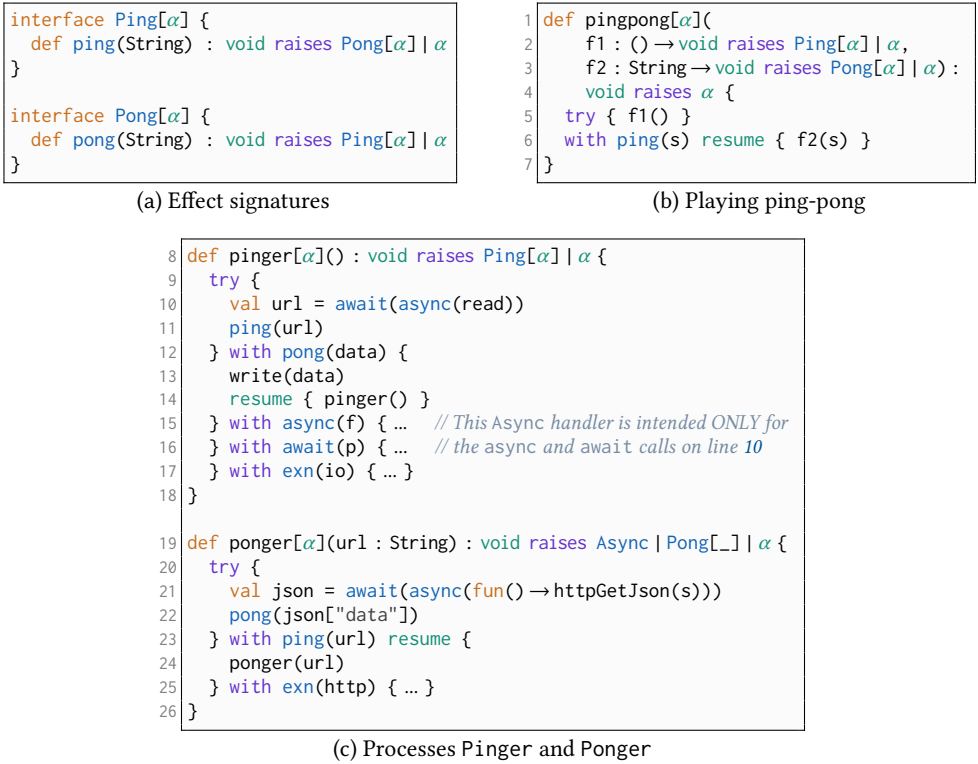

\centering
\newbox\boxa
\newbox\boxb
\newbox\boxc
\begin{lrbox}{\boxa}{\begin{minipage}{.422\textwidth}
\begin{lstlisting}[linewidth=\textwidth,lineskip=.27ex]
interface Ping(*\EVAR*) {
  def ping((*\STRING*)) : (*\VOID\RAISES*)Pong(*\EVAR*)(*\,*)|(*\,*)α
}

interface Pong(*\EVAR*) {
  def pong((*\STRING*)) : (*\VOID\RAISES*)Ping(*\EVAR*)(*\,*)|(*\,*)α
}
\end{lstlisting}
\end{minipage}
 }      \end{lrbox}
\begin{lrbox}{\boxb}{\begin{minipage}{.408\textwidth}
\begin{lstlisting}[linewidth=\textwidth,numbers=left,lineskip=.27ex]
def pingpong(*\EVAR*)(
    f1(*\;*):(*\;*)()(*\,*)→(*\,*)(*\VOID\RAISES*)Ping(*\EVAR*)(*\,*)|(*\,*)α,
    f2(*\;*):(*\;*)(*\STRING*)(*\,*)→(*\,*)(*\VOID\RAISES*)Pong(*\EVAR*)(*\,*)|(*\,*)α)(*\,*):(*\label{line:accident-pingpong-f2-type}*)
    (*\VOID\RAISES*)α {
  try { f1() }
  with ping(s) resume { f2(*\label{line:accident-pingpong-call-ponger}*)(s) }
}
\end{lstlisting}
\end{minipage}
 } \end{lrbox}
\begin{lrbox}{\boxc}{\begin{minipage}{.585\textwidth}
\begin{lstlisting}[linewidth=\textwidth,numbers=left,lineskip=.27ex,firstnumber=8]
def pinger(*\EVAR*)()(*\;*):(*\;*)(*\VOID\RAISES*)Ping[α](*\,*)|(*\,*)α(*\;*){
  try {
    val url = await(async(*\label{line:accident-pinger-call-async-await}*)(read(*\label{line:accident-readLine}*)))
    ping(*\label{line:accident-pinger-call-ping}*)(url)
  } with pong(data) {
    write(data)
    resume { pinger() }
  } with async(f) {(*\;\dadada\;*)  (*\codecomment{This \texttt{Async} handler is intended ONLY for}*)(*\label{line:accident-pinger-async-handler-begin}*)
  } with await(p) {(*\;\dadada\;*)  (*\codecomment{the \texttt{async} and \texttt{await} calls on line \ref{line:accident-pinger-call-async-await}}*)(*\label{line:accident-pinger-async-handler-end}*)
  } with exn(io) {(*\;\dadada\;*)}
}(*\Suppressnumber*)
(*\Reactivatenumber*)
def ponger(*\EVAR*)(url(*\;*):(*\;*)(*\STRING*))(*\;*):(*\;*)(*\VOID\RAISES*)Async(*\,*)|(*\,*)Pong[_](*\,*)|(*\,*)α(*\;*){
  try {
    val json = await(async(fun()(*\,*)→(*\,*)httpGetJson(s)))(*\label{line:accident-httpGet}*)
    pong(json["data"])
  } with ping(url) resume {
    ponger(*\label{line:accident-ponger-call-ponger}*)(url)
  } with exn(http) {(*\;\dadada\;*)}
}
\end{lstlisting}
\end{minipage}
 }\end{lrbox}
\def\figa{\usebox\boxa}
\def\figb{\usebox\boxb}
\def\figc{\usebox\boxc}
\def\capa{Effect signatures}
\def\capb{Playing ping-pong}
\def\capc{Processes "Pinger" and "Ponger"}
\savestack{\capfiga}{\subcaptionbox{\capa\label{fig:accident-sig}}{\figa}}\savestack{\capfigb}{\subcaptionbox{\capb\label{fig:accident-pingpong}}{\figb}}\savestack{\capfigc}{\subcaptionbox{\capc\label{fig:accident-processes}}{\figc}}\capfiga
\hspace{3em}
\capfigb
\bigskip\\
\capfigc
\caption{
Program \textls[-20]{"pingpong(pinger\!,ponger)"} risks
accidental handling under the signature-based semantics
}
\label{fig:accident-code}
\end{figure}

Suppose we want to use the ping-pong protocol to define two
processes where one process fetches webpages for HTTP URLs it
receives from the other process.
Both processes may make asynchronous queries.
To this end, effect signatures "Ping" and "Pong" are modified
(\cref{fig:accident-sig}) to carry a payload and to be
parameterized by an effect variable ranging over the extra
effects the processes may have.
A pair of processes with these two effects are composed using
method "pingpong" defined in \cref{fig:accident-pingpong}.
It is effect-polymorphic, allowing the processes to have
additional effects.

Whereas major platforms supporting async--await dispatch
asynchronous jobs using a single dispatcher in the runtime, the
ability to treat async--await as a regular algebraic
effect enables software components to handle their own "Async"
effects.
This ability is useful, for example, when a special event loop
is wanted, or when "Async" effects must be monitored~\cite{koka-async}.

The processes are defined in \cref{fig:accident-processes}.
Process "pinger" reads URLs asynchronously from an input source
(line~\ref{line:accident-readLine}) and handles its own "Async"
effect (lines~\ref{line:accident-pinger-async-handler-begin}--\ref{line:accident-pinger-async-handler-end}).
Process "ponger" issues asynchronous HTTP GET requests
(line~\ref{line:accident-httpGet}) but chooses to let an outer
event loop to handle its "Async" effects; hence "Async" appears
in its "raises" clause.
Underscores are placeholders for inferred effects.
The signature-based semantics would infer, via unification, the
placeholder effect in the definition of "ponger" to be "Async|α".

Program "pingpong(pinger\!,\,ponger)" starts the bidirectional
communication.
The signature-based semantics would instantiate the effect
variable α of "pingpong" to be "Async", that of "pinger" to be
"Async", and that of "ponger" to be the empty effect.
The program as a whole would have effect "Async"---the programmer
expects an outer event loop to handle "ponger"'s "Async" effects.

However, under the signature-based semantics, the program would
not execute in the expected way.
When "ponger" invokes either operation "async" or "await", the
dynamically closest enclosing handler for "Async" would intercept
and handle it:
climbing up the call chain, "ponger" is called by "ping"
(lines~\ref{line:accident-pingpong-call-ponger} and \ref{line:accident-ponger-call-ponger}),
which is invoked on line~\ref{line:accident-pinger-call-ping},
which is enveloped by the "Async" handler intended only for
the "async" and "await" calls on line~\ref{line:accident-pinger-call-async-await}.
The programmer is in for a surprise.

This phenomenon of handler resumptions accidentally handling effects
is new; key ingredients of the example include bidirectionality
and effect signatures parameterized over effects, which are
missing in previous work addressing accidental handling~\cite{zm18,bpps2018}.
Still, we can rely on the prior work to help us understand
the crux of the problem.

\subsection{A Loss of Parametricity}
\label{sec:loss-of-parametricity}

An insight of prior work~\cite{zm18,bpps2018} is that accidental
handling reflects a loss of parametricity.
Reynolds' Abstraction Theorem for System~F~\cite{reynolds83} implies
that parametricity of type polymorphism relies on polymorphic
functions not being able to make decisions based on the types
instantiating the type parameters.
Analogously, parametricity of effect polymorphism requires that
effect-polymorphic functions not make decisions based on the effects
they are instantiated with.
The signature-based semantics runs afoul of this requirement.
In the example above, "pinger" is a function polymorphic over an
abstract effect~α. But under the signature-based semantics, it
would be able to inspect, at run time, the signatures of
propagating effects otherwise statically denoted by~α, causing
accidental handling.

A loss of parametricity is a loss of modular reasoning.
The signature-based semantics means, for example, that one cannot
reason modularly about the program context
"pingpong(pinger,$\bullet$)" by just looking at the types and
without knowing how "pinger" is implemented.
The hole expects a function of a type as shown on
line~\ref{line:accident-pingpong-f2-type} of
\cref{fig:accident-pingpong}, which does not speak of "Async".
Yet filling the hole with a function with effect "Async" would
lead to surprise.

\subsection{Tunneling via Lexically Scoped Handlers}
\label{sec:tunneling}

To restore parametricity, we adapt the idea of tunneled algebraic
effects~\cite{zm18}. 
Tunneling echoes the modular reasoning requirement that handlers
should only handle effects they are locally ``aware'' of;
otherwise, effects tunnel through handlers.
For example, the definition of "pinger" is polymorphic over
the effect variable α it binds, so it ought to be ``oblivious'' to
any propagating effects that correspond to α at run time---these effects tunnel through the handler in "pinger".
By contrast, the call site "pingpong(pinger,ponger)" is ``aware''
that "ponger" may raise "Async"---the call site is thus required
to handle this effect.

This modular reasoning requirement suggests that the tunneling
semantics choose handlers _lexically_.
This lexical scoping of handlers generalizes naturally to
bidirectional algebraic effects, with
handler bindings brought into the lexical scope in three ways:

\begin{enumerate}[label={(\arabic*)},topsep=.5ex,]
\item
a "try"--"with" statement binds an identifier, corresponding to
the handler following "with", for use in the "try"-block
computation,
\item
a "raises" clause binds a set of identifiers, each corresponding
to an effect signature in the clause, for use in the method body or
handler body, and
\item
a handler definition binds an identifier named "self",
corresponding to the current handler, for use in the handler
body.
\end{enumerate}

\noindent
The first two ways are a straightforward adaptation of the
tunneling semantics.
The third way explains why a bidirectional handler can demand its own
effects to be handled by itself (\cref{sec:deep-handlers-are-recursive}).

The approach of \citet{zm18} is that programs written in the usual syntax are
desugared to give handler bindings explicit names.
Handlers for effectful computations are then chosen by resolving an
omitted handler to the lexically closest enclosing binding.
Because handlers are resolved lexically, effects appear to
tunnel to handlers without allowing dynamically enclosing handlers
to intercept them, even if they have identical signatures.

\begin{figure}
\begin{minipage}{.842\textwidth}
\begin{lstlisting}[linewidth=\textwidth,numbers=left,lineskip=.30ex]
def pinger(*\EVAR*)[(*\HVAR{Pi}\;:\;*)Ping(*\EVAR*)]()(*\;*):(*\;*)(*\VOID\RAISES*)(*\HVAR{Pi}*)(*\,*)|(*\,*)α(*\;*){
  try {
    val url = (*\HVAR{A}*).await[(*\HVAR{E}*)]((*\HVAR{A}*).async(read))(*\label{line:tunnel-pinger-call-async}*)
    (*\HVAR{Pi}*).ping(*\label{line:tunnel-call-ping}*)[(*\HVAR{Po}*)](url)
  } with (*\HVAR{Po}*)(*\;*):(*\;*)Pong(*\EVAR*) = {
    def pong[(*\HVAR{Pi}\;:\;*)Ping(*\EVAR*)](data(*\;*):(*\;\STRING*))(*\;*):(*\;*)(*\VOID\RAISES*)(*\HVAR{Pi}*)(*\,*)|(*\,*)α(*\;*){
      write[(*\HVAR{E}*)](data)(*\label{line:tunnel-call-write}*)
      resume { pinger(*\EVAR*)[(*\HVAR{Pi}*)]() }(*\label{line:tunnel-recursive-call-pinger}*)
    }
  } with (*\HVAR{A}*)(*\;*):(*\;*)Async = {
    def async[X,Y](f(*\;*):(*\;*)Fun2[X,Y])(*\;*):(*\;*)Promise[X,Y](*\;*){(*\;\dadada\;*)}
    def await[X,Y][(*\HVAR{E}\;:\;*)Exn[Y]](p(*\;*):(*\;*)Promise[X,Y])(*\;*):(*\;*)X(*\RAISES\HVAR{E}*)(*\;*){(*\;\dadada\;*)}
  } with (*\HVAR{E}*)(*\;*):(*\;*)Exn[IO] = {(*\;\dadada\;*)}
}(*\Suppressnumber*)
(*\Reactivatenumber*)
(*\DEF*) (*\textls[-20]{ponger}*)(*\EVAR[80]*)[(*\HVAR{A}\;:\;\ASYNC*)][(*\HVAR{Po}\;:\;\PONG*)[(*\HVAR{A}*)|α]](url(*\;:\;\STRING*))(*\;*):(*\;\VOID\RAISES*)(*\HVAR{A}*)(*\,*)|(*\,*)(*\HVAR{Po}*)(*\,*)|(*\,*)α {(*\;\dadada\;*)}
\end{lstlisting}
\vspace{-.7ex}

\begin{lstlisting}[linewidth=\textwidth]
(*\codecomment{Client code}*)
pingpong[(*\HVAR{A2}*)](pinger[(*\HVAR{A2}*)], ponger[(*\nil*)][(*\HVAR{A2}*)])
\end{lstlisting}
\end{minipage}
 \caption{
Desugaring \textls[-30]{"pinger"}, \textls[-30]{"ponger"}, and
the client program
}
\label{fig:desugaring}
\end{figure}

As an example,
\cref{fig:desugaring} shows the desugaring of "pinger"
(\cref{fig:accident-processes}).
It makes explicit all handler bindings and references to the
bindings.
Desugaring the \textls[-53]{"raises\;Ping\EVar"} clause
introduces an identifier \HVar{Pi} into the "pinger" body; it
denotes the handler used to handle "Ping\EVar" effects raised by
"pinger".
Invoking operation "ping" (line~\ref{line:tunnel-call-ping}, in
the "try" block) requires a "Ping\EVar" handler to be provided.
This handler is resolved to \HVar{Pi}, the lexically closest enclosing
binding for the signature.
The invocation also requires a handler for "Pong\EVar" to be
provided, because operation "ping" is defined to raise
"Pong\EVar".
This handler is resolved to \HVar{Po}, the binding introduced
into the "try" block by one of the surrounding handler
definitions.
The locally defined "Async" handler, denoted by \HVar{A}, is used
to handle the invocations of "async" and "await"
(line~\ref{line:tunnel-pinger-call-async}).

\cref{fig:desugaring} also shows the desugaring of the client
program "pingpong(pinger\!,ponger)".
Assuming \HVar{A2} denotes a surrounding "Async" handler, it
instantiates the effect variable of "pingpong" to be \HVar{A2},
that of "pinger" to be \HVar{A2}, and that of "ponger" to be the
empty effect.
The client program as a whole has effect \HVar{A2}, indicating it
uses the outer "Async" handler to handle "Async" effects raised
by "ponger".

\subsubsection{Lifetimes as effects}

Effect handlers obey a stack discipline.
A handler's lifetime begins when the corresponding "try"-block is
entered, and ends when the "try"-block computation is done:
the handler cannot outlive the lexical structure binding it.
The desugaring outlined above makes it explicit that references to
handlers are passed as lexically scoped arguments.  But lexical
scoping alone does not prevent closures that outlive handlers from
capturing them, making them dangling references.

To prevent such dangling references, the type system follows
prior work~\cite{zm18,effekt-2020,bpps2020} in treating handler lifetimes as
computational effects, in a similar way to region-based type
systems~\cite{lg88-effects,tt97,cyclone-regions}.
A computation that uses handlers to handle algebraic effects is
typed with the lifetimes of those handlers as its computational
effects.

Such lifetime effects, denoted by handler identifiers, are
written in "raises" clauses of desugared types.
For example, desugaring the \textls[-40]{"raises\;Ping\EVar"}
clause not only introduces the handler binding \HVar{Pi}, but
also means the method has lifetime effect \HVar{Pi}.
Because the method body is typed with the lifetime of \HVar{Pi},
it cannot outlive \HVar{Pi}; it is thus allowed to use the
non-dangling reference \HVar{Pi} to invoke "ping".

\snd{
The desugaring that introduces explicit handler bindings can thus
be understood as also assigning default lifetimes automatically
as follows: an ordinary object is not lifetime-bounded, since its lifetime is the
same as the heap region; the lifetime of a handler is bounded by
the "try"–"with" stack region to which it is attached; and a method
with a "raises" clause is lifetime-polymorphic.
Assigning default lifetimes is present in
Cyclone (_default regions_~\cite{cyclone-regions}) and
Rust (_lifetime elision_~\cite{rust-book}).
We do not require the complexity of these full-blown region-based
type systems, however, because only handlers are
lifetime-bounded and because handler lifetimes are restricted to
stack regions.
}

\if 0
Notice that computations passed to "resume" are the only places
that make sense for handler variables (introduced by "raises"
clauses) to be in scope.
As described in \cref{sec:baf-examples}, computations passed
to "resume" propagate effects listed in the "raises" clause to
the call site of the effect operation.
The resumption at the call site, represented by "resume",
contains handlers (in the form of "try"--"with") for those
effects; the handler variables are instantiated by these handlers
at run time.
Therefore, the lifetimes of these handlers are dominated by that
of the resumption.
By contrast, the type system prevents handler variables from
being used (to invoke effect operations) outside computations
passed to "resume", because otherwise the lifetime effects
triggered by such invocations could not be discharged.

This scoping rule does not affect tail-resumptive
handlers or ordinary method definitions.
In fact, an ordinary method definition can be viewed as an effect
handler where the entire method body is passed to a tail
resumption. So handler variables of the method definition are
always in scope in the entire method body.
\fi

\subsubsection{Fixpoint handlers}
\label{sec:fixpoint-handlers}

\begingroup
\BeforeBeginEnvironment{wrapfigure}{\setlength{\intextsep}{0.0ex}}\begin{wrapfigure}[6]{r}{.41\textwidth}
\begin{minipage}{.41\textwidth}
\begin{lstlisting}
try { pinger[(*\HVAR{Pi}*)]() }
with (*\HVAR{Pi}\,*):(*\,*)Ping = {
  def ping[(*\HVAR{Po}\,*):(*\,*)Pong]()(*\;*):(*\;*)(*\VOID\RAISES\HVAR{Po}*) {
    resume { ponger[(*\HVAR{Po}*)][(*\SELFVAR*)]() }
  }
}
\end{lstlisting}
\end{minipage}
\end{wrapfigure}
The availability of a "self" handler makes it precise that
bidirectional (deep) handlers are fixpoint definitions: the
fixpoint is taken of a handler definition quantifying over "self".
Because the resumption to a deep handler is enveloped by the same
handler, the computation passed to "resume" cannot outlive the
handler and can thus safely use it to handle its effects.
The figure above shows how \cref{fig:comm-client} is
desugared using "self", assuming "ponger" has the signature as
shown in \cref{sec:deep-handlers-are-recursive}.

Since being self-referential is also a key characteristic of
objects, it seems that objects and handlers are almost unifiable.
\cref{sec:core} makes this unification precise for a core
language.

\endgroup

\section{A Core Language}
\label{sec:core}

We study the formal foundation of bidirectional algebraic effects
using a core language, \lamlang, that captures the key aspects of
the language mechanism introduced in
\cref{sec:baf-examples,sec:abstraction-safety}.

\lamlang is both functional and object-oriented. Like a lambda
calculus, it does not support imperative state. Like an
object-oriented language, it supports the separation between
objects and interfaces, and objects in \lamlang are
self-referential.
\lamlang supports effect signatures and handlers using the
same constructs for object interfaces and objects.
The effect-handling construct "try"--"with" is captured by
something that resembles delimited control~\cite{cupto},
following prior work~\cite{zm18,effekt-2020}.

\lamlang is intended to capture the essence of the language
mechanism. It makes simplifications similar to existing
formalisms of algebraic effects with lexically scoped effect
handlers~\cite{bpps2020,zm18}:
it is assumed that handlers are always given explicitly for effectful
computations (rather than resolving elided handlers to the closest
lexically enclosing binding) and that effect signatures contain
exactly one effect operation. Lifting these restrictions is
straightforward but adds syntactic complexity that obscures the key
issues.
Because of its recursive interface definitions and fixpoint
handlers, \lamlang is Turing-complete.

\setlength{\belowdisplayskip}{.8ex} \setlength{\belowdisplayshortskip}{0pt}
\setlength{\abovedisplayskip}{.8ex} \setlength{\abovedisplayshortskip}{0pt}

\subsection{Syntax}
\label{sec:syntax}

\begin{figure}
\small \renewcommand\VSEP{\vspace{.75ex}\\}\renewcommand\VSep{\vspace{0ex}\\}\begin{align*}
\begin{array}{@{}l@{\hspace{1.1ex}}c@{\ \ }c@{\ \ \ }l@{}}
\textit{programs} & \prog & ::= &
\lst{\isig}\,;\,\DOWN{\lbl}{\tm}
\VSEP
\textit{interface definitions} & \isig & ::= &
\MkISig{\iname}{\lst\evar}{\msig}
\VSEP
\textit{operation signatures} & \msig, \msig[S] & ::= &
\efunty{\evar}{\msig} \bnf
\lfunty{\lvar}{\msig} \bnf
\fnty{\ty}{\msig} \bnf
\raises{\ty}{\eff}
\VSEP
\textit{types} & \ty,\ty[S] & ::= &
\Unit \bnf
\itype{\iname}{\lst\eff}{\life} \bnf
\mtype{\msig}{\life} \bnf
\tycont{\ty_1}{\eff[1]}{\ty_2}{\eff[2]}
\VSEP
\textit{composite effects} & \eff & ::= &
\nil \bnf
\eff,\,\ef
\VSEP
\textit{atomic effects} & \ef & ::= &
\evar \bnf
\life
\VSEP
\textit{lifetimes} & \life & ::= &
\lvar \bnf
\lbl
\VSEP
\textit{\textls[-30]{operation implementations}} & \mdef & ::= &
\efun{\evar}{\mdef} \bnf
\lfun{\lvar}{\mdef} \bnf
\lam{\x}{\mdef} \bnf
\lam{\x[k]}{\tm}
\VSEP
\textit{values} & \val,\val[u] & ::= &
\x \bnf
\unit \bnf
\fixpoint{\self}{\obj{\mdef}{\lbl}} \bnf
\obj{\mdef}{\lbl} \bnf
\cont{\EC}
\VSEP
\textit{terms} & \tm,\tm[s] & ::= &
\val \bnf
\unroll{\tm} \bnf
\eapp{\tm}{\eff} \bnf
\lapp{\tm}{\life} \bnf
\app{\tm}{\tm[s]} \bnf
\DOWN{\lbl}{\tm} \bnf
\up{\tm} \bnf
\letexp{\x}{\tm}{\tm[s]} \bnf
\throw{\tm}{\tm[s]}
\VSEP
\textit{evaluation contexts} & \EC & ::= &
\hole \bnf
\unroll{\EC} \bnf
\eapp{\EC}{\eff} \bnf
\lapp{\EC}{\life} \bnf
\app{\EC}{\tm} \bnf
\app{\val}{\EC} \bnf
\up{\EC} \bnf
\letexp{\x}{\EC}{\tm} \bnf
\throw{\EC}{\tm}
\VSEP
\multicolumn{4}{@{}l@{}}{
\textls[-10]{\textit{effect variables}}   \ \,\evar
\ \ \ \,
\textls[-10]{\textit{lifetime variables}} \ \,\lvar
\ \ \ \,
\textls[-10]{\textit{lifetime constants}} \ \,\lbl
\ \ \ \,
\textls[-10]{\textit{value variables}}    \ \,\x,\x[k],\x[H],\self,\textnormal{...}
\ \ \ \,
\textls[-10]{\textit{interface names}}    \ \,\iname
}
\end{array}
\end{align*}

 \caption[]{Syntax of \lamlang}
\label{fig:syntax}
\end{figure}

\cref{fig:syntax} presents the syntax of \lamlang.
Metavariables standing for identifiers have a lighter
color.
An overline denotes a (possibly empty) sequence of syntactic
objects.
For instance, {\lst\ef} denotes a sequence of effects;
an empty sequence is \nil.
Effect sequences, or _composite effects_, are denoted by
{\eff}.
\ifreport
The notation~\lsti{\bullet} denotes the $i$-th element in a
sequence.
\fi

The type system tracks handler lifetimes as effects.
An effect~{\ef} is either an effect variable~{\evar} or a
lifetime~{\life}, which is either a lifetime variable~{\lvar} or a
lifetime constant~{\lbl}.
Lifetime effects compose easily, since effect sequences are
essentially sets---the order and multiplicity of effects in a sequence
are irrelevant.
Substituting an effect sequence~{\lst\ef} for an effect variable~{\evar}
in another effect sequence works by flattening~{\lst\ef} and
replacing~{\evar} with the flattened effects.
Substituting a lifetime~{\life} for a lifetime variable~{\lvar} works
in the usual way.

A value~{\val} is either
a variable~{\x},
the unit value~{\unit},
a handler value \fixpoint{\self}{\obj{\mdef}{\lbl}},
an operation value {\obj{\mdef}{\lbl}},
or a continuation~{\cont{\EC}}.
Continuations are represented by evaluation contexts~{\EC}.
In a term of form {\throw{\tm}{\tm[s]}}, term~{\tm} must evaluate to a
continuation, after which~{\tm[s]} is placed in the evaluation context
representing the continuation.
 
While in \cref{sec:tunneling} lifetimes are identified by
handler bindings, in \lamlang lifetimes are decoupled from
handlers and form a separate syntactic category:
handler values (subsuming objects) are of form
\fixpoint{\self}{\obj{\mdef}{\lbl}},
consisting of an operation implementation~{\mdef} and the
lifetime~{\lbl} of the value.
Handler values are fixpoints; the {\x[self]} variable is
bound in~{\mdef}.
An operation value is of form {\obj{\mdef}{\lbl}}; it is the
result of unrolling the fixpoint definition of a handler value
to extract the operation.

Lifetime constants~{\lbl} are declared by, and bound in, \freset-terms.
\lamlang encodes the "try"--"with" construct using \freset-terms.
The computation~{\tm} guarded by a {\DOWN{\lbl}{\!}} may have lifetime
effect~{\lbl}.
While \freset-terms bind and discharge lifetime effects, \shift-terms
invoke handlers and thus introduce lifetime effects.

A term has either
the unit type~{\Unit},
a handler type (a.k.a.\ an interface type)~{\itype{\iname}{\lst\eff}{\life}},
an operation type~{\mtype{\msig}{\life}},
or a continuation type~{\tycont{\ty_1}{\lst{\ef_1}}{\ty_2}{\lst{\ef_2}}}.
Handler values have handler types, while operation values have
operation types.
A term of form {\unroll{\tm}} extracts the operation value from a
handler value, by unrolling the fixpoint handler definition.
A continuation of type~{\tycont{\ty_1}{\lst{\ef_1}}{\ty_2}{\lst{\ef_2}}}
can be applied to a computation of type~{\raises{\ty_1}{\lst{\ef_1}}}.
 
An operation implementation~{\mdef} is possibly polymorphic over effect variables
(\efun{\evar}{\mdef}), lifetime variables (\lfun{\lvar}{\mdef}), and
value variables (\lam{\x}{\mdef}).
Correspondingly, the operation signature~{\msig} of a handler can be
effect-polymorphic (\efunty{\evar}{\msig}), lifetime-polymorphic
(\lfunty{\lvar}{\msig}), and value-polymorphic (\fnty{\ty}{\msig}).
The last parameter~{\x[k]} of an operation implementation is a
continuation---to wit, the handler resumption.
An operation has a result type~{\raises{\ty}{\eff}}; the handler
resumption is able to discharge effects in~{\eff}.
Unlike previous work that gives algebraic effects operational
meanings~\cite{lmm2017,koka-lang,zm18,bpps2019}, handler
resumptions in \lamlang are evaluation contexts taking as input
(possibly effectful) computations, instead of functions that take
only pure values.

An \lamlang program consists of a set of interface definitions
and a ``main'' term to be evaluated.
The interfaces are mutually recursive and can be parameterized by
effect variables.
The ``main'' term is guarded by a~{\freset}, which binds a lifetime
constant. This lifetime constant is used as the lifetime of
handler values that correspond to ordinary objects; they exist for
the full lifetime of the program and hence need not obey the usual stack
discipline imposed on other handlers.

\paragraph{Example}

\lamlang is less removed from the informal surface language used in
\cref{sec:baf-examples,sec:abstraction-safety} than its syntax
might suggest.
Below we encode function "pinger" (\cref{fig:desugaring}) in
\lamlang.
This example demonstrates how \lamlang encodes
various language constructs including
handler bindings, the "try"--"with" construct, functions
and function calls, and handlers and handler invocations.

The (recursive) function "pinger" is encoded as a
(self-referential) object, which is
expressed as a handler value that applies its resumption to its
body $\tm_{\mathsf{pinger}}$:
\begin{gather*}
\fixpoint{\x[pinger]}{\obj{
  \efun{\evar}{\lfun{\x[{\lvar_{\!Pi}}]}{\lam{\x[{H_{Pi}}]}{\lam{\x[k]}{
    \throw{\x[k]}{\tm_{\mathsf{pinger}}}
  }}}}
}{\lbl[0]}}
\end{gather*}

\noindent
In the informal language of \cref{sec:abstraction-safety},
a handler binding denotes both the handler and also its lifetime.
In \lamlang, a handler binding is modeled with two bindings,
one for a lifetime and the other for a value variable:
the value variable stands for a handler of the given lifetime.
For example, "Pinger\EVar" in the "raises" clause of
"pinger" is modeled in \lamlang by a handler binding
{\x[{H_{Pi}}]} and a lifetime binding {\x[{\lvar_{\!Pi}}]}, where
{\x[{H_{Pi}}]} has lifetime {\x[{\lvar_{\!Pi}}]}.

Function "pinger", encoded as an object, need not obey a
stack discipline, so it has lifetime $\lbl[0]$, which is assumed
to be the lifetime constant declared by the~{\freset} guarding
the ``main'' program.
Term $\tm_{\mathsf{pinger}}$, that is, the body of "pinger",
looks as follows, where the bindings {\evar}, {\x[{\lvar_{\!Pi}}]},
{\x[{H_{Pi}}]}, {\x[k]}, and the self reference
{\x[pinger]}, are in scope:
\begin{gather*}
\begin{array}{@{}l@{\,}l@{\ }l@{\ }l@{\ }l@{\ }l@{}}
\tm_{\mathsf{pinger}} \defeq
\freset^{\lbl} &
  \KEYWORD{let} & \x[{H_E}] & = & 
  \fixpoint{\x[self]}{\obj{\dadada}{\lbl}} \ \KEYWORD{in}
\\
  &
  \KEYWORD{let} & \x[{H_A}] & = & 
  \fixpoint{\x[self]}{\obj{\dadada}{\lbl}} \ \KEYWORD{in}
\\
  &
  \KEYWORD{let} & \x[{H_{Po}}] & = & 
  \fixpoint{\x[self]}{
    \obj{\mdef_{\mathsf{pong}}}{\lbl}
  } \ \KEYWORD{in}
\\
  &
  \KEYWORD{let} & \x[url] & = & 
  \up{\app{\app{\lapp{\unroll{\x[H_{A}]}}{\lbl}}{\x[H_{E}]}}{\x[read]}} \ \KEYWORD{in}
  &
  \text{\small\ \ (line~\ref{line:tunnel-pinger-call-async})}
\\
  &
  \multicolumn{4}{@{}l@{}}{
    \up{\app{\app{\lapp{\unroll{\x[H_{Pi}]}}{\lbl}}{\x[H_{Po}]}}{\left(
        \x[url]     
    \right)}}
  }
  &
  \text{\small\ \ (line~\ref{line:tunnel-call-ping})}
\end{array}
\hspace{1.7ex}
\begin{array}{@{}l@{\,}l@{\ }l@{\ }l@{}}
\mdef_{\mathsf{pong}} \defeq &
  \lfun{\x[{\lvar_{\!Pi}}]}{\lam{\x[{H_{Pi}}]}{\lam{\x[data]}{\lam{\x[k]}{}}}}
\\
  &
  \letexp{\_}{\up{\app{\app{\lapp{\unroll{\x[write]}}{\lbl}}{\x[{H_E}]}}{\x[data]}}}{}
  &
  \text{\small(line~\ref{line:tunnel-call-write})}
\\
  &
  \throw{\x[k]}{\left(\up{\app{\lapp{\eapp{\unroll{\x[pinger]}}{\evar}}{\x[{\lvar_{\!Pi}}]}}{\x[{H_{Pi}}]}}\right)}
  &
  \text{\small(line~\ref{line:tunnel-recursive-call-pinger})}
\\
~
\\
~
\end{array}
\end{gather*}

The body of "pinger" is a "try" block followed by a series of
handlers.
A "try"--"with" statement is encoded in \lamlang as
$\DOWN{\lbl}{\letexp{\x[H]}{\fixpoint{\x[self]}{\obj{\mdef}{\lbl}}}{\tm}}$,
where~{\mdef} is the implementation of the effect operation,
and~{\tm} is the "try"-block computation, which may invoke
handler~{\x[H]}.
A handler should not outlive its "try"--"with" statement, so
the handler value has the same lifetime as declared by the
\freset-term.
In term $\tm_{\mathsf{pinger}}$ above, all three handlers,
{\x[H_E]}, {\x[H_A]}, and {\x[H_{Po}]},
have the same lifetime~{\lbl} because there is only one "try".

Operation invocations on handlers (subsuming method calls on
objects) are encoded as \shift-terms.
For example, the "try"-block computation
"\HVar{Pi}.ping[\HVar{Po}](\dadada)"
(line~\ref{line:tunnel-call-ping}, \cref{fig:desugaring})
invokes "\HVar{Pi}",
so it is encoded in \lamlang as
{\up{\app{\app{\lapp{\unroll{\x[H_{Pi}]}}{\lbl}}{\x[H_{Po}]}}{\dadada}}},
where {\lbl} is the lifetime of {\x[H_{Po}]}.
Similarly, the recursive call to "pinger"
(line~\ref{line:tunnel-recursive-call-pinger})
is also encoded as a \shift-term.
(For simplicity, in the encoding above we have assumed the two operations
of "Async" are combined into one.)

\subsection{Operational Semantics}
\label{sec:operational-semantics}

\begin{figure}
\small
\begin{subfigure}{.68\linewidth}
\judgebox{
\steps{\lst{\lbl[1]}}{\tm_1}{\lst{\lbl[2]}}{\tm_2}
}
\begin{mathpar}
\Rule[name=ktx]{
    \steps{\lst{\lbl[1]}}{\tm_1}{\lst{\lbl[2]}}{\tm_2}
}{
    \steps{\lst{\lbl[1]}}{\appctx{\EC}{\tm_1}}{\lst{\lbl[2]}}{\appctx{\EC}{\tm_2}}
}

\Rule[name=let]{}{
    \steps
    {\lst{\lbl}}
    {
        \letexp{\x}{\val}{\tm}
    }
    {\lst{\lbl}}
    {
        \subst{\tm}{\val}{\x}
    }
}

\Rule[name=op]{}{
    \def\fixobj{\fixpoint{\self}{\obj{\mdef}{\lbl[0]}}}\steps
    {\lst{\lbl}}
    {\unroll{\left(\fixobj\right)}}
    {\lst{\lbl}}
    {\obj{\subst{\mdef}{\fixobj}{\self}}{\lbl[0]}}
}

\Rule[name=eapp]{}{
    \steps
    {\lst{\lbl}}
    {
        \eapp
        {\left( \obj{\Lam{\evar}{\mdef}}{\lbl[0]} \right)}
        {\eff}
    }
    {\lst{\lbl}}
    {\subst{\obj{\mdef}{\lbl[0]}}{\eff}{\evar}}
}

\Rule[name=lapp]{}{
    \steps
    {\lst{\lbl}}
    {
        \lapp
        {\left( \obj{\Lam{\lvar}{\mdef}}{\lbl[0]} \right)}
        {\lbl[1]}
    }
    {\lst{\lbl}}
    {\subst{\obj{\mdef}{\lbl[0]}}{\lbl[1]}{\lvar}}
}

\Rule[name=app]{}{
    \steps
    {\lst{\lbl}}
    {
        \app
        {\left( \obj{\lam{\x}{\mdef}}{\lbl[0]} \right)}
        {\val} 
    }
    {\lst{\lbl}}
    {\subst{\obj{\mdef}{\lbl[0]}}{\val}{\x}}
}

\Rule[name=throw]{}{
    \steps
    {\lst{\lbl}}
    {
        \throw{
            \left(
            \cont{\EC}
            \right)
        }{\tm}
    }
    {\lst{\lbl}}
    {
        \appctx{\EC}{\tm}
    }
}

\Rule[name=down]{}{
    \steps
    {\lst{\lbl}}
    {
        \DOWN{\lbl[1]}{\tm}
    }
    {\lst{\lbl},\,\lbl[2]}
    {
        \down{\lbl[2]}{\subst{\tm}{\lbl[2]}{\lbl[1]}}
    }
    \;
    \left(
    \lbl[2] \notin \lst{\lbl}
    \right)
}

\Rule[name=downval]{}{
    \steps
    {\lst{\lbl}}
    {
        \down{\lbl[0]}{\val}
    }
    {\lst{\lbl}}
    {
        \val
    }
}

\Rule[name=downup]{}{
    \steps{\lst{\lbl}}
    {
        \down{\lbl[0]}{
            \appctx{\EC}{
                \up{\obj{\lam{\x[k]}{\tm}}{\lbl[0]}}
            }
        }
    }
    {\lst{\lbl}}
    {
        \subst{
            \tm
        }{
            \cont{\down{\lbl[0]}{\!\EC}}
        }{\x[k]}
    }
    \,
    \left(
    \tunnels{\lbl[0]}{\EC}
    \right)
}

\end{mathpar}
 \end{subfigure}
\hfill
\begin{subfigure}{.29\linewidth}
\judgebox{
\tunnels{\lbl}{\EC}
}
\begin{mathpar}
\Rule{}{
\tunnels{\lbl}{\hole}
}

\Rule{
\tunnels{\lbl}{\EC}
}{
\tunnels{\lbl}{\app{\EC}{\eff}}
}

\Rule{
\tunnels{\lbl}{\EC}
}{
\tunnels{\lbl}{\app{\EC}{\lbl[']}}
}

\Rule{
\tunnels{\lbl}{\EC}
}{
\tunnels{\lbl}{\app{\EC}{\tm}}
}

\Rule{
\tunnels{\lbl}{\EC}
}{
\tunnels{\lbl}{\app{\val}{\EC}}
}

\Rule{
\tunnels{\lbl}{\EC}
}{
\tunnels{\lbl}{\up{\EC}}
}

\Rule{
\tunnels{\lbl}{\EC}
}{
\tunnels{\lbl}{\letexp{\x}{\EC}{\tm}}
}

\Rule{
\tunnels{\lbl}{\EC}
}{
\tunnels{\lbl}{\throw{\EC}{\tm}}
}

\Rule{
\tunnels{\lbl}{\EC}
}{
\tunnels{\lbl}{\unroll{\EC}}
}

\Rule{
\tunnels{\lbl}{\EC}
\\
\lbl \neq \lbl[0]
}{
\tunnels{\lbl}{\down{\lbl[0]}{\EC}}
}

\end{mathpar}
 \end{subfigure}
\caption[]{Operational semantics of \lamlang}
\label{fig:operational}
\end{figure}

To give an operational semantics to \lamlang, terms and evaluation
contexts in \cref{fig:syntax} are extended with a \reset-construct:
\begin{align*}
\small
\begin{array}{@{}l@{\hspace{0.6em}}c@{\ }c@{\ }l@{\hspace{2.5em}}
l@{\hspace{0.6em}}c@{\ }c@{\ }l@{}}
\textit{terms} & \tm,\tm[s] & ::= &
... \bnf
\down{\lbl}{\tm}
&
\textit{evaluation contexts} & \EC & ::= &
... \bnf
\down{\lbl}{\EC}
\end{array}
\end{align*}

\noindent
\cref{fig:operational} defines the small-step operational semantics.
Individual reduction steps take the form
$\steps{\lst{\lbl[1]}}{\tm_1}{\lst{\lbl[2]}}{\tm_2}$,
meaning that term $t_1$ steps to term $t_2$ while the set of
freshly created lifetime constants possibly grows from $\lst{\lbl[1]}$ to
$\lst{\lbl[2]}$.
Per rule~\ruleref{down}, a lifetime constant~{\lbl[1]} declared by a
\freset-term is replaced by a fresh copy~{\lbl[2]} when the \freset-term is reduced to a
\reset-term.
While \freset-terms lexically bind lifetime constants,
\reset-terms are non-binding constructs; evaluation contexts of
form~{\down{\lbl}{\EC}} serve as stack delimiters.
This use of freshness is analogous to how calculi with reference
cells allocate fresh memory locations;
closed terms can mention fresh identifiers.
The distinction between~{\freset} and~{\reset} is not apparent
in \citet{zm18}, albeit present in their Coq formalization;
\citet{bpps2020} clarify the distinction.

Rule~\ruleref{op} exposes the operation implementation of a handler by
unrolling the fixpoint handler value.
Rule \ruleref{downup} handles operation invocations.
To execute the operation's implementation, a resumption must be
found to substitute for the free variable~{\x[k]}.
Because the operation value {\obj{\lam{\x[k]}{\tm}}{\lbl[0]}} has
lifetime~$\lbl[0]$,
the surrounding evaluation context is searched for a stack
delimiter~{\down{\lbl[0]}{\!}}.
The part of the evaluation context delimited
by~{\down{\lbl[0]}{\!}} is then used as the resumption.
In comparison, prior formalisms of lexically scoped handlers would use
the value {\lam{\x[y]}{\down{\lbl[0]}{\appctx{\EC}{\x[y]}}}}, rather than
{\cont{\down{\lbl[0]}{\EC}}}, as the resumption.
Because handlers are deep, the resumption
has~{\down{\lbl[0]}{\!}} at its outermost layer.
When the operation value corresponds to an ordinary function, its lifetime
$\lbl[0]$ must have been introduced by evaluating the {\DOWN{\lbl[0]}{\!}} guarding the
``main'' term.
So what \ruleref{downup} does in this case is essentially calling
the function with the ``current continuation'' (in Scheme parlance).

\subsection{Static Semantics}
\label{sec:static}

\begin{figure}
\small
\begin{minipage}{.21\textwidth}
\judgebox{
\wftm{\EffEnv}{\LifeEnv}{\TmEnv}{\LEnv}{\tm}{\ty}{\eff}
}
\end{minipage}
\hfill
\begin{minipage}{.75\textwidth}
$
\EffEnv\ ::=\ 
\nil \bnf
\EffEnv,\,\evar
\hspace{1.35em}
\LifeEnv\ ::=\ 
\nil \bnf
\LifeEnv,\,\lvar
\hspace{1.35em}
\TmEnv\ ::=\ 
\nil \bnf
\TmEnv,\,\x \hasty \ty
\hspace{1.35em}
\LEnv\ ::= \ 
\nil \bnf
\LEnv,\,\lbl \hasty \raises{\ty}{\eff}
$
\end{minipage}
\begin{mathpar}
\Rule[name=t-unit]{}{
    \wftm{\EffEnv}{\LifeEnv}{\TmEnv}{\LEnv}{\unit}{\Unit}{\nil}
}

\Rule[name=t-var]{
    \lookup{\TmEnv}{\x} = \ty
}{
    \wftm{\EffEnv}{\LifeEnv}{\TmEnv}{\LEnv}{\x}{\ty}{\nil}
}

\Rule[name=t-up]{
    \wftm{\EffEnv}{\LifeEnv}{\TmEnv}{\LEnv}{
        \tm
    }{
        \mtype{\raises{\ty}{\eff[1]}}{\lbl}
    }{
        \eff[2]
    }
}{
    \wftm{\EffEnv}{\LifeEnv}{\TmEnv}{\LEnv}{
        \up{\tm}
    }{
        \ty
    }{
       \eff[1],\,\eff[2],\,\lbl
    }
}

\Rule[name=t-op]{
    \wftm{\EffEnv,\,\lst{\evar}}{\LifeEnv}{\TmEnv}{\LEnv}{
        \tm
    }{
        \itype{\iname}{\lst{\eff[1]}}{\lbl}
    }{
        \eff[2]
    }
    \\\\
    \msigof{\iname} = \MkISig{\iname}{\lst{\evar}}{\msig}
}{
    \wftm{\EffEnv}{\LifeEnv}{\TmEnv}{\LEnv}{
        \unroll{\tm}
    }{
        \mtype{\subst{\msig}{\lst{\eff[1]}}{\lst{\evar}}}{\lbl}
    }{
        \eff[2]
    }
}

\Rule[name=t-fix]{
    \msigof{\iname} = \MkISig{\iname}{\lst{\evar}}{\msig}
    \\\\
    \wftm{\EffEnv}{\LifeEnv}{
        \TmEnv,\,
        \self \hasty {\itype{\iname}{\lst\eff}{\lbl}}
    }{\LEnv}{
        \obj{\mdef}{\lbl}
    }{
        \mtype{\subst{\msig}{\lst\eff}{\lst\evar}}{\lbl}
    }{
        \nil
    }
}{
    \wftm{\EffEnv}{\LifeEnv}{\TmEnv}{\LEnv}{
        \fixpoint{\self}{\obj{\mdef}{\lbl}}
    }{
        \itype{\iname}{\lst\eff}{\lbl}
    }{
        \nil
    }
}

\Rule[name=t-elam]{
    \wftm
    {\EffEnv,\,\evar}{\LifeEnv}{\TmEnv}{\LEnv}
    {\obj{\mdef}{\lbl}}
    {\mtype{\msig}{\lbl}}
    {\nil}
}{
    \wftm
    {\EffEnv}{\LifeEnv}{\TmEnv}{\LEnv}
    {\obj{\Lam{\evar}{\mdef}}{\lbl}}
    {\mtype{\efunty{\evar}{\msig}}{\lbl}}
    {\nil}
}

\Rule[name=t-llam]{
    \wftm
    {\EffEnv}{\LifeEnv,\,\lvar}{\TmEnv}{\LEnv}
    {\obj{\mdef}{\lbl}}
    {\mtype{\msig}{\lbl}}
    {\nil}
}{
    \wftm
    {\EffEnv}{\LifeEnv}{\TmEnv}{\LEnv}
    {\obj{\Lam{\lvar}{\mdef}}{\lbl}}
    {\mtype{\lfunty{\lvar}{\msig}}{\lbl}}
    {\nil}
}

\Rule[name=t-lam]{
    \wftm
    {\EffEnv}{\LifeEnv}{\TmEnv,\,\x\hasty\ty}{\LEnv}
    {\obj{\mdef}{\lbl}}
    {\mtype{\msig}{\lbl}}
    {\nil}
}{
    \wftm
    {\EffEnv}{\LifeEnv}{\TmEnv}{\LEnv}
    {\obj{\lam{\x}{\mdef}}{\lbl}}
    {\mtype{\fnty{\ty}{\msig}}{\lbl}}
    {\nil}
}

\Rule[name=t-klam]{
    \lookup{\LEnv}{\lbl}=\raises{\ty_2}{\eff[2]}
    \\\\
    \wftm
    {\EffEnv}{\LifeEnv}
    {
        \TmEnv,\,
        \x[k] \hasty
        \tycont{\ty_1}{\eff[1],\lbl}{\ty_2}{\eff[2]}
    }
    {\LEnv}
    {\tm}
    {\ty_2}{\eff[2]}
}{
    \wftm
    {\EffEnv}{\LifeEnv}{\TmEnv}{\LEnv}
    {\obj{\lam{\x[k]}{\tm}}{\lbl}}
    {\mtype {\raises{\ty_1}{\eff[1]}} {\lbl}}
    {\nil}
}

\Rule[name=t-eapp,sep=1em]{
    \wftm{\EffEnv}{\LifeEnv}{\TmEnv}{\LEnv}{
        \tm
    }{
        \mtype{\efunty{\evar}{\msig}}{\lbl}
    }{
        \eff[2]
    }
    \\
    \wfef{\EffEnv}{\LifeEnv}{\LEnv}{\eff[1]}
}{
    \wftm{\EffEnv}{\LifeEnv}{\TmEnv}{\LEnv}{
        \eapp{\tm}{\eff[1]}
    }{
        \mtype{
            \subst{
                \msig
            }{\eff[1]}{\evar}
        }{\lbl}
    }{
        \eff[2]
    }
}

\Rule[name=t-lapp,sep=1em]{
    \wftm{\EffEnv}{\LifeEnv}{\TmEnv}{\LEnv}{
        \tm
    }{
        \mtype{\efunty{\lvar}{\msig}}{\lbl}
    }{
        \eff
    }
    \\
    \wflife{\LifeEnv}{\LEnv}{\life}
}{
    \wftm{\EffEnv}{\LifeEnv}{\TmEnv}{\LEnv}{
        \eapp{\tm}{\life}
    }{
        \mtype{
            \subst{
                \msig
            }{\life}{\lvar}
        }{\lbl}
    }{
        \eff
    }
}

\Rule[name=t-app]{
    \wftm{\EffEnv}{\LifeEnv}{\TmEnv}{\LEnv}{
        \tm
    }{
        \mtype{\fnty{\ty}{\msig}}{\lbl}
    }{
        \eff
    }
    \\\\
    \wftm{\EffEnv}{\LifeEnv}{\TmEnv}{\LEnv}{
        \tm[s]
    }{
        \ty
    }{
        \eff
    }
}{
    \wftm{\EffEnv}{\LifeEnv}{\TmEnv}{\LEnv}{
        \app{\tm}{\tm[s]}
    }{
        \mtype{\msig}{\lbl}
    }{
        \eff
    }
}

\Rule[name=t-down]{
    \wftm{\EffEnv}{\LifeEnv}{\TmEnv}{\LEnv,\,\lbl\hasty\raises{\ty}{\eff}}
    {\tm}
    {\ty}{\eff,\,\lbl}
    \\\\
    \wfty{\EffEnv}{\LifeEnv}{\LEnv}{\ty}
    \\
    \wfef{\EffEnv}{\LifeEnv}{\LEnv}{\eff}
}{
    \wftm{\EffEnv}{\LifeEnv}{\TmEnv}{\LEnv}
    {\DOWN{\lbl}{\tm}}
    {\ty}{\eff}
}

\Rule[name=t-cont]{
    \wfktx{\EffEnv}{\LifeEnv}{\TmEnv}{\LEnv}
    {\EC}
    {\ty_1}{\eff[1]}{\ty_2}{\eff[2]}
}{
    \wftm{\EffEnv}{\LifeEnv}{\TmEnv}{\LEnv}{
        \cont{\EC}
    }{
        \tycont{\ty_1}{\eff[1]}{\ty_2}{\eff[2]}
    }{\nil}
}

\Rule[name=t-let]{
    \wftm{\EffEnv}{\LifeEnv}{\TmEnv}{\LEnv}
    {\tm[s]}
    {\ty[S]}{\eff}
    \\\\
    \wftm{\EffEnv}{\LifeEnv}{\TmEnv,\,\x\hasty\ty[S]}{\LEnv}
    {\tm}
    {\ty}{\eff}
}{
    \wftm{\EffEnv}{\LifeEnv}{\TmEnv}{\LEnv}
    {\letexp{\x}{\tm[s]}{\tm}}
    {\ty}{\eff}
}

\Rule[name=t-throw]{
    \wftm{\EffEnv}{\LifeEnv}{\TmEnv}{\LEnv}
    {\tm}
    {\tycont{\ty_1}{\eff[1]}{\ty_2}{\eff[2]}}{\eff[2]}
    \\\\
    \wftm{\EffEnv}{\LifeEnv}{\TmEnv}{\LEnv}
    {\tm[s]}
    {\ty_1}{\eff[1]}
}{
    \wftm{\EffEnv}{\LifeEnv}{\TmEnv}{\LEnv}
    {\throw{\tm}{\tm[s]}}
    {\ty_2}{\eff[2]}
}

\Rule[name=t-sub]{
    \wftm{\EffEnv}{\LifeEnv}{\TmEnv}{\LEnv}{\tm}{\ty_1}{\eff[1]}
    \\\\
    \subeff{\EffEnv}{\LifeEnv}{\LEnv}{\eff[1]}{\eff[2]}
    \\
    \subty{\EffEnv}{\LifeEnv}{\LEnv}{\ty_1}{\ty_2}
}{
    \wftm{\EffEnv}{\LifeEnv}{\TmEnv}{\LEnv}{\tm}{\ty_2}{\eff[2]}
}

\end{mathpar}
 \caption[]{Rules for typing \lamlang terms}
\label{fig:static-semantics-selected}
\end{figure}

\cref{fig:static-semantics-selected} presents the term-typing rules of
\lamlang.
Term typing rules have form
\wftm{\EffEnv}{\LifeEnv}{\TmEnv}{\LEnv}{\tm}{\ty}{\lst{\ef}}, where
{\EffEnv}, {\LifeEnv}, {\TmEnv}, and {\LEnv} are binding
contexts.
The judgment form says that under these environments term~{\tm} has
type~{\ty} and effects~{\lst\ef}.
Rule~\ruleref{t-up} types operation invocations. The effects of
this term include the operation value's own lifetime and the
effects in the operation's result type.

Rule~\ruleref{t-down} shows that the lifetime constant~{\lbl} declared 
by a {\DOWN{\lbl}{\tm}} term can appear in the effects of the
computation guarded by~{\DOWN{\lbl}{\!}}:
to type-check~{\tm}, the environment~{\LifeEnv} of lifetime constants is
augmented with~{\lbl}. The environment also tracks the type and
effects \raises{\ty}{\eff} of the entire computation~\DOWN{\lbl}{\tm}.
Importantly, however, {\lbl} must _not_ occur free in
{\raises{\ty}{\eff}}; 
while~{\tm} has lifetime~{\lbl}, \DOWN{\lbl}{\tm} lives beyond~{\lbl}.
Notice that we do not give a typing rule for the
auxiliary~\reset-form, which only emerges when evaluating a \lamlang
program.

Rule~\ruleref{t-fix} type-checks a handler value. The operation
of the handler is type-checked with the self reference in scope.
The self reference has the same type
{\itype{\iname}{\lst\eff}{\lbl}}
as the handler value.
Using the self reference triggers the lifetime effect~\lbl.

Rule~\ruleref{t-klam} type-checks an operation value whose
parameters (except for the resumption) are already in scope.
A salient difference from previous work is that the type of the
resumption, apart from being a continuation type, does not have
to exactly match the result type and effect
{\raises{\ty_1}{\eff[1]}} of the operation:
the continuation can be applied to a computation with the
additional lifetime effect~{\lbl} that is the lifetime of the
current value.
A consequence is that the computation the resumption is applied
to is allowed to use the self reference to handle its effects.

Rules for type-level well-formation and orderings
\ifreport
    can be found in \cref{sec:static-semantics-supplement}.
\else
    are deferred to the accompanying technical report~\cite{ufo-tr}.
\fi
Because composite effects are sets, the subeffecting relation
$\subeff{\EffEnv}{\LifeEnv}{\LEnv}{\eff[1]}{\eff[2]}$
simply says $\eff[1]$ is a subset of $\eff[2]$.

\section{Establishing Parametricity for a Logical-Relations Model}
\label{sec:results}

As \cref{sec:loss-of-parametricity} argues, the key property to
establish about \lamlang should be parametricity.
To this end,
this section develops a logical-relations model for~\lamlang, and
shows it satisfies parametricity and is sound with respect to
contextual equivalence.
These results, fully mechanized in Coq, provide a rigorous
account for abstraction safety and also imply type safety.

\subsection{A Logical-Relations Model for \lamlang}
\label{sec:logrel}

\begin{figure*}
\newcommand\LABEL[1]{\text{\itshape#1}}
\begin{minipage}{\linewidth}
\judgebox{\LABEL{Semantic Types}}
\begin{align*}
\renewcommand\VSEP{\vspace{1ex}\\}\renewcommand\VSep{\vspace{0ex}\\}\begin{array}{@{}r@{\ \ \ }c@{\ \ \ }l@{}}
\inTerR
    {\W}{\val_1}{\val_2}
    {\SemTy{\Unit}{\lsub}{\esub}}
& \defeq &
\val_1=\unit \AND \val_2=\unit
\VSEP
\inTerR
    {\W}{\val_1}{\val_2}
    {\SemTy{\itype{\iname}{\lst\eff}{\life}}{\lsub}{\esub}}
& \defeq &
    \exists \msig.\ 
    \msigof{\iname}=\MkISig{\iname}{\lst{\evar}}{\msig}
    \AND
    \VSep & &
    \exists \lbl[1], \mdef_1, \lbl[2], \mdef_2.\ 
    \val_i = \fixpoint{\self}{\obj{\mdef_i}{\lbl[i]}} \onetwo
    \AND
    \VSep & &
    \later{
        \inTerR
        {\W}{\obj{\mdef_1}{\lbl[1]}}{\obj{\mdef_2}{\lbl[2]}}
        {\SemTy{\mtype{\subst{\msig}{\lst\eff}{\lst{\evar}}}{\life}}{\lsub}{\esub}}
    }
\VSEP
\inTerR
    {\W}{\val_1}{\val_2}
    {\SemTy{\mtype{\msig}{\life}}{\lsub}{\esub}}
& \defeq &
    \exists \lbl[1], \mdef_1, \lbl[2], \mdef_2.\ 
    \lsubst{i}{\life} = \lbl[i]
    \AND
    \val_i = \obj{\mdef_i}{\lbl[i]} \onetwo
    \AND
    \VSep & &
    \inQuaR
    {\W}{\life}
    {\mdef_1}{\mdef_2}
    {\SemTy{\msig}{\lsub}{\esub}}
\VSEP
\inTerR
    {\W}{\val_1}{\val_2}
    {\SemTy{\tycont{\ty_1}{\eff[1]}{\ty_2}{\eff[2]}}{\lsub}{\esub}}
& \defeq &
    \exists \EC_1, \EC_2.\ 
    \val_i = \cont{\EC_i} \onetwo
    \AND
    \inTerR
    {\W}{\EC_1}{\EC_2}
    {\DContR{\ty_1}{\eff[1]}{\ty_2}{\eff[2]}{\lsub}{\esub}}
\VSEP
\end{array}
\end{align*}

\judgebox{\LABEL{Semantic Operation Signatures}}
\begin{align*}
\begin{array}{@{}r@{\ \ \ }c@{\ \ \ }l@{}}
\inQuaR
    {\W}{\life}{\mdef_1}{\mdef_2}
    {\SemTy{\efunty{\evar}{\msig}}{\lsub}{\esub}}
& \defeq &
    \exists \mdef_1', \mdef_2'.\ 
    \mdef_i = \lfun{\evar}{\mdef_i'}\onetwo
    \AND
    \forall \W['], \lst{\lbl[1]}, \lst{\lbl[2]}, \rel[\psi].\ 
    \extW{\W}{\W[']}
    \implies
    \VSep & &
    \extW{\rel[\psi]}{\W[']}
    \implies
    \inQuaR
    {\W'}
    {\life}
    {\subst{\mdef_1'}{\lst{\lbl[1]}}{\evar[1]}}
    {\subst{\mdef_2'}{\lst{\lbl[2]}}{\evar[2]}}
    {\SemTy{\msig}{\lsub}{\esubext{\evar}{\lst{\lbl[1]}}{\lst{\lbl[2]}}{\rel[\psi]}}{\esub}}
\VSEP
\inQuaR
    {\W}{\life}{\mdef_1}{\mdef_2}
    {\SemTy{\lfunty{\lvar}{\msig}}{\lsub}{\esub}}
& \defeq &
    \exists \mdef_1', \mdef_2'.\ 
    \mdef_i = \lfun{\lvar}{\mdef_i'}\onetwo
    \AND
    \forall \W['], \lbl[1], \lbl[2], \rel'.\ 
    \extW{\W}{\W[']}
    \implies
    \VSep & &
    \lbl[i] \in \W['i] \onetwo
    \implies
    \inQuaR
    {\W'}
    {\life}
    {\subst{\mdef_1'}{\lbl[1]}{\lvar[1]}}
    {\subst{\mdef_2'}{\lbl[2]}{\lvar[2]}}
    {\SemTy{\msig}{
        \lsub,\lsubext{\lvar}{\lbl[1]}{\lbl[2]}{\rel'}
    }{\esub}}
\VSEP
\inQuaR
    {\W}{\life}{\mdef_1}{\mdef_2}
    {\SemTy{\fnty{\ty}{\msig}}{\lsub}{\esub}}
& \defeq &
    \exists \mdef_1', \mdef_2'.\ 
    \mdef_i = \lam{\x}{\mdef_i'} \onetwo
    \AND
    \forall \W['], \val_1, \val_2.\ 
    \extW{\W}{\W[']}
    \implies
    \VSep & &
    \inTerR
      {\W[']}{\val_1}{\val_2}
      {\SemTy{\ty}{\lsub}{\esub}}
    \implies
    \inQuaR
      {\W[']}
      {\life}
      {\subst{\mdef_1'}{\val_1}{\x}}
      {\subst{\mdef_2'}{\val_2}{\x}}
      {\SemTy{\msig}{\lsub}{\esub}}
\VSEP
\inQuaR
    {\W}{\life}{\mdef_1}{\mdef_2}
    {\SemTy{\raises{\ty}{\eff}}{\lsub}{\esub}}
& \defeq &
    \exists \tm_1, \tm_2.\ 
    \mdef_i = \lam{\x[k]}{\tm_i} \onetwo
    \AND
    \forall \W['], \EC_1, \EC_2.\ 
    \extW{\W}{\W[']}
    \implies
    \VSep & &
    \inTerR
        {\W[']}{\EC_1}{\EC_2}
        {
            \DContRRaw
            {\TmR{\ty}{\eff,\life}{\lsub}{\esub}}
            {\SemLife{\life}{\lsub}{\esub}}
        }
    \implies
    \inTerR
        {\W[']}
        {\subst{\tm_1}{\cont{\EC_1}}{\x[k]}}
        {\subst{\tm_2}{\cont{\EC_2}}{\x[k]}}
        {\SemLife{\life}{\lsub}{\esub}}
\VSEP
\end{array}
\end{align*}

\judgebox{\LABEL{Semantic Effects}}
\begin{align*}
\begin{array}{@{}r@{\ \ \ }c@{\ \ \ }l@{}}
\inSenR
    {\W}{\tm_1}{\tm_2}{\rel}{\lst{\lbl[1]}}{\lst{\lbl[2]}}
    {\SemEff{\evar}{\lsub}{\esub}}
& \defeq &
    \exists \rel[\psi].\ 
    \lookup{\esub}{\evar}=\rel[\psi]
    \AND
    \inSenR
        {\W}{\tm_1}{\tm_2}{\rel}{\lst{\lbl[1]}}{\lst{\lbl[2]}}
        {\rel[\psi]}
\VSEP
\inSenR
    {\W}{\tm_1}{\tm_2}{\rel}{\lbl[1]}{\lbl[2]}
    {\SemEff{\life}{\lsub}{\esub}}
& \defeq &
    \exists \lbl[1],\lbl[2],\tm[s]_1,\tm[s]_2.\ 
    \lsubst{i}{\life}=\lbl[i]
    \AND
    \tm_i=\up{\obj{\lam{\x[k]}{\tm[s]_i}}{\lbl[i]}}\onetwo
    \AND
    \VSep & &
    \forall \W['],\EC_1,\EC_2.\ 
    \extW{\W}{\W[']}
    \implies
    \later{
        \inTerR
            {\W[']}{\EC_1}{\EC_2}
            {\DContRRaw{\rel}{\SemLife{\life}{\lsub}{\esub}}}
    }
    \implies
    \VSep & &
    \later{
        \inTerR
            {\W[']}
            {\subst{\tm[s]_1}{\cont{\EC_1}}{\x[k]}}
            {\subst{\tm[s]_2}{\cont{\EC_2}}{\x[k]}}
            {\SemLife{\life}{\lsub}{\esub}}
    }
\VSEP
\inSenR
    {\W}{\tm_1}{\tm_2}{\rel}{\lst{\lbl[1]}}{\lst{\lbl[2]}}
    {\SemEff{\eff}{\lsub}{\esub}}
& \defeq &
    \exists \ef \in \eff.\
    \inSenR
        {\W}{\tm_1}{\tm_2}{\rel}{\lst{\lbl[1]}}{\lst{\lbl[2]}}
        {\SemEff{\ef}{\lsub}{\esub}}
\VSEP
\end{array}
\end{align*}

\judgebox{
  \LABEL{Auxiliary Relations}
}
\begin{align*}
\begin{array}{@{}r@{\ \ \ }c@{\ \ \ }l@{}}
\inTerR
    {\W}{\EC_1}{\EC_2}
    {\DContRRaw{\rel_1}{\rel_2}}
& \defeq &
\forall \W['], \tm_1, \tm_2.\ 
\extW{\W}{\W[']}
\implies
\inTerR
    {\W[']}
    {\tm_1}
    {\tm_2}
    {\rel_1}
\implies
\inTerR
    {\W[']}
    {\EC_1[\tm_1]}
    {\EC_2[\tm_2]}
    {\rel_2}
\VSEP
\inTerR
    {\W}{\EC_1}{\EC_2}
    {\DContR{\ty_1}{\eff[1]}{\ty_2}{\eff[2]}{\lsub}{\esub}}
& \defeq &
\inTerR
    {\W}{\EC_1}{\EC_2}{
        \DContRRaw
        {\TmR{\ty_1}{\eff[1]}{\lsub}{\esub}}
        {\TmR{\ty_2}{\eff[2]}{\lsub}{\esub}}
    }
\VSEP
\inTerR
    {\W}{\tm_1}{\tm_2}
    {\SemLife{\lvar}{\lsub}{\esub}}
& \defeq &
\exists \rel.\ 
\lookup{\lsub}{\lvar} = \rel
\AND
\inTerR
    {\W}{\tm_1}{\tm_2}
    {\rel}
\VSEP
\inTerR
    {\W}{\tm_1}{\tm_2}
    {\SemLife{\lbl}{\lsub}{\esub}}
& \defeq &
\exists \ty, \eff.\ 
\lookup{\LEnv}{\lbl}=\raises{\ty}{\eff}
\AND
\inTerR
    {\W}{\tm_1}{\tm_2}
    {\TmR{\ty}{\eff}{\lsub}{\esub}}
\VSEP
\extW{\rel[\psi]}{\W}
& \defeq &
  \forall \W['],\tm_1,\tm_2,\rel,\lst{\lbl[1]},\lst{\lbl[2]}.\ 
  \inSenR
  {\W[']}{\tm_1}{\tm_2}{\rel}{\lst{\lbl[1]}}{\lst{\lbl[2]}}
  {\rel[\psi]}
  \implies
  \extW{\lst{\lbl[i]}}{\W[i]} \onetwo
\VSEP
\end{array}
\end{align*}
 \end{minipage}
\caption{
Relational interpretations of types, operation signatures, and
effects.
(The definitions are implicitly indexed by $\EffEnv$, $\LifeEnv$,
and $\LEnv$.)
}
\label{fig:semtypes}
\end{figure*}

Technical devices used in the definition include
step indexing~\cite{step-indexing,ahmed2006},
possible worlds, and biorthogonality~\cite{biorthogonality}
to deal with challenges such as Turing-completeness,
freshly generated lifetimes, and delimited continuations.
\cref{fig:semtypes} presents the semantic interpretation of
various type-level entities.
Other definitions---including observational refinement,
the relation on closed terms
\vcrunch{\inTerR{\W}{\tm_1}{\tm_2}{\TmR{\ty}{\eff}{\lsub}{\esub}}}, and
its lifting to open terms
\vcrunch{\logref{\EffEnv}{\LifeEnv}{\TmEnv}{\LEnv}{\tm_1}{\tm_2}{\ty}{\eff}}---are largely standard and
\ifreport
    can be found in \cref{sec:logical-relations-supplement}.
\else
    for space are deferred to the technical report.
\fi

A logical-relations model for a typed language
interprets types as relations.
As is standard, the relational interpretations in
\cref{fig:semtypes} are indexed by substitutions {\esub} and
{\lsub} that provides---in addition to syntactic substitution functions (denoted by
$\esubst{i}{}$ and $\lsubst{i}{}$ where $i=1,2$) for free effect
variables and lifetime variables in the type-level entity being
interpreted---semantic interpretation of the free variables.
The logical relations are also indexed by a world $\W$ of freshly
created lifetime constants;
{\extW{\W}{\W[']}} means world {\W[']} is a future world of
(i.e., extends) {\W}.

Language features like recursion make it difficult
to define these relations inductively on the structure of types.
Fixpoint handlers and mutually recursive interfaces in
\lamlang pose a similar challenge.
The technique of step indexing~\cite{step-indexing,ahmed2006}
addresses this challenge.
Our logical relation is step-indexed; the logical relation is
defined using a double induction, first on a step index, and
second on the structure of types.
The definition is given in terms of a logic equipped with the
``later'' modality~{\latersym}, which offers a clean abstraction
of step indexing~\cite{amrv2007,lslr}.
For example, the relational interpretation of an interface type
\vcrunch{\SemTy{\itype{\iname}{\lst\eff}{\life}}{\lsub}{\esub}}
is defined as that of an operation type
\vcrunch{\SemTy{\mtype{\subst{\msig}{\lst\eff}{\lst{\evar}}}{\life}}{\lsub}{\esub}}
guarded by~{\latersym}.
Although {\msig} may recursively mention~{\iname}, the use
of~{\latersym} ensures the definition remains well-founded.

The relational interpretation of continuation types
\vcrunch{\SemTy{\tycont{\ty_1}{\eff[1]}{\ty_2}{\eff[2]}}{\lsub}{\esub}}
is defined in terms a relation on evaluation contexts
\vcrunch{\DContR{\ty_1}{\eff[1]}{\ty_2}{\eff[2]}{\lsub}{\esub}}.
The latter relation is a standard auxiliary definition in
logical-relations proofs.
Two evaluation contexts~$\EC_1$ and~$\EC_2$ are in this relation
if applying them to terms~$\tm_1$ and~$\tm_2$
related by~\vcrunch{{\TmR{\ty_1}{\eff[1]}{\lsub}{\esub}}} implies
the resulting terms {\appctx{\EC_1}{\tm_1}}
and {\appctx{\EC_2}{\tm_2}} are related
by \vcrunch{\TmR{\ty_2}{\eff[2]}{\lsub}{\esub}}.

The interpretation of an operation type
\vcrunch{\SemTy{\mtype{\msig}{\life}}{\lsub}{\esub}}
is defined in terms of that of the operation signature
\vcrunch{\SemTy{\msig}{\lsub}{\esub}}, indexed
on the lifetime~{\life}.
Of particular interest is the interpretation
\vcrunch{\SemTy{\raises{\ty}{\eff}}{\lsub}{\esub}}.
It relates two operation implementations
{\lam{\x[k]}{\tm_1}}
and~{\lam{\x[k]}{\tm_2}} in which~$\tm_1$ and~$\tm_2$ are related
when the free variables~{\x[k]} standing for resumptions are
replaced by related continuations.
The continuations are allowed to be related by 
\vcrunch{
  \DContRRaw
  {\TmR{\ty}{\eff,\life}{\lsub}{\esub}}
  {\SemLife{\life}{\lsub}{\esub}}
}, where the occurrence of {\life} in addition to {\eff}
indicates recursive handling by the fixpoint handler,
corresponding to the premise of typing rule \ruleref{t-klam}.

The semantic interpretation of atomic or composite lifetime effects
\vcrunch{\SemEff{\ef}{\lsub}{\esub}} or
\vcrunch{\SemEff{\eff}{\lsub}{\esub}}
follows techniques developed in \citet{bpps2018} and
\citet{zm18}.

\subsection{Results}
\label{sec:theorems}

Parametricity is a strong indicator that abstraction is
preserved~\cite{reynolds83,dreyer2018}.
It implies that effect-polymorphic functions behave uniformly,
irrespective of the choice of effects with which they are
instantiated.

\begin{restatable}
[Parametricity,
a.k.a.\ Abstraction Theorem, a.k.a.\ Fundamental Property]
{theorem}{parametricity}
If {\LEnv} and~{\TmEnv} are well-formed, then
$
    \wftm
    {\EffEnv}{\LifeEnv}{\TmEnv}{\LEnv}
    {\tm}{\ty}{\eff}
$
implies
$
    \logref
    {\EffEnv}{\LifeEnv}{\TmEnv}{\LEnv}
    {\tm}{\tm}{\ty}{\eff}
$.
\end{restatable}

Type safety means that well-typed \lamlang programs
do not get stuck---they either reduce to values or diverge.
Type safety follows from parametricity.

\begin{restatable}
[Type Safety]
{theorem}{typesafety}
If $
    \wftm
    {\nil}{\nil}{\nil}{\nil}
    {\tm}{\ty}{\nil}
$ and $
    \stepsrt{\nil}{\tm}{\lst{\lbl[']}}{\tm'}
$,
then either there exists $\val$ such that $\tm'=\val$
or there exists $\lst{\lbl['']}$ and $\tm''$ such that
$\steps{\lst{\lbl[']}}{\tm'}{\lst{\lbl['']}}{\tm''}$.
\end{restatable}

Abstraction is preserved when no clients can distinguish between
implementations of the same abstraction.
The gold standard of indistinguishability is _contextual
equivalence_~\cite{jhmorris-thesis}, whose definition in the
context of \lamlang can be found in
\ifreport
    \cref{sec:contextual-eq-supplement}.
\else
    the technical report.
\fi
If the logical-relations model is sound, in the sense that logically
related terms are contextually equivalent, then
indistinguishability can be established through logical relatedness.

\begin{restatable}[Soundness w.r.t.\ contextual equivalence]
{theorem}{soundness}
\label{thm:soundness}\ \\
$
    \logref
    {\EffEnv}{\LifeEnv}{\TmEnv}{\LEnv}
    {\tm_1}{\tm_2}{\ty}{\eff}
    \implies
    \ctxref
    {\EffEnv}{\LifeEnv}{\TmEnv}{\LEnv}
    {\tm_1}{\tm_2}
    {\ty}{\eff}
$.
\end{restatable}

These results prove our claim that the type system of
\lamlang upholds strong abstraction boundaries.

\subsection{Formalization in Coq}

The formal results above have been mechanized using the Coq proof
assistant, in 17,800 lines of code
architected similarly to prior work~\cite{zm18,bpps2018}.
We also wrote
a 400-line extension of the IxFree
library~\cite{IxFree}---a shallow embedding of a logic
with the \latersym~modality~\cite{lslr}---to implement the
logical relations as dependently typed fixpoint functions.
Cofinite quantification~\cite{LN-cofinite} makes it
easy to generate fresh lifetime constants.
This Coq implementation is available at
\url{https://github.com/yizhouzhang/olaf-coq}.

\section{Implementation Issues}
\label{sec:compile}

While we leave a full-featured compiler to future work, we
discuss two key compilation issues.

\subsection{Tail-Resumption Optimization}

Being able to apply the tail-resumption optimization is important
because calling resumptions at the tail position is probably the
most common way handler resumptions are used in practice---in the
preceding examples, all handlers except those for "exn" or
"await" are tail-resumptive.
The optimization avoids having to capture the handler resumption
as a first-class value, which would otherwise involve copying
stack frames, a rather expensive operation.

Bidirectional handlers can benefit from this optimization too.
Per rule \ruleref{downup}, a tail-resumptive handler in \lamlang
is reduced as follows:
\begin{gather*}
\stepconfig{\lst{\lbl}}{
    \down{\lbl[0]}{
        \appctx{\EC}{
            \up{\obj{\lam{\x[k]}{\throw{\x[k]}{\tm}}}{\lbl[0]}}
        }
    }
}
\step
\stepconfig{\lst{\lbl}}{
    \throw{\left(
        \cont{\down{\lbl[0]}{\!\EC}}
    \right)}{\tm}
}
\step
\stepconfig{\lst{\lbl}}{
    \down{\lbl[0]}{
        \appctx{\EC}{\tm}
    }
}
\end{gather*}
\noindent
There is no need to reify the delimited continuation
{\cont{\down{\lbl[0]}{\!\EC}}}.
It remains as the surrounding evaluation context after two
steps of reduction.

\subsection{Translation into Unidirectional Effect Handlers}
\label{sec:translation}

An obvious compilation target for bidirectional handlers is a
language with deep, ordinary effect handlers, and
an obvious approach to this compilation is as follows:
effect operations are translated to return callbacks, and
invocations of operations are translated to call the callbacks.
For example, effects "Ping" and "Pong" from \cref{fig:comm-sig}
are translated into these signatures:

\begin{centered}
\begin{minipage}{50ex}
\begin{lstlisting}[
xleftmargin=0ex,morekeywords={effect},
frame=none,backgroundcolor=\color{white},
]
effect Ping {(*\;*)def ping() : ()(*\;→\;*)(*\VOID\RAISES*)Pong(*\,|\,*)Ping(*\;*)}
effect Pong {(*\;*)def pong() : ()(*\;→\;*)(*\VOID\RAISES*)Ping(*\,|\,*)Pong(*\;*)}
\end{lstlisting}
\end{minipage}
\end{centered}

\noindent
The type of the returned callback additionally includes the effect
being translated, so that the callback computation can raise
effects that are to be handled by the (deep) handler being
defined.
Whether such a type-preserving translation is feasible in general
should be examined per target language---_macro expressivity_~\cite{felleisen-expressivity} in the context
of control-flow mechanisms is
sensitive to the precise set of cross-cutting features
under consideration~\cite{fklp2017}.

Importantly, the translation outlined above bears unpleasant performance
implications:
in a tail-resumptive setting, the cost of communicating through
callbacks could be avoided if a handler computation were
allowed to directly raise effects to the surrounding evaluation
context.

\newcommand\ucpp{$\mu$C++\xspace}

To understand this cost empirically, we use a modified
implementation of \ucpp~\cite{uC++2019}.
The \ucpp language extends C++~\cite{C++}
with effect handlers that are either abortive or tail-resumptive.
The modified version allows tail-resumptive effect handlers
to directly raise effects to their resumptions.

Historically, in the absence of a static effect system,
bidirectional control flow has been banned
because it is considered too complex to reason about or use.
For example, Mesa \cite{Mitchell79}, one of the few
languages with resumable exceptions, forbids recursive handling.
Similarly, \ucpp does not check effects statically, and the
original implementation uses an extra run-time check to prevent
effects raised by a handler from
being handled by the handler resumption \cite[§5.5]{uC++2019}.
Our type system addresses this concern by offering reasoning
principles for bidirectional, recursive effect handling.

We performed two hand translations of the ping-pong
communication example (\cref{sec:interprocess}, with "ponger" implemented
as in \cref{sec:deep-handlers-are-recursive}) into \ucpp,
with one of the two using callbacks.
This program was chosen because it
exercises high-frequency bidirectional control transfer.
We ran the hand-translated code using the modified \ucpp
implementation with the extra run-time check disabled,
and measured the running time on a 3.2GHz Intel Xeon
Gold processor, averaging 500 runs.
The translation relying on callbacks incurred a 2.1$\times$ slowdown:
42.6 ms vs.\@ 19.8 ms.
This result argues for bidirectional handlers as a
first-class language feature:
obtaining bidirectionality via a desugaring into callbacks is 
less efficient.
(In the other way around, compiling callbacks away into efficient
bidirectionality would involve a complex interprocedural
analysis.)

\section{Related Work}
\label{sec:related}

\paragraph{Control effects}

Effect handlers~\cite{pp2013,eff-lang} offer a form of _delimited
control_~\cite{evalctxts,shiftreset} (or _coroutining_~\cite{hfw86}),
together with a nice separation
between the syntax of effects and their semantics.
Whereas \citet{fklp2017} show effect handlers and (a particular
variant of) delimited-control operators fail to
macro-express~\cite{felleisen-expressivity} one another while
preserving typing,
\citet{pps2019} show they are equally expressive when their type
systems support polymorphic operations and answer-type
polymorphism~\cite{ak2007}, respectively.
The core language \lamlang further blurs the boundaries:
key elements of algebraic effects (i.e., effect
signatures and handlers) and those of delimited control (i.e., a
pair of control operators) coexist and play complementary roles
in \lamlang.

Bidirectionality is possible with effect handlers or
delimited-control operators using, for example, callbacks
(\cref{sec:translation})---however, as we discuss later, \lamlang is likely not
macro-expressible by recent formalisms that also lexically
scope effect handlers~\cite{zm18,bpps2020}.
Bidirectional handlers inherit the appeal of algebraic effects,
and address bidirectional control flow with
a straightforward programming style,
an economy of language constructs,
efficient compilation,
and strong reasoning principles.

\paragraph{Applications of bidirectional handlers}

Interruptible iterators~\cite{jmatch2} generalize the expressive
power of generators to allow concurrent updates to the
underlying collection being iterated over.
The example in Section~\ref{sec:coroutine-iterators-bidirectional} shows that bidirectional
algebraic effects capture the expressive power of interruptible
iterators but as part of a single unified effect mechanism with
formally defined semantics, rather than by introducing interrupts
as a separate mechanism.

A key motivation for promises and async--await as
language features was to enable better exception
handling; the state of a promise indicates if an exception
occurred asynchronously.  However, the lack of static checking on
asynchronous exceptions makes software error-prone~\cite{azmt2018}.
Previous encodings of async--await as algebraic effects are unsatisfactory:
as discussed in \cref{sec:async-exceptions}, they
either do not express asynchronous exceptions as an algebraic
effect~\cite{koka-async}, or
require ad hoc constructs and
do not track exceptions statically~\cite{dehmsw2017}.

Session types~\cite{session-types} are a behavioral-typing
discipline for communication protocols.
The possibility of encoding session types using algebraic effects
has been hypothesized before~\cite{flmd19}, and bidirectional
effects make this connection more substantial.
The encoding does not yet offer the full power of session types,
though; it enforces weaker session fidelity and does not prevent
deadlocks.
Linear effect handlers are a promising future direction.

\paragraph{Preventing accidental handling}

Accidental handling of algebraic effects in the presence of
effect polymorphism is a known problem.
Tunneling (lexically scoped, lifetime-bounded handlers)
as a way to avoid accidental handling of
exceptions was introduced by \citet{exceptions-pldi16};
follow-on work adapted it to explicit effect polymorphism and
proved parametricity~\cite{zm18}.
\citet{effekt-2020,bso2018} implement tunneled algebraic effects
in a Scala library, called Effekt, that encodes lifetime effects
through Scala's intersection types and path-dependent types.

\citet{bpps2020} distinguish between an _open_ and a
_generative_ semantics of lexically scoped handlers; generativity
is critical to ensuring parametricity when effect operations can
be effect-polymorphic.
\lamlang uses the generative semantics---its operational
semantics creates fresh lifetimes.
Generativity is occasionally found in prior work on control
effects (e.g., \cite{eff-lang,cupto}), but without effects
being statically tracked by a type-and-effect system.

We conjecture that \lamlang cannot be macro-expressed by prior calculi
of lexically scoped, generative handlers
(an inexpressivity proof
like that of \citet{fklp2017} is beyond the scope of this paper):
the calculus of \citet{zm18} does not support
effect-parameterized signatures (\cref{sec:accidental-handling})
or effect-polymorphic operations \cite{bpps2020},
either of which could help cause accidental handling;
and \citet{bpps2020} do not support recursively defined
signatures, needed to mimic fixpoint handlers
(\cref{sec:translation}).
Previous parametricity results do not carry over.

Another way to avoid accidental handling---in languages where composite effects are _rows_~\cite{wand1991}
rather than sets---is
directives to signal that effects should bypass the dynamically
closest enclosing handler.
These directives include
the _inject_\/ function of \citet{koka-lang-2014}, 
_lift_ and~_coercions_ of \citet{bpps2018,bpps2019},
and _adaptors_ of \citet{clmm2020}.
For example, the semantics of applying the lift construct
$\hole_{\mathit{A}}$ to a computation whose effect is a
polymorphic row~$\alpha$ is as follows: statically, the lifted computation
has effects~$\mathit{A}|\alpha$; dynamically, an~$\mathit{A}$
effect raised by the original compu\-ta\-tion is handled by the
dynamically _second_ closest enclosing $\mathit{A}$ handler.
An enclosing handler for $\mathit{A}$ thus cannot 
intercept $\mathit{A}$ effects raised by the
computation, because its effect $\alpha$ is not considered as a
subeffect of the row $\mathit{A}|\alpha$;
an explicit _lift_ is needed to please the type checker.

\paragraph{Generalizing algebraic effects}

\citet{lmm2017} treat Frank's handling construct
as a higher-order function that pattern-matches on
the effects of its computation-typed arguments.
This design decision is orthogonal to ours:
while we generalize effect signatures and handlers, Frank
generalizes the "try"--"with" construct.
It is plausible that either idea can be adapted to the other's setting,
though Frank's shallow-handling semantics is incompatible with
handlers being fixpoints.

\section{Conclusion}
\label{sec:conclusions}

This paper proposes a new design for effect handlers, 
in which a handler can raise effects and have its resumption---including
the handler itself---handle the effects.
As our examples show, these ideas address a need common
to assorted programming challenges for better
bidirectional communication between software components.
The expressive power falls out naturally when effect operations
and handlers are unified with methods and objects; however, the
ideas also generalize to non-object-oriented languages.
We captured the essence of the new mechanism in a core language
with some distinctive features such as fixpoint handlers and
the ability to treat handler resumptions as evaluation contexts.
Bidirectionality exposes previously unidentified ways to
accidentally handle effects that propagate per the usual,
signature-based semantics;
hence, we make bidirectional handlers lexically scoped and
focused on a convincing proof that they are
compatible with strong abstraction boundaries.
While a complete implementation is left to future work,
experiments suggest bidirectional handlers can be compiled
efficiently.

The recent flurry of language designs for advanced control-flow
features show that modern software needs language-based
support for complex control flow.
Bidirectional algebraic effects address this challenge in 
two important ways.
\snd{
First, they unify various previously separately proposed language
features (interruptible iterators, exceptional async/await, etc.)
via a natural generalization of effect handlers. This unification
should help increase
confidence in language metatheories and lower the hurdle for
use of powerful control-flow features.
Second, the guarantees that no effects are unhandled or accidentally handled
are critical to writing safer code.
They help the programmer manage the control-flow complexity via a
type system; static typing offers guidance on where to apply
effect handling, and the
parametricity guarantee enables truly compositional reasoning.
While these consequences are entirely anticipated, future
software-engineering studies could assess the empirical
effectiveness of bidirectional algebraic effects in achieving
these goals.
}

Together, these contributions offer an appealing way
to support complex control flow in mainstream programming languages.

\ifacknowledgments
\section*{Acknowledgments}

We thank 
Jonathan Brachth\"{a}user,
Peter Buhr,
Filip Sieczkowski,
and the anonymous reviewers
for their valuable feedback.
We also thank Peter Buhr for patching \ucpp.

This work was supported by NASA grant NNX16AB09G.  The views and
opinions expressed are those of the authors and do not
necessarily reflect the position of any government agency.
\fi

\ifreport\relax\else
    \appendix
    \section{ADT example}

\begin{figure}[h]
\centering
\newbox\boxa
\newbox\boxb
\begin{lrbox}{\boxa}\begin{minipage}{33.0ex}
\lstset{
    classoffset=3,morekeywords={data},
    classoffset=0,
}
\begin{lstlisting}[linewidth=\textwidth,lineskip=.346ex]
// The algebraic data type
data YieldResult[X] =
| ToContinue
| ToReplace(X)
| ToBehead

effect Yield[X] {
  def yield(X) : YieldResult[X]
}

(*\codecomment{The need for a {\tt Behead} effect cannot be}*)
(*\codecomment{easily dismissed: the {\tt iter}\! code has to raise}*)
(*\codecomment{it to the caller and wait for control to}*)
(*\codecomment{come back.}*)
effect Behead {
  def behead() : void
}
\end{lstlisting}
\end{minipage}
 \end{lrbox}
\begin{lrbox}{\boxb}\begin{minipage}{54.5ex}
\lstset{
  morekeywords={skip}
}
\begin{lstlisting}[linewidth=\textwidth]
class Node[X] {
  var head : X
  var tail : Node[X]
  (*\dadada*)
  def iter()(*\;*):(*\;*)void(*\RAISES*)Yield[X](*\;|\;*)Behead {
    match yield(head) {
    | ToContinue ⇒ skip
    | ToReplace(x) ⇒ head = x
    | ToBehead ⇒ behead() // convert ADT value to algebraic effect
    }
    if (tail != null)
      try { tail.iter() }
      with behead() {
        tail = tail.tail
        resume()
      }
  }
}
\end{lstlisting}
\end{minipage}
  \end{lrbox}
\def\figa{\usebox\boxa}
\def\figb{\usebox\boxb}
\def\capa{ADT definition and effect signatures}
\def\capb{
Iterator pattern-matches the result of \texttt{yield}
(cf.\ \cref{fig:interrupt-iter})
}
\savestack{\capfiga}{\subcaptionbox{\capa\label{fig:ADT-sig}}{\figa}}\savestack{\capfigb}{\subcaptionbox{\capb\label{fig:ADT-client}}{\figb}}\capfiga
\hfill
\capfigb
\setlength{\abovecaptionskip}{.5ex}
\caption{
Using an ADT to encode iterator interrupts.
By comparison, bidirectional algebraic effects allow for more
concise code.
\label{fig:ADT}
}
\end{figure}
 \fi

\ifreport\relax\else
    \newpage
\fi
\setlength{\bibsep}{.5ex}
\bibliographystyle{abbrvnat}
\bibliography{bibtex/pm-master}

\appendix
\ifreport
\newpage
\appendix

\section{Static Semantics: A Supplement to Section \ref{sec:static}}
\label{sec:static-semantics-supplement}

\cref{fig:wfty,fig:wfktx}
complement \cref{fig:static-semantics-selected} by presenting
the rest of the static semantics of \lamlang.

\begin{figure}[h]
\judgebox{
    \wfmsig \EffEnv \LifeEnv \LEnv \msig
}
\begin{mathpar}
\Rule{
    \wfmsig
    {\EffEnv,\, \evar}
    \LifeEnv
    \LEnv 
    \msig
}{
    \wfmsig {\EffEnv} \LifeEnv \LEnv {\efunty{\evar}{\msig}}
}

\Rule{
    \wfmsig
    \EffEnv
    {\LifeEnv,\, \lvar}
    \LEnv 
    \msig
}{
    \wfmsig {\EffEnv} \LifeEnv \LEnv {\lfunty{\lvar}{\msig}}
}

\Rule[]{
    \wfty \EffEnv \LifeEnv \LEnv \ty
    \\
    \wfmsig \EffEnv \LifeEnv \LEnv \msig
}{
    \wfmsig {\EffEnv} \LifeEnv \LEnv {\fnty{\ty}{\msig}}
}

\Rule[]{
    \wfty \EffEnv \LifeEnv {\LEnv} {\ty}
    \\
    \wfef \EffEnv \LifeEnv {\LEnv} {\eff}
}{
    \wfmsig {\EffEnv} \LifeEnv \LEnv {\raises{\ty}{\eff}}
}

\end{mathpar}

\judgebox{
    \wfty \EffEnv \LifeEnv \LEnv \ty
}
\begin{mathpar}
\Rule{}{
    \wfty \EffEnv \LifeEnv \LEnv \Unit
}

\Rule{
    \wflife \LifeEnv \LEnv \life
    \\\\
    \forall \eff_0 \in \lst\eff,\,
    \wfef \EffEnv \LifeEnv {\LEnv} {\eff_0}
}{
    \wfty \EffEnv \LifeEnv \LEnv {\itype{\iname}{\lst\eff}{\life}}
}

\Rule[]{
    \wfmsig \EffEnv \LifeEnv \LEnv \msig
    \\\\
    \wflife \LifeEnv \LEnv \life
}{
    \wfty \EffEnv \LifeEnv \LEnv {\mtype{\msig}{\life}}
}

\Rule[]{
    \wfty \EffEnv \LifeEnv \LEnv {\ty_1}
    \\
    \wfef \EffEnv \LifeEnv \LEnv {\eff[1]}
    \\\\
    \wfty \EffEnv \LifeEnv \LEnv {\ty_2}
    \\
    \wfef \EffEnv \LifeEnv \LEnv {\eff[2]}
}{
    \wfty \EffEnv \LifeEnv \LEnv
    {\tycont{\ty_1}{\eff[1]}{\ty_2}{\eff[2]}}
}

\end{mathpar}

\judgebox{
    \wfef \EffEnv \LifeEnv \LEnv {\eff}
}
\begin{mathpar}
\Rule{}{
    \wfef \EffEnv \LifeEnv \LEnv {\nil}
}

\Rule{
    \wfef \EffEnv \LifeEnv \LEnv {\eff}
    \\
    \evar \in \EffEnv
}{
    \wfef \EffEnv \LifeEnv \LEnv {\eff,\,\evar}
}

\Rule{
    \wfef \EffEnv \LifeEnv \LEnv {\eff}
    \\
    \wflife \LifeEnv \LEnv {\life}
}{
    \wfef \EffEnv \LifeEnv \LEnv {\eff,\,\life}
}

\end{mathpar}

\judgebox{
    \wflife \LifeEnv \LEnv {\life}
}
\begin{mathpar}
\Rule{
    \lvar \in \LifeEnv
}{
    \wflife \LifeEnv \LEnv \lvar
}

\Rule{
    \lbl \in \dom{\LEnv}
}{
    \wflife \LifeEnv \LEnv \lbl
}

\end{mathpar}

\judgebox{
    \wflenv \EffEnv \LifeEnv {\LEnv}
}
\begin{mathpar}
\Rule{}{
    \wflenv \EffEnv \LifeEnv \nil
}

\Rule{
    \wflenv \EffEnv \LifeEnv {\LEnv}
    \\
    \wfty \EffEnv \LifeEnv \LEnv \ty
    \\
    \wfef \EffEnv \LifeEnv {\LEnv} {\eff}
}{
    \wflenv \EffEnv \LifeEnv {\LEnv,\,\lbl\hasty\raises{\ty}{\eff}}
}

\end{mathpar}

\judgebox{
    \wftmenv \EffEnv \LifeEnv {\LEnv} \TmEnv
}
\begin{mathpar}
\Rule{}{
    \wftmenv \EffEnv \LifeEnv {\LEnv} \nil
}

\Rule{
    \wftmenv \EffEnv \LifeEnv {\LEnv} \TmEnv
    \\
    \wfty \EffEnv \LifeEnv \LEnv \ty
}{
    \wftmenv \EffEnv \LifeEnv \LEnv {\TmEnv,\,\x\hasty\ty}
}

\end{mathpar}
 \judgebox{
  \subty \EffEnv \LifeEnv \LEnv {\ty_1} {\ty_2}
}
\begin{mathpar}
\Rule{}{
    \subty \EffEnv \LifeEnv \LEnv \Unit \Unit
}

\Rule{}{
    \subty \EffEnv \LifeEnv \LEnv
    {\itype{\iname}{\lst\eff}{\life}}
    {\itype{\iname}{\lst\eff}{\life}}
}

\Rule{
    \subty \EffEnv \LifeEnv \LEnv {\msig_1} {\msig_2}
}{
    \subty \EffEnv \LifeEnv \LEnv
    {\mtype{\msig_1}{\life}}
    {\mtype{\msig_2}{\life}}
}

\Rule{
    \subty \EffEnv \LifeEnv \LEnv {\ty_{21}} {\ty_{11}}
    \\
    \subeff \EffEnv \LifeEnv \LEnv {\eff_{21}} {\eff_{11}}
    \\
    \subty \EffEnv \LifeEnv \LEnv {\ty_{12}} {\ty_{22}}
    \\
    \subeff \EffEnv \LifeEnv \LEnv {\eff_{12}} {\eff_{22}}
}{
    \subty \EffEnv \LifeEnv \LEnv
    {\tycont{\ty_{11}}{\eff_{11}}{\ty_{12}}{\eff_{12}}}
    {\tycont{\ty_{21}}{\eff_{21}}{\ty_{22}}{\eff_{22}}}
}

\Rule{
  \subty \EffEnv \LifeEnv \LEnv {\ty_1} {\ty_2}
  \\
  \subty \EffEnv \LifeEnv \LEnv {\ty_2} {\ty_3}
}{
  \subty \EffEnv \LifeEnv \LEnv {\ty_1} {\ty_3}
}

\end{mathpar}

\judgebox{
  \subty \EffEnv \LifeEnv \LEnv {\msig_1} {\msig_2}
}
\begin{mathpar}
\Rule{
    \subty {\EffEnv,\evar} \LifeEnv \LEnv {\msig_1} {\msig_2}
}{
    \subty \EffEnv \LifeEnv \LEnv
    {\efunty{\evar}{\msig_1}}
    {\efunty{\evar}{\msig_2}}
}

\Rule{
    \subty \EffEnv {\LifeEnv,\lvar} \LEnv {\msig_1} {\msig_2}
}{
    \subty \EffEnv \LifeEnv \LEnv
    {\lfunty{\lvar}{\msig_1}}
    {\lfunty{\lvar}{\msig_2}}
}

\Rule{
    \subty \EffEnv \LifeEnv \LEnv {\ty_2} {\ty_1}
    \\
    \subty \EffEnv \LifeEnv \LEnv {\msig_1} {\msig_2}
}{
    \subty \EffEnv \LifeEnv \LEnv
    {\fnty{\ty_1}{\msig_1}}
    {\fnty{\ty_2}{\msig_2}}
}

\Rule{
    \subty \EffEnv \LifeEnv \LEnv {\ty_{1}} {\ty_{2}}
    \\
    \subeff \EffEnv \LifeEnv \LEnv {\lst{\ef_{1}}} {\lst{\ef_{2}}}
}{
    \subty \EffEnv \LifeEnv \LEnv
    {\raises{\ty_{1}}{\lst{\ef_{1}}}}
    {\raises{\ty_{2}}{\lst{\ef_{2}}}}
}

\Rule{
  \subty \EffEnv \LifeEnv \LEnv {\msig_1} {\msig_2}
  \\
  \subty \EffEnv \LifeEnv \LEnv {\msig_2} {\msig_3}
}{
  \subty \EffEnv \LifeEnv \LEnv {\msig_1} {\msig_3}
}

\end{mathpar}

\judgebox{
  \subeff \EffEnv \LifeEnv \LEnv {\eff[1]} {\eff[2]}
}
\begin{mathpar}
\Rule{
    \forall \ef \in {\eff[1]},\, \ef \in {\eff[2]}
}{
    \subeff \EffEnv \LifeEnv \LEnv {\eff[1]} {\eff[2]}
}

\end{mathpar}
 \caption{Type-level well-formedness and orderings\vspace{-3em}}
\label{fig:wfty}
\end{figure}

\begin{figure}
\judgebox{
  \wfktx{\EffEnv}{\LifeEnv}{\TmEnv}{\LEnv}{\EC}{\ty_1}{\eff[1]}{\ty_2}{\eff[2]}
}
\begin{mathpar}
\Rule{}{
\wfktx{\EffEnv}{\LifeEnv}{\TmEnv}{\LEnv}{\hole}{\ty}{\eff}{\ty}{\eff}
}

\Rule{
    \wfktx{\EffEnv}{\LifeEnv}{\TmEnv}{\LEnv}{\EC}{\ty'}{\eff[']}{
        \itype{\iname}{\lst\eff}{\lbl}
    }{
        \eff''
    }
    \\\\
    \msigof{\iname} = \MkISig{\iname}{\lst{\evar}}{\msig}
}{
    \wfktx{\EffEnv}{\LifeEnv}{\TmEnv}{\LEnv}{
        \unroll{\EC}
    }{\ty'}{\eff[']}{
        \mtype{\subst{\msig}{\lst\eff}{\lst{\evar}}}{\lbl}
    }{
        \eff''
    }
}

\Rule{
    \wfktx{\EffEnv}{\LifeEnv}{\TmEnv}{\LEnv}{\EC}{\ty'}{\eff[']}{
        \mtype{\efunty{\evar}{\msig}}{\lbl}
    }{
        \eff[2]
    }
    \\\\
    \wfef{\EffEnv}{\LifeEnv}{\LEnv}{\eff[1]}
}{
    \wfktx{\EffEnv}{\LifeEnv}{\TmEnv}{\LEnv}{
        \eapp{\EC}{\eff[1]}
    }{\ty'}{\eff[']}{
        \mtype{
            \subst{
                \msig
            }{\eff[1]}{\evar}
        }{\lbl}
    }{
        \eff[2]
    }
}

\Rule{
    \wfktx{\EffEnv}{\LifeEnv}{\TmEnv}{\LEnv}{\EC}{\ty'}{\eff[']}{
        \mtype{\efunty{\lvar}{\msig}}{\lbl}
    }{
        \eff
    }
    \\\\
    \wflife{\LifeEnv}{\LEnv}{\life}
}{
    \wfktx{\EffEnv}{\LifeEnv}{\TmEnv}{\LEnv}{
        \eapp{\EC}{\life}
    }{\ty'}{\eff[']}{
        \mtype{
            \subst{
                \msig
            }{\life}{\lvar}
        }{\lbl}
    }{
        \eff
    }
}

\Rule{
    \wfktx{\EffEnv}{\LifeEnv}{\TmEnv}{\LEnv}{\EC}{\ty'}{\eff[']}{
        \mtype{\fnty{\ty}{\msig}}{\lbl}
    }{
        \eff
    }
    \\\\
    \wftm{\EffEnv}{\LifeEnv}{\TmEnv}{\LEnv}{
        \tm[s]
    }{
        \ty
    }{
        \eff
    }
}{
    \wfktx{\EffEnv}{\LifeEnv}{\TmEnv}{\LEnv}{
        \app{\EC}{\tm[s]}
    }{\ty'}{\eff[']}{
        \mtype{\msig}{\lbl}
    }{
        \eff
    }
}

\Rule{
    \wftm{\EffEnv}{\LifeEnv}{\TmEnv}{\LEnv}{
        \val
    }{
        \mtype{\fnty{\ty}{\msig}}{\lbl}
    }{
        \nil
    }
    \\\\
    \wfktx{\EffEnv}{\LifeEnv}{\TmEnv}{\LEnv}{\EC}{\ty'}{\eff[']}{
        \ty
    }{
        \eff
    }
}{
    \wfktx{\EffEnv}{\LifeEnv}{\TmEnv}{\LEnv}{
        \app{\val}{\EC}
    }{\ty'}{\eff[']}{
        \mtype{\msig}{\lbl}
    }{
        \eff
    }
}

\Rule{
    \wfktx{\EffEnv}{\LifeEnv}{\TmEnv}{\LEnv}{\EC}{\ty'}{\eff[']}{
        \mtype{\raises{\ty}{\eff[1]}}{\lbl}
    }{
        \eff[2]
    }
}{
    \wfktx{\EffEnv}{\LifeEnv}{\TmEnv}{\LEnv}{
        \up{\EC}
    }{\ty'}{\eff[']}{
        \ty
    }{
       \eff[1],\,\eff[2],\,\lbl
    }
}

\Rule{
    \wfktx{\EffEnv}{\LifeEnv}{\TmEnv}{\LEnv}{\EC}{\ty'}{\eff[']}
    {\ty[S]}{\eff}
    \\\\
    \wftm{\EffEnv}{\LifeEnv}{\TmEnv,\,\x\hasty\ty[S]}{\LEnv}
    {\tm}
    {\ty}{\eff}
}{
    \wfktx{\EffEnv}{\LifeEnv}{\TmEnv}{\LEnv}
    {\letexp{\x}{\EC}{\tm}}
    {\ty'}{\eff[']}
    {\ty}{\eff}
}

\Rule{
    \wfktx{\EffEnv}{\LifeEnv}{\TmEnv}{\LEnv}{\EC}{\ty'}{\eff[']}
    {\tycont{\ty_1}{\eff[1]}{\ty_2}{\eff[2]}}{\eff[2]}
    \\
    \wftm{\EffEnv}{\LifeEnv}{\TmEnv}{\LEnv}
    {\tm[s]}
    {\ty_1}{\eff[1]}
}{
    \wfktx{\EffEnv}{\LifeEnv}{\TmEnv}{\LEnv}
    {\throw{\EC}{\tm[s]}}
    {\ty'}{\eff[']}
    {\ty_2}{\eff[2]}
}

\end{mathpar}
 \caption{Well-formed evaluation contexts in \lamlang}
\label{fig:wfktx}
\end{figure}

\clearpage

\section{Logical-Relations Model: A Supplement to Section \ref{sec:logrel}}
\label{sec:logical-relations-supplement}

\cref{fig:tmrel}
complements \cref{fig:semtypes} by presenting
the rest of the logical-relations definitions.

\begin{figure}[h]
\newcommand\LABEL[1]{\text{\itshape#1}}
\judgebox{\LABEL{Observational Refinement}}
\begin{align*}
\begin{array}{@{}r@{\ \ \ }c@{\ \ \ }l@{}}
\inTerR{\W}{\tm_1}{\tm_2}{\ObsR}
& \defeq &
\left(
    \exists \W['], \val_1, \val_2.\ 
    \W['1] = \W[1] \AND
    \tm_1 = \val_1 \AND
    \stepsrt{\W[2]}{\tm_2}{\W['2]}{\val_2}
\right)
\OR
\VSep & &
\left(
    \exists \W['], \tm_1'.\ 
    \W['2] = \W[2] \AND
    \steps{\W[1]}{\tm_1}{\W['1]}{\tm_1'}
    \AND
    \later{\inTerR{\W[']}{\tm_1'}{\tm_2}{\ObsR}}
\right)
\end{array}
\end{align*}

\judgebox{\LABEL{Logical Relation on Closed Terms (via Biorthogonality)}}
\begin{align*}
\begin{array}{@{}r@{\ \ \ }c@{\ \ \ }l@{}}
\inTerR
    {\W}
    {\tm_1}
    {\tm_2}
    {
        \TmR
        {\ty}{\eff}{\lsub}{\esub}
    }
& \defeq &
\forall \EC_1,
        \EC_2.\ 
\inTerR
    {\W}
    {\EC_1}
    {\EC_2}
    {
        \ContR
        {\ty}{\eff}{\lsub}{\esub}
    }
\implies
\inTerR
    {\W}
    {\EC_1[\tm_1]}
    {\EC_2[\tm_2]}
    {\ObsR}
\VSEP
\inTerR
    {\W}
    {\EC_1}
    {\EC_2}
    {
        \ContR
        {\ty}{\eff}{\lsub}{\esub}
    }
& \defeq &
\forall \W['].\ 
\extW{\W}{\W[']}
\implies
\VSep & &
\left(
\forall \val_1, \val_2.\ 
\inTerR
    {\W[']}
    {\val_1}
    {\val_2}
    {
        \SemTy
        {\ty}{\lsub}{\esub}
    }
\implies
\inTerR
    {\W[']}
    {\EC_1[\val_1]}
    {\EC_2[\val_2]}
    {\ObsR}
\right)
\AND
    \VSep & &
\left(
\forall \tm_1, \tm_2.\ 
\inTerR
    {\W[']}
    {\tm_1}
    {\tm_2}
    {
        \STmR
        {\ty}{\eff}{\lsub}{\esub}
    }
\implies
\inTerR
    {\W[']}
    {\EC_1[\tm_1]}
    {\EC_2[\tm_2]}
    {\ObsR}
\right)
\VSEP
\inTerR
    {\W}
    {\appctx{\EC_1}{\tm_1}}
    {\appctx{\EC_2}{\tm_2}}
    {\STmR{\ty}{\eff}{\lsub}{\esub}}
& \defeq &
\exists \rel,
        \lst{\lbl[1]},
        \lst{\lbl[2]}.\ 
\inSenR
    {\W}
    {\tm_1}
    {\tm_2}
    {\rel}
    {\lst{\lbl[1]}}
    {\lst{\lbl[2]}}
    {\SemEff{\eff}{\lsub}{\esub}}
\AND
\left(
\forall j.\
\tunnels {\lsti[j]{\lbl[i]}} {\EC_i}
\right)
\onetwo
\AND
\VSep & &
\forall \W['], \tm_1', \tm_2'.\ 
\extW{\W}{\W[']}
\implies
\inTerR
    {\W[']}
    {\tm_1'}
    {\tm_2'}
    {\rel}
\implies
\inTerR
    {\W[']}
    {\appctx{\EC{}_1}{\tm_1'}}
    {\appctx{\EC{}_2}{\tm_2'}}
    {
        \later{
            \TmR{\ty}{\eff}{\lsub}{\esub}
        }
    }
\end{array}
\end{align*}
 \def\tightcomma{\hskip -2pt,\hskip -1pt}

\judgebox{\LABEL{Logical Relation on Open Terms}}
\begin{gather*}
\begin{array}{@{}rcl@{}}
\logref{\EffEnv}{\LifeEnv}{\TmEnv}{\LEnv}{\tm_1}{\tm_2}{\ty}{\eff}
& \defeq &
\forall \W,\esub,\lsub,\vsub.\ 
\extW{\dom{\LEnv}}{\W[i]}\onetwo
\implies
\VSep & &
\inBinR{\W}{\esub}{\SemEffEnv{\EffEnv}}
\implies
\inBinR{\W}{\lsub}{\SemEffEnv{\LifeEnv}}
\implies
\inBinR{\W}{\vsub}{\SemTmEnv{\TmEnv}{\lsub}{\esub}}
\implies
\VSep & &
\inTerR{\W}
    {\esubst{1}{\lsubst{1}{\vsubst{1}{\tm_1}}}}
    {\esubst{2}{\lsubst{2}{\vsubst{2}{\tm_2}}}}
    {\TmR{\ty}{\eff}{\lsub}{\esub}}    
\VSEP
\logrefk{\EffEnv}{\LifeEnv}{\TmEnv}{\LEnv}{\EC_1}{\EC_2}
{\raises{\ty}{\eff}}{\raises{\ty'}{\eff[']}}
& \defeq &
\forall \W,\esub,\lsub,\vsub.\ 
\extW{\dom{\LEnv}}{\W[i]}\onetwo
\implies
\VSep & &
\inBinR{\W}{\esub}{\SemEffEnv{\EffEnv}}
\implies
\inBinR{\W}{\lsub}{\SemEffEnv{\LifeEnv}}
\implies
\inBinR{\W}{\vsub}{\SemTmEnv{\TmEnv}{\lsub}{\esub}}
\implies
\VSep & &
\inTerR{\W}
    {\esubst{1}{\lsubst{1}{\vsubst{1}{\EC_1}}}}
    {\esubst{2}{\lsubst{2}{\vsubst{2}{\EC_2}}}}
    {\DContR{\ty}{\eff}{\ty'}{\eff[']}{\lsub}{\esub}}    
\end{array}
\end{gather*}

\begin{gather*}
\begin{array}
{@{}r@{\ \ }c@{\ \ }l@{\hspace{1.0em}}r@{\ \ }c@{\ \ }l@{}}
\inBinR{\W}{\esub}{\SemEffEnv{\nil}}
& \defeq &
\esub = \nil
&
\inBinR{\W}
    {\esub}
    {
        \SemEffEnv
        {\EffEnv,\evar}
    }
& \defeq &
\esub =
\esub'\tightcomma
\esubext
    {\evar}
    {\lst{\lbl[1]}}
    {\lst{\lbl[2]}}
    {\rel[\psi]}
\AND
\extW{\rel[\psi]}{\W}
\AND
\inBinR{\W}{\esub'}{\SemEffEnv{\EffEnv}}
\VSEP
\inBinR{\W}{\lsub}{\SemLifeEnv{\nil}}
& \defeq &
\lsub = \nil
&
\inBinR
    {\W}{\lsub}
    {\SemLifeEnv{\LifeEnv,\,\lvar} }
& \defeq &
\lsub =
\lsub'\tightcomma
    \lsubext
        {\lvar}
        {\lbl[1]}
        {\lbl[2]}
        {\rel}
\AND
{\lbl[i]} \in {\W}\onetwo
\AND
\inBinR{\W}{\lsub'}{\SemLifeEnv{\LifeEnv}}
\VSEP
\inBinR{\W}{\vsub}{
    \SemTmEnv
    {\nil}
    {\lsub}
    {\esub}
}
& \defeq &
\vsub = \nil
&
\inBinR
    {\W}{\vsub}
    {
        \SemTmEnv
        {\TmEnv,\x\hasty\ty}
        {\lsub}
        {\esub}
    }
& \defeq &
\vsub = \vsub'\tightcomma\vsubext{\x}{\val_1}{\val_2}
\AND
\inTerR{\W}{\val_1}{\val_2}{
    \SemTy{\ty}{\lsub}{\esub}
}
\AND
\inBinR{\W}{\vsub'}{
    \SemTmEnv{\TmEnv}{\lsub}{\esub}
}
\end{array}
\end{gather*}

 \caption{Observational refinement,
biorthogonal term relation, and its lifting to open terms}
\label{fig:tmrel}
\end{figure}

\clearpage

\section{Contextual Equivalence: A Supplement to Section
\ref{sec:theorems}}
\label{sec:contextual-eq-supplement}

This section defines contextual equivalence precisely.

A program context is a term with a hole in it.
In \lamlang, they are defined as follows:

\noindent
\begin{minipage}{\textwidth}
\begin{align*}
\begin{array}{@{}c@{\ \ }c@{\ \ }l@{}}
\Ctx & ::= &
\hole \bnf
\appctx{\Ctx}{\eapp{\hole}{\eff}} \bnf
\appctx{\Ctx}{\lapp{\hole}{\life}} \bnf
\appctx{\Ctx}{\app{\hole}{\tm}} \bnf
\appctx{\Ctx}{\app{\tm}{\hole}} \bnf
\appctx{\Dtx}{\lam{\x[k]}{\hole}} \bnf
\appctx{\Ctx}{\up{\hole}} \bnf
\appctx{\Ctx}{\DOWN{\lbl}{\hole}}
\VSep & &
\appctx{\Ctx}{\letexp{\x}{\hole}{\tm}} \bnf
\appctx{\Ctx}{\letexp{\x}{\tm}{\hole}} \bnf
\appctx{\Ctx}{\throw{\tm}{\hole}} \bnf
\appctx{\Ctx}{\throw{\hole}{\tm}} \bnf
\VSEP
\Dtx & ::= &
\appctx{\Ctx}{\obj{\fixpoint{\self}{\hole}}{\lbl}} \bnf
\appctx{\Ctx}{\obj{\hole}{\lbl}} \bnf
\appctx{\Dtx}{\lam{\evar}{\hole}} \bnf
\appctx{\Dtx}{\lam{\lvar}{\hole}} \bnf
\appctx{\Dtx}{\lam{\x}{\hole}}
\end{array}
\end{align*}
 \end{minipage}
\\

Judgments of context well-formedness have form 
\wfctx{\Ctx}{\EffEnv}{\LifeEnv}{\TmEnv}{\LEnv}{\ty}{\eff}{\ty[S]},
meaning that filling the hole in~{\Ctx} with a term~{\tm} which
has a typing judgment
{\wftm{\EffEnv}{\LifeEnv}{\TmEnv}{\LEnv}{\tm}{\ty}{\eff}}
results in a program~{\appctx\Ctx\tm} which has typing judgment
{\wftm{\nil}{\nil}{\nil}{\nil}{\appctx\Ctx\tm}{\ty[S]}{\nil}}.

Contextual equivalence \CTXEQ is defined in terms of
contextual refinement \CTXREF, as follows:

\noindent
\begin{minipage}{\textwidth}
\begin{gather*}
\begin{array}{@{}l@{\ \ \ }c@{\ \ \ }l@{}}
\ctxref{\EffEnv}{\LifeEnv}{\TmEnv}{\LEnv}{\tm_1}{\tm_2}{\ty}{\eff}
& \defeq &
\forall \Ctx,\,\ty[S].\ 
\wfctx{\Ctx}{\EffEnv}{\LifeEnv}{\TmEnv}{\LEnv}{\ty}{\eff}{\ty[S]}
\implies
\\ & &
\forall \lst{\lbl_1},\,\val_1.\ 
\stepsrt {\nil} {\appctx{\Ctx}{\tm_1}} {\lst{\lbl_1}} {\val_1}
\implies
\exists \lst{\lbl_2},\,\val_2.\ 
\stepsrt {\nil} {\appctx{\Ctx}{\tm_2}} {\lst{\lbl_2}} {\val_2}
\medskip\\
\ctxeq{\EffEnv}{\LifeEnv}{\TmEnv}{\LEnv}{\tm_1}{\tm_2}{\ty}{\eff}
& \defeq &
\ctxref{\EffEnv}{\LifeEnv}{\TmEnv}{\LEnv}{\tm_1}{\tm_2}{\ty}{\eff}
\AND
\ctxref{\EffEnv}{\LifeEnv}{\TmEnv}{\LEnv}{\tm_2}{\tm_1}{\ty}{\eff}
\end{array}
\end{gather*}
\end{minipage}
\\

\noindent
In this definition, the observable behavior is whether a
computation terminates. This seemingly weak observation power
does not weaken the discriminating power of the definition.
Because \lamlang is Turing-complete, when two terminating
computations step to different values, one can always construct a
program context that exhibits different termination behaviors when
plugged with these computations.

Contextual equivalence is hard to establish directly because of
the universal quantification over contexts.
\cref{thm:soundness} makes it possible to establish contextual
equivalence through logical relatedness.

 \section{ADT example}

\begin{figure}[h]
\centering
\newbox\boxa
\newbox\boxb
\begin{lrbox}{\boxa}\begin{minipage}{33.0ex}
\lstset{
    classoffset=3,morekeywords={data},
    classoffset=0,
}
\begin{lstlisting}[linewidth=\textwidth,lineskip=.346ex]
// The algebraic data type
data YieldResult[X] =
| ToContinue
| ToReplace(X)
| ToBehead

effect Yield[X] {
  def yield(X) : YieldResult[X]
}

(*\codecomment{The need for a {\tt Behead} effect cannot be}*)
(*\codecomment{easily dismissed: the {\tt iter}\! code has to raise}*)
(*\codecomment{it to the caller and wait for control to}*)
(*\codecomment{come back.}*)
effect Behead {
  def behead() : void
}
\end{lstlisting}
\end{minipage}
 \end{lrbox}
\begin{lrbox}{\boxb}\begin{minipage}{54.5ex}
\lstset{
  morekeywords={skip},
}
\begin{lstlisting}[linewidth=\textwidth]
class Node[X] {
  var head : X
  var tail : Node[X]
  (*\dadada*)
  def iter()(*\;*):(*\;*)void(*\RAISES*)Yield[X](*\;|\;*)Behead {
    match yield(head) {
    | ToContinue ⇒ skip
    | ToReplace(x) ⇒ head = x
    | ToBehead ⇒ behead() // convert ADT value to algebraic effect
    }
    if (tail != null)
      try { tail.iter() }
      with behead() {
        tail = tail.tail
        resume()
      }
  }
}
\end{lstlisting}
\end{minipage}
  \end{lrbox}
\def\figa{\usebox\boxa}
\def\figb{\usebox\boxb}
\def\capa{ADT definition and effect signatures}
\def\capb{
Iterator pattern-matches the result of \texttt{yield}
(cf.\ \cref{fig:interrupt-iter})
}
\savestack{\capfiga}{\subcaptionbox{\capa\label{fig:ADT-sig}}{\figa}}\savestack{\capfigb}{\subcaptionbox{\capb\label{fig:ADT-client}}{\figb}}\capfiga
\hfill
\capfigb
\setlength{\abovecaptionskip}{.5ex}
\caption{
Using an ADT to encode iterator interrupts.
By comparison, bidirectional algebraic effects allow for more
concise code.
\label{fig:ADT}
}
\end{figure}
 \fi

 \end{document}